\documentclass[aps,prd,preprintnumbers,superscriptaddress,showpacs,showkeys,floatfix,amsmath,amssymb,amsfonts,twocolumn,nofootinbib]{revtex4-2}

\usepackage{graphicx}
\usepackage{multirow}
\usepackage{xparse} 
\usepackage{float}
\usepackage{xcolor}
\usepackage{makecell}
\usepackage{enumitem}
\usepackage{braket}
\usepackage{dsfont}

\NewDocumentCommand\Nf{mgg}{N\textsubscript{f}=#1\IfNoValueTF{#2}{}{+#2}\IfNoValueTF{#3}{}{+#3}}
\NewDocumentCommand\vol{mg}{#1\textsuperscript{3}\IfNoValueTF{#2}{}{×#2}}

\newcommand{\tins}{t_\mathrm{ins}}

\usepackage[normalem]{ulem}

\usepackage[colorlinks=true,linkcolor=blue,urlcolor=blue,citecolor=blue,anchorcolor=blue]{hyperref}
\usepackage[capitalise]{cleveref}
\Crefname{section}{Sec.}{Secs.}
\crefrangelabelformat{equation}{(#3#1#4--#5#2#6)}
\usepackage{slashed}

\newcommand{\vp}{\vec{p}}
\newcommand{\vpp}{\vec{p}^{\,\prime}}

\newcommand{\JN}{\mathcal{J}_N}
\newcommand{\JNbar}{\bar{\mathcal{J}}_N}
\newcommand{\ts}{t_s}

\begin{document}

\title{Nucleon unpolarized second Mellin moments    using lattice QCD ensembles with  physical quark masses and in the continuum limit}

\author{Constantia Alexandrou} \affiliation{Department of Physics, University of Cyprus, P.O. Box 20537, 1678 Nicosia, Cyprus}\affiliation{Computation-based Science and Technology Research Center, The Cyprus Institute, 20 Kavafi Str., Nicosia 2121, Cyprus}
\author{Simone Bacchio} \affiliation{Computation-based Science and Technology Research Center, The Cyprus Institute, 20 Kavafi Str., Nicosia 2121, Cyprus}
\author{Jacob Finkenrath}\affiliation{Department of Physics, Bergische Universität Wuppertal, Gaußstraße 20, Wuppertal, 42119, Germany}
\author{Christos Iona}\affiliation{Department of Physics, University of Cyprus, P.O. Box 20537, 1678 Nicosia, Cyprus} \affiliation{Computation-based Science and Technology Research Center, The Cyprus Institute, 20 Kavafi Str., Nicosia 2121, Cyprus}
\author{Giannis Koutsou} \affiliation{Computation-based Science and Technology Research Center, The Cyprus Institute, 20 Kavafi Str., Nicosia 2121, Cyprus}
\author{Christian Kummer} \affiliation{Department of Physics, University of Cyprus, P.O. Box 20537, 1678 Nicosia, Cyprus}\affiliation{Technical University of Berlin, Berlin, Germany}
\author{Yan Li} \affiliation{Computation-based Science and Technology Research Center, The Cyprus Institute, 20 Kavafi Str., Nicosia 2121, Cyprus}
\author{Bhavna Prasad}\affiliation{Computation-based Science and Technology Research Center, The Cyprus Institute, 20 Kavafi Str., Nicosia 2121, Cyprus}
\author{Gregoris Spanoudes}\affiliation{Department of Physics, University of Cyprus, P.O. Box 20537, 1678 Nicosia, Cyprus}

\date{\today}

\begin{abstract}
	We compute the matrix elements of the energy-momentum tensor of the nucleon using four ensembles of twisted mass clover-improved fermions with the  up, down, strange and charm quark masses tuned to approximately their physical values. 
The four ensembles have similar physical volume and lattice spacings  $a=0.080$~fm, $0.068$~fm, $0.057$~fm, and $0.049$ fm, allowing us to take the continuum limit directly at the physical pion mass point. 
We compute both connected and disconnected quark contributions as well as gluon contributions. 
All renormalization functions, including the mixing of the quark singlet with the gluon,  are determined  non-perturbatively. We extract the  gravitational form factors  in the continuum limit  at $Q^2=0$ and evaluate the contribution of quarks and gluons to   the momentum and  angular momentum of the proton. Using the values of the intrinsic quark spin computed using the same gauge ensembles we also determine the  orbital angular momentum for each quark flavor.

\centerline{\includegraphics[width=0.2\linewidth]{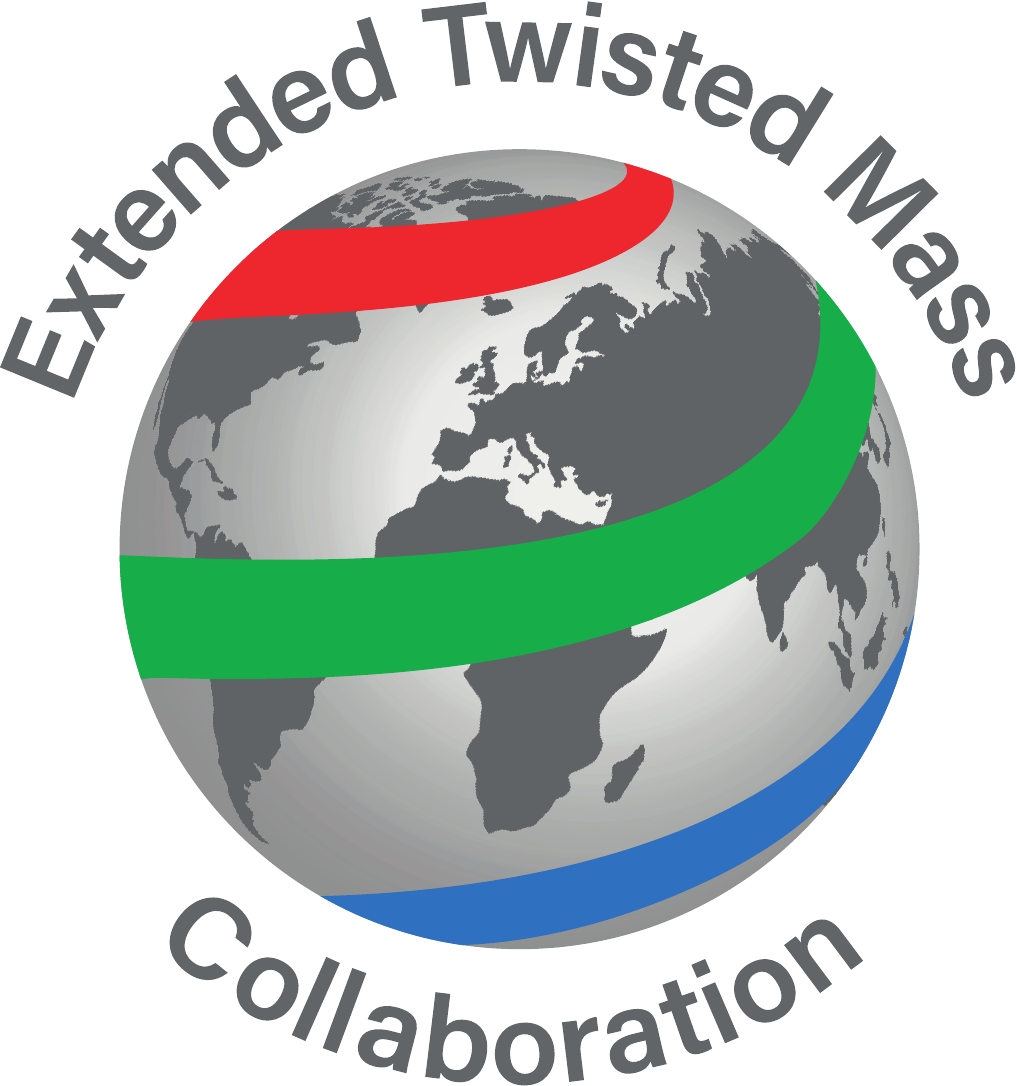}}
\end{abstract}

\maketitle
\section{Introduction}

Understanding the internal structure of the nucleon in terms of its quark and gluon constituents remains one of the central challenges in quantum chromodynamics (QCD)~\cite{Ji:1997gm,Polyakov:2002yz,Diehl:2003ny}. While the proton is composed of quarks and gluons, the manner in which these fundamental degrees of freedom combine to generate its observable properties, including its mass, momentum, and spin, continues to be the subject of extensive theoretical~\cite{Belitsky:2005qn} and experimental investigation~\cite{Aidala:2012mv,STAR:2014wox,Accardi:2012qut,SpinMuon:1998eqa,HERMES:2006jyl,COMPASS:2016xvm}. A particularly powerful framework for addressing these questions is provided by the matrix elements of the QCD energy--momentum tensor (EMT), whose matrix elements encode essential information about the mechanical and dynamical structure of hadrons~\cite{Shanahan:2018nnv,Yang:2018nqn}.

The matrix elements of the symmetric, gauge-invariant energy--momentum tensor determine the gravitational form factors of the nucleon. These form factors are accessible through  lattice QCD calculations and phenomenological analyses. At zero momentum transfer, the forward matrix elements of the quark and gluon components of the EMT provide direct access to the momentum fraction carried by each constituent, while the corresponding generalized form factors satisfy Ji's sum rule~\cite{Ji:1997pf}, relating them to the  angular momentum carried by quarks and gluons. Together, these quantities offer a rigorous decomposition of the nucleon momentum and spin within QCD. 

Over the past decade, significant progress has been achieved in determining the generalized form factors from lattice QCD using simulations at or near the physical pion mass. Calculations have addressed both the connected and disconnected quark contributions, leading to increasingly precise determinations of the  momentum fraction  and angular momentum~\cite{Alexandrou:2017oeh,Alexandrou:2020okk,Bali:2018zgl,Hackett:2023rif}. The gluon contribution remains considerably more challenging because of the poor signal-to-noise ratio of gluonic operators and the necessity of treating quark--gluon operator mixing under renormalization~\cite{Alexandrou:2016ekb,Shanahan:2018pib}.

A complete understanding of the nucleon structure requires resolving not only the total quark contribution but also its decomposition into individual quark flavor. The flavor dependence of the EMT matrix elements reveals how the up, down, strange, and heavier quarks share the nucleon momentum and angular momentum, providing valuable insight into the interplay between valence and sea quarks. Likewise, determining the gluon contribution is essential for establishing the complete momentum and spin budgets of the nucleon and for understanding the role of gluonic degrees of freedom in nonperturbative QCD. These decompositions are also of direct relevance to global parton distribution function analyses and to precision studies at current and future facilities, including the Electron-Ion Collider.

From a theoretical perspective, the renormalization of the quark and gluon energy-momentum tensors presents additional challenges. Since the corresponding operators mix under renormalization, a consistent determination of their individual matrix elements requires careful treatment of operator mixing and the associated renormalization constants. This is particularly important in lattice QCD, where discretization effects and the construction of improved operators influence the extraction of physical observables. Achieving a precise flavor and gluon decomposition therefore relies on both accurate nonperturbative calculations of the relevant matrix elements and a robust renormalization procedure.

In this work, we present a lattice QCD determination of the nucleon matrix elements of the quark and gluon energy-momentum tensors using four gauge ensembles generated with \Nf{2}{1}{1} twisted-mass clover-improved fermions at the physical pion mass. This enables the first extrapolation of these quantities to the continuum limit directly at physical pion mass, eliminating any systematic errors due to the chiral extrapolation. We determine the generalized form factors of the individual quark flavours and  the gluon, taking into account the renormalization and mixing of the singlet quark and gluon operators fully nonperturbatively. From these matrix elements, we extract the quark flavor and gluon contributions to the nucleon momentum and  angular momentum, and examine the corresponding momentum and spin sum rules. These results provide a comprehensive first-principles determination of the quark and gluon decomposition of the nucleon and contribute to a quantitative understanding of its internal structure.

The rest of the paper is structured as follows: Sec.~\ref{sec:matrix_elements} introduces the matrix elements of the energy-momentum tensor, Sec.~\ref{sec:lattice_setup} gives our lattice ensembles and procedure, and Secs.~\ref{sec:analysis 2pt} and \ref{sec:bgff} describe the analysis of two- and three-point functions for the extraction of the bare generalized form factors (GFFs) $A_{20}(Q^2)$ and $B_{20}(Q^2)$ including a discussion of the $Q^2$ dependence of the GFFs and the extrapolation to $Q^2=0$. Sec.~\ref{sec:renormalization} details the computation of the renormalization functions including the mixing between quark singlet and gluon,  Sec.~\ref{sec:continuum} discusses the continuum limit and the various systematics, and Secs.~\ref{sec:discussion} and \ref{sec:conclusion}  present our results, and provide our summary and conclusions, respectively.

\section{Matrix elements of the energy-momentum tensor}
\label{sec:matrix_elements}
The quark and gluon contributions to the nucleon gravitational form factors  can be evaluated from the matrix elements of the QCD  energy-momentum tensor. 
We consider the traceless symmetric gauge-invariant EMT $T^{\mu\nu}$, decomposed into quark and gluon contributions as 
\begin{align}\label{eq:EMT}
    T^{\mu\nu}=T^{\mu\nu}_q+T^{\mu\nu}_g =  \bar{\psi}i\gamma^{\{\mu} \overleftrightarrow{D}^{\nu\}} \psi  + F^{\rho\{\mu} F^{\nu\}}_{\ \rho} \,,
\end{align}
where $F^{\mu\nu}$ is the gluon field-strength tensor and the notation $\{\cdots\}$ means symmetrization over the indices in the curly brackets  and subtraction of the trace. Summation over color indices is implied. The symmetrized  covariant derivative $\overleftrightarrow{D}$ is defined as $(\overrightarrow{D}-\overleftarrow{D})/2$. 

Going to Euclidean space and following the conventions of Ref.~\cite{Hagler:2003jd}, the nucleon matrix element of  EMT is parameterized in terms of the three GFFs as 
\begin{align}\label{eq:me2ff}
& \braket{N(p^\prime, s^\prime) | T^{\mu\nu}_{q,g} | N(p, s) } = 
\bar{u}_N(p^\prime, s^\prime) \Bigg[
{i}A^{q,g}_{20}(Q^2)\, \gamma^{\{\mu} P^{\nu\}} \nonumber\\
&+B^{q,g}_{20}(Q^2)\, \frac{P^{\{\mu} \sigma^{\nu\}\rho} q_\rho}{2m_N} 
+\, C^{q,g}_{20}(Q^2)\, \frac{q^{\{\mu} q^{\nu\}}}{m_N}
\Bigg] u_N(p, s),
\end{align}
where $N(p,s)$ denotes the nucleon state with initial (final) momentum $p$ ($p^\prime$) and spin $s$ ($s^\prime$), $u_N$ is the Euclidean nucleon spinor, $m_N$ is the nucleon mass, $P=(p^\prime+p)/2$, $q=p-p^\prime$, and $Q^2=-q^2$.

In the forward limit, the form factor $A^{q,g}_{20}(0)$ yields the momentum fraction carried by quarks and gluons,
\begin{align}
\braket{x}_{q,g} = A^{q,g}_{20}(0),
\end{align}
which should satisfy the momentum sum rule $\braket{x}_q+\braket{x}_g=1$. The total spin carried by quarks and gluons is obtained via \cite{Ji:1996ek}
\begin{align}\label{eq:J}
J_{q,g} = \frac{1}{2}\left[A^{q,g}_{20}(0) + B^{q,g}_{20}(0)\right].
\end{align}

The quark spin $J_q$ can be further decomposed into the sum of the intrinsic quark spin $\frac{1}{2}\Delta \Sigma_q$ and the orbital angular momentum $L_q$ \cite{Ji:1996ek} as
\begin{align}
J_q = \frac{1}{2}\Delta \Sigma_q + L_q.
\end{align}
The intrinsic quark spin contribution is obtained from the axial charge $g_A^q$ via
\begin{align}
	\frac{1}{2}\Delta \Sigma_q=\frac{1}{2} g_A^q \,,
\end{align}
and the orbital angular momentum can then be determined from
\begin{align}
	L_q = J_q - \frac{1}{2}\Delta \Sigma_q.
\end{align}

\section{Lattice setup}\label{sec:lattice_setup}

We use twisted mass clover-improved fermion ensembles simulated with two degenerate up and down quarks, a strange quark and a charm quark (\Nf{2}{1}{1}), with quark masses tuned approximately to their physical values. This formalism allows for automatic ${\cal O}(a)$ improvement without requiring further improvement of the operators resulting in physical observables~\cite{Frezzotti:2000nk,Frezzotti:2003ni}. The parameters of the  ensembles are given in \cref{tab:ens}.

\begin{table}[h]
	\caption{Parameters of the four \Nf{2}{1}{1} ensembles analyzed in
		this work. From the leftmost to the rightmost columns, we provide
		the name of the ensemble and its abbreviation, the lattice volume,  the lattice spacing, the pion
		mass, and the value of $m_\pi L$. Lattice spacings and pion
		masses are taken from
Refs.~\cite{ExtendedTwistedMass:2022jpw,ExtendedTwistedMass:2024nyi}.}
	\label{tab:ens}
	\centering
	\begin{tabular}{ccccccc}
		\hline\hline
		Ensemble          & $(\frac{L}{a})^3{\times}(\frac{T}{a})$ &  \makecell[c]{$a$                   \\$[$fm$]$} & \makecell[c]{$m_\pi$\\ $[$MeV$]$}  & $m_\pi L$ \\
		\hline
		\texttt{cB211.072.64} (B64) & $64^3 {\times} 128$                      & 0.07948(11)      & 140.2(2) & 3.62 \\
		\texttt{cC211.060.80} (C80) & $80^3 {\times} 160$                      & 0.06819(14)      & 136.7(2) & 3.78 \\
		\texttt{cD211.054.96} (D96) & $96^3 {\times} 192$                       & 0.056850(90)      & 140.8(2) & 3.90 \\
		\texttt{cE211.044.112} (E112)& $112^3 {\times} 224$                      & 0.04892(11)      & 136.5(2) & 3.79 \\
		\hline
	\end{tabular}
\end{table}

In order to determine the nucleon matrix elements of  EMT, we compute the nucleon two- and three-point functions
\begin{align}
    &\quad C_{\rm 2pt}(\vp,t)= \text{Tr}\left[\Gamma_0 \braket{\JN(\vp,t) \JNbar(\vp,0)}\right] \,, \\
	    &\quad C_{\rm 3 pt}^{\mu\nu}(\Gamma_\alpha,\vpp,\vp,\ts,\tins)
	= \nonumber \\
    & \qquad \qquad \quad \text{Tr}\left[\Gamma_\alpha\braket{\JN(\vpp,t_s)\, T^{\mu\nu}(\vec{q},t_{\rm ins}) \JNbar(\vp,0) } \right] \,,
\end{align}
where $\Gamma_0=\frac{1}{2}(1+\gamma_4)$ and $\Gamma_k=i\gamma_5\gamma_k\Gamma_0$.
The nucleon interpolating field is given by
\begin{align}
    \JN(\vec{x},t) = \epsilon_{abc} \,  [u^{a}(\vec{x},t)^\top \, \mathcal{C}\gamma_5 \,  d^b(\vec{x},t)]\, u^c(\vec{x},t) \,.
\end{align}
We apply Gaussian smearing~\cite{Gusken:1989ad,Alexandrou:1992ti} to the quark fields $\psi=u,d$ entering the nucleon interpolating field,
\begin{align}
  \psi^a_{\rm smear}(\vec{x};t)
  &=
  \sum_{\vec{y}}
  F^{ab}(\vec{x},\vec{y}; t ) \,
  \psi^b(\vec{y},t),
    \label{eq:smear}
\end{align}
with smearing kernel
\begin{align}
    F
    &=
    (\mathds{1}+\alpha_G\,H)^{N_G},
    \label{eq:hopping} \\
    H(\vec{x},\vec{y}; t )
    &=
    \sum_{i=1}^{3}
    \left[
    U_i(\vec{x}, t)\, \delta_{ \vec{x},\vec{y}-\hat{i}}
    +
    U^\dagger_i(\vec{x}-\hat{i}, t)\, \delta_{\vec{x},\vec{y}+\hat{i}}
    \right].
    \nonumber
\end{align}
The smearing parameters are tuned to approximately reproduce a nucleon root mean square radius of $0.5~{\rm fm}$. For the gauge links entering the hopping matrix $H$ in \cref{eq:hopping}, we apply APE smearing~\cite{APE:1987ehd} in order to reduce ultraviolet fluctuations and improve the statistical signal.
The smearing parameters are summarized in \cref{tab:smearing}.

\begin{table}[!ht]
    \caption{Number of Gaussian smearing iterations, $N_G$, and Gaussian smearing parameter $a_G$ used for each ensemble. We also provide the number of APE-smearing iterations $n_{\mathrm{APE}}$ and parameter $\alpha_{\mathrm{APE}}$ applied to the links that enter the Gaussian smearing hopping matrix.
    }\label{tab:smearing}
    \centering
    \begin{ruledtabular}
    \begin{tabular}{ccccc}
Ensemble & $N_G$ & $\alpha_G$ & $n_{\mathrm{APE}}$ & $\alpha_{\mathrm{APE}}$ \\
\hline
B64   & 125 & 0.2 & 50 & 0.5 \\
C80   & 140 & 1.0 & 60 & 0.5 \\
D96   & 200 & 1.0 & 60 & 0.5 \\
E112  & 250 & 1.0 & 60 & 0.5 \\
    \end{tabular}
    \end{ruledtabular}
\end{table}

For the gauge links entering the gluon EMT operator $T_g^{\mu\nu}$, we apply stout smearing~\cite{Morningstar:2003gk} with parameter $\rho=0.129$~\cite{Alexandrou:2016ekb}. Different choices of the number of stout-smearing steps, $n_s$, correspond to different lattice definitions of the gluon EMT operator and may therefore lead to different finite lattice spacing   artifacts. These differences are expected to vanish after renormalization and in the continuum limit. We will investigate this dependence using $n_s=7,10,13,15,$ and $20$.

\begin{table*}[t]
\centering

\setlength{\tabcolsep}{3pt}
\renewcommand{\arraystretch}{1.1}
\caption{
 We give per ensemble listed in the first column, the number of gauge configurations 
$N_{\rm conf}$, the number of source positions used for the nucleon two-point functions, $n_{\rm src}^{\rm 2pt}$, the source-sink separations  in lattice units, $t_s/a$, and physical units, $t_s$, together with the number of source positions used for the connected three-point functions, $n_{\rm src}^{\rm conn}$.
For the E112 ensemble, we use  501 configurations with the exceptions of the connected three-point functions with $t_s/a<29$, for which 225 configurations are used. }\label{tab:stats}
\begin{tabular}{cccccc}
\hline\hline
Ensemble &
$N_{\rm conf}$ &
$n_{\rm src}^{\rm 2pt}$ &
$t_s/a$ &
$t_s$ [fm] &
$n_{\rm src}^{\rm conn}$ \\
\hline
B64 & 725 & 349 &
\begin{tabular}{@{}c@{}}
8, 10, 12, 14\\
16, 18, 20
\end{tabular} &
\begin{tabular}{@{}c@{}}
0.64, 0.80, 0.96, 1.12\\
1.28, 1.44, 1.60
\end{tabular} &
\begin{tabular}{@{}c@{}}
1, 2, 5, 10\\
32, 112, 128
\end{tabular} \\
\hline
C80 & 400 & 650 &
\begin{tabular}{@{}c@{}}
6, 8, 10, 12, 14\\
16, 18, 20, 22
\end{tabular} &
\begin{tabular}{@{}c@{}}
0.41, 0.55, 0.69, 0.82, 0.96\\
1.10, 1.24, 1.37, 1.51
\end{tabular} &
\begin{tabular}{@{}c@{}}
1, 2, 4, 10, 22\\
48, 45, 116, 246
\end{tabular} \\
\hline
D96 & 493 & 368 &
\begin{tabular}{@{}c@{}}
8, 10, 12, 14, 16\\
18, 20, 22, 24, 26
\end{tabular} &
\begin{tabular}{@{}c@{}}
0.46, 0.57, 0.68, 0.80, 0.91\\
1.03, 1.14, 1.25, 1.37, 1.48
\end{tabular} &
\begin{tabular}{@{}c@{}}
1, 2, 4, 8, 16\\
32, 64, 16, 32, 64
\end{tabular} \\
\hline
E112 & $501$ & 311 &
\begin{tabular}{@{}c@{}}
8, 11, 14, 17\\
20, 23, 26, 29
\end{tabular} &
\begin{tabular}{@{}c@{}}
0.39, 0.54, 0.69, 0.83\\
0.98, 1.13, 1.27, 1.42
\end{tabular} &
\begin{tabular}{@{}c@{}}
1, 2, 4, 8\\
16, 32, 64, 128
\end{tabular} \\
\hline\hline
\end{tabular}
\end{table*}

\begin{table*}
\centering

\setlength{\tabcolsep}{3pt}
\renewcommand{\arraystretch}{1.1}
\caption{
For the computation of the quark disconnected loops, we give,  for each ensemble listed in the first row and flavor given in the first column,   the number of deflated modes $N_{\rm def}$ and the number of vectors inverted per configuration, $n_{\rm vec}=n_{\rm dil}N_{\rm Had}N_r$, where $n_{\rm dil}=12$ comes from the spin-color dilutions, $N_{\rm Had}$ is the number of Hadamard vectors, and $N_r$ is the number of stochastic noise vectors. The number of configurations used are the same as that given in Table~\ref{tab:stats}, except for the E112 ensembles where the light-quark disconnected three-point functions are computed using  461 configurations. Three-point functions are constructed by combining  the quark and gluon loops with the two-point functions given in Table~\ref{tab:stats}.
}
\label{tab:stats-disc}

\begin{tabular}{c|cc|cc|cc|cc}
\hline\hline
&
\multicolumn{2}{c|}{B64} &
\multicolumn{2}{c|}{C80} &
\multicolumn{2}{c|}{D96} &
\multicolumn{2}{c}{E112}
\\
Flavor &
$N_{\rm def}$ & $n_{\rm vec}$ &
$N_{\rm def}$ & $n_{\rm vec}$ &
$N_{\rm def}$ & $n_{\rm vec}$ &
$N_{\rm def}$ & $n_{\rm vec}$
\\
\hline
light
& 200 & $12\times512\times1$
& 450 & $12\times512\times1$
& 0 & $12\times512\times8$
& 530 & $12\times512\times1$
\\
s
& 0 & $12\times512\times2$
& 0 & $12\times512\times4$
& 0 & $12\times512\times4$
& 0 & $12\times512\times2$
\\
c
& 0 & $12\times32\times12$
& 0 & $12\times512\times1$
& 0 & $12\times512\times1$
& 0 & $12\times512\times1$
\\
\hline\hline
\end{tabular}
\end{table*}
The statistics used in the calculation are summarized in
\cref{tab:stats} for the two-point and connected three-point functions and in Table~\ref{tab:stats-disc} for the disconnected quark loops.  
\begin{figure}[!ht]
    \centering
    \includegraphics[width=\columnwidth]{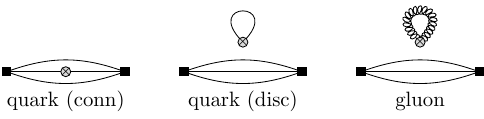}
    \caption{Diagrammatic representation of the quark connected (left), quark disconnected (middle), and gluon (right) for the nucleon three-point function.}
    \label{fig:diagrams_EMT}
\end{figure}

The nucleon ETM three-point function receives quark-connected,
quark-disconnected, and gluon contributions, as illustrated in
\cref{fig:diagrams_EMT}. The connected three-point functions are
computed using the sequential-inversion-through-sink method. A new sequential
inversion is required for each source-sink time separation $t_s$ and each
choice of sink momentum. Therefore, we fix $\vec{p}\,'=\vec{0}$ and use
the source-sink time separations listed in \cref{tab:stats}. The number of
source positions is increased with $t_s$ in order to make the statistical uncertainties approximately constant for all vales of $t_s$.
For the E112 ensemble,  for most of the computations, i.e. for 
two-point and connected three-point functions at  $t_s/a=29$ as well as  disconnected
strange, charm, and gluon contributions  we use $N_{\rm conf}=501$. The
exceptions are the connected three-point functions with source-sink time separations, $t_s/a<29$,
which are computed using $N_{\rm conf}=225$ configurations, and the light-quark loops
 which are computed with $N_{\rm conf}=461$. Since these E112 datasets are
defined on different sets of configurations, they are
combined using the superjackknife method, as explained in Appendix~\ref{app:superjackknife}.

The disconnected quark contributions are constructed by correlating
the highest statistics nucleon two-point function with the corresponding quark loops.
The quark loops are estimated using stochastic sources together with
the one-end trick~\cite{McNeile:2006bz}, spin-color dilution, and
hierarchical probing~\cite{Stathopoulos:2013aci}. Low-mode
deflation~\cite{Gambhir:2016uwp} is additionally used for the
light-quark loops for all ensembles except for the D96 where more stochastic sources are used. The implementation of these
noise-reduction techniques for disconnected diagrams with twisted-mass
fermions is discussed in
Ref.~\cite{Abdel-Rehim:2016pjw}. The detailed
statistics for the computation of the quark loops are given in  \cref{tab:stats-disc}.

 The gluon loops are computed using stout-smeared gauge links to construct the gluon energy-momentum tensor insertion.  We  combine the gluon loops with
the  nucleon two-point functions computed using the gauge configurations  listed in \cref{tab:stats}
to construct the gluon three-point correlation functions.

\begin{figure*}[!ht]
    \centering
    \includegraphics[width=0.49\textwidth]{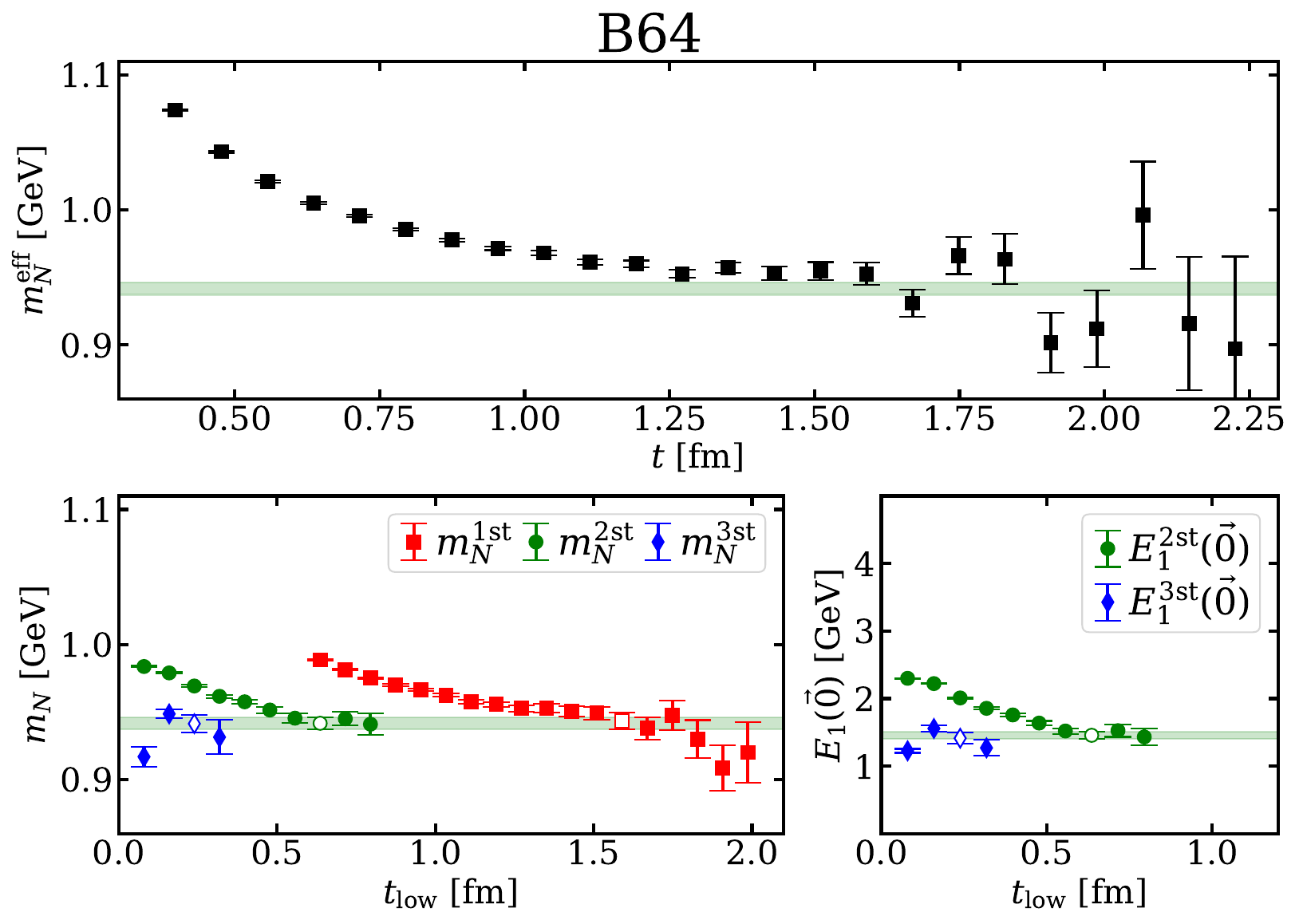} 
    \includegraphics[width=0.49\textwidth]{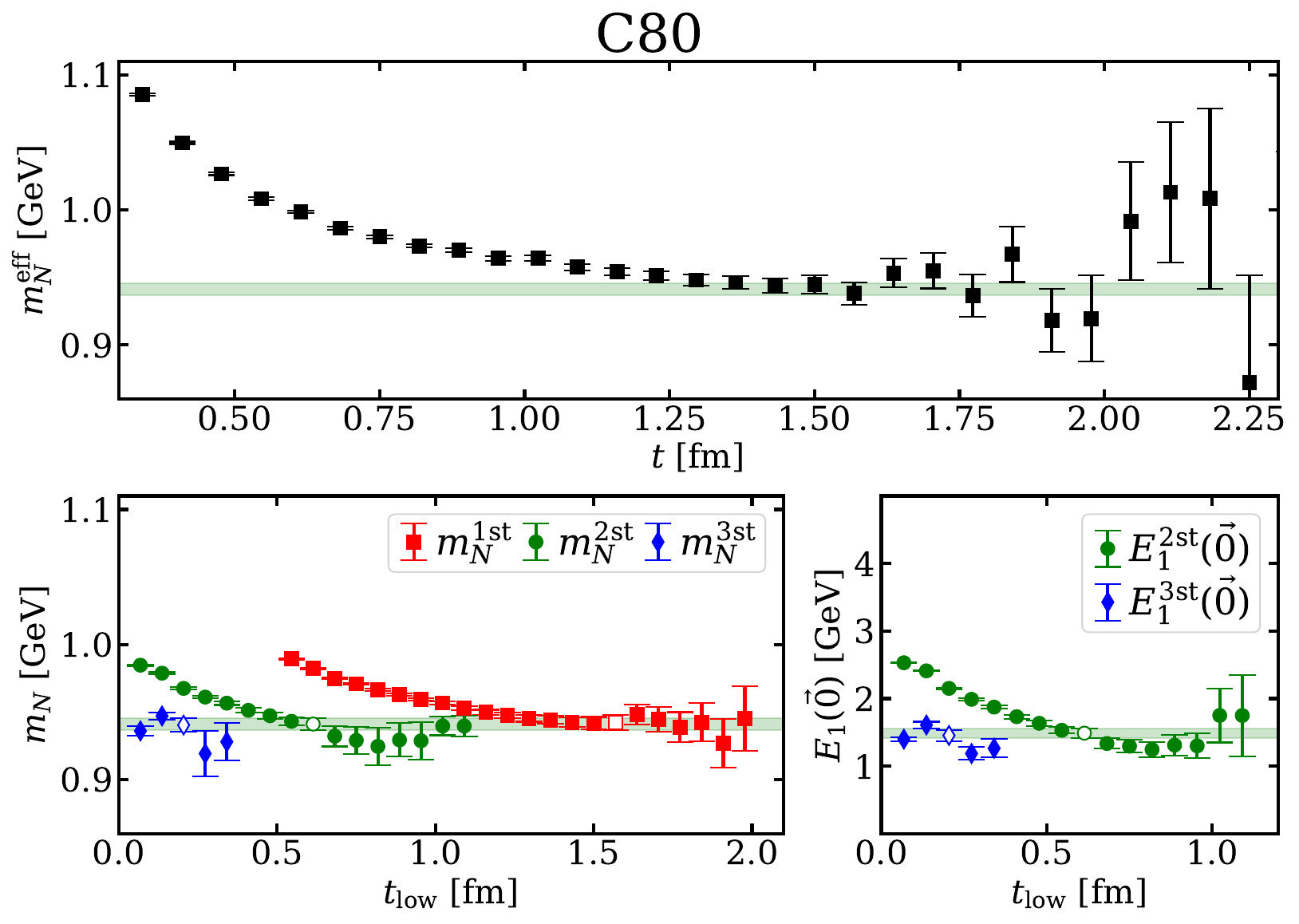} \\
    \includegraphics[width=0.49\textwidth]{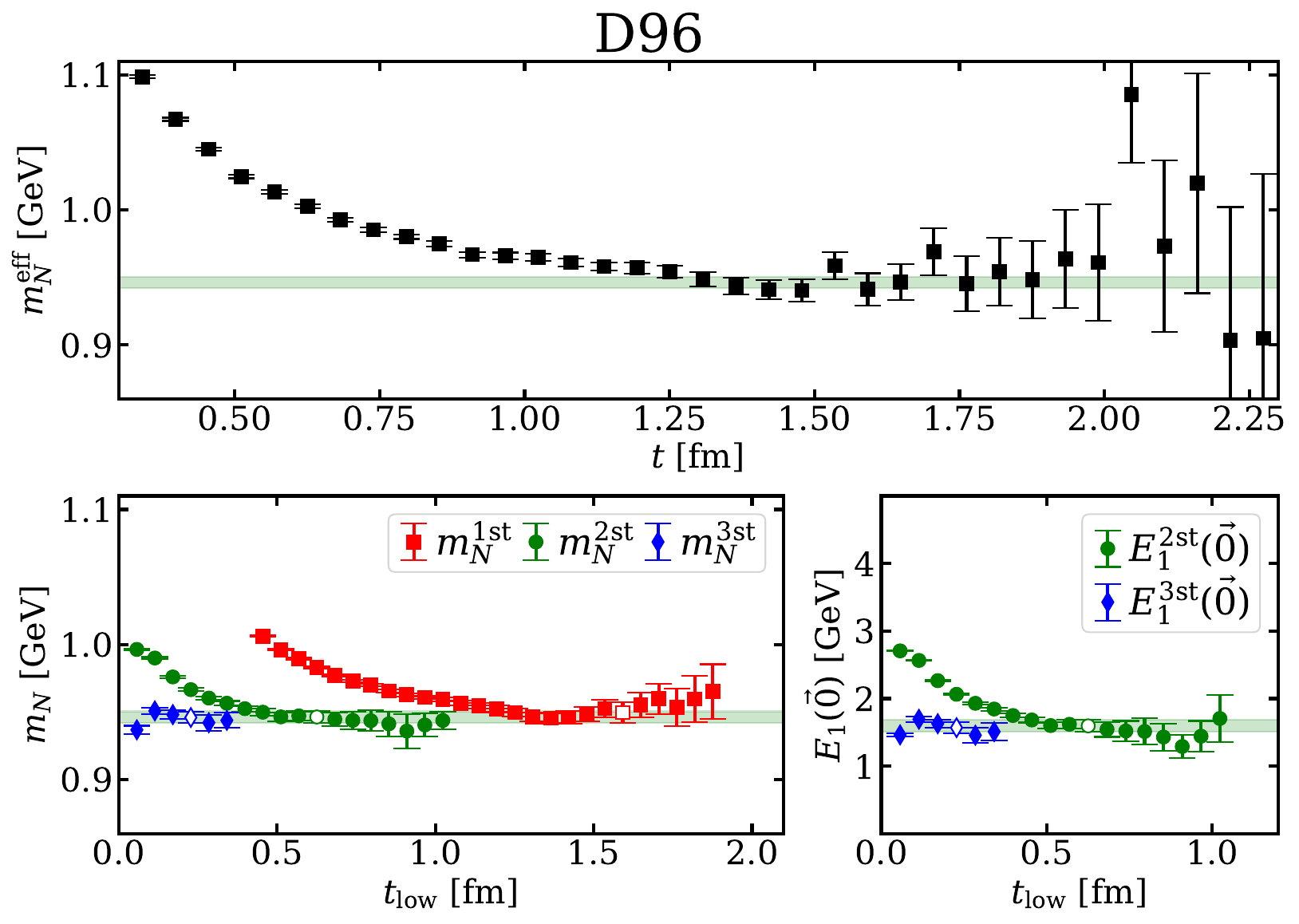} 
    \includegraphics[width=0.49\textwidth]{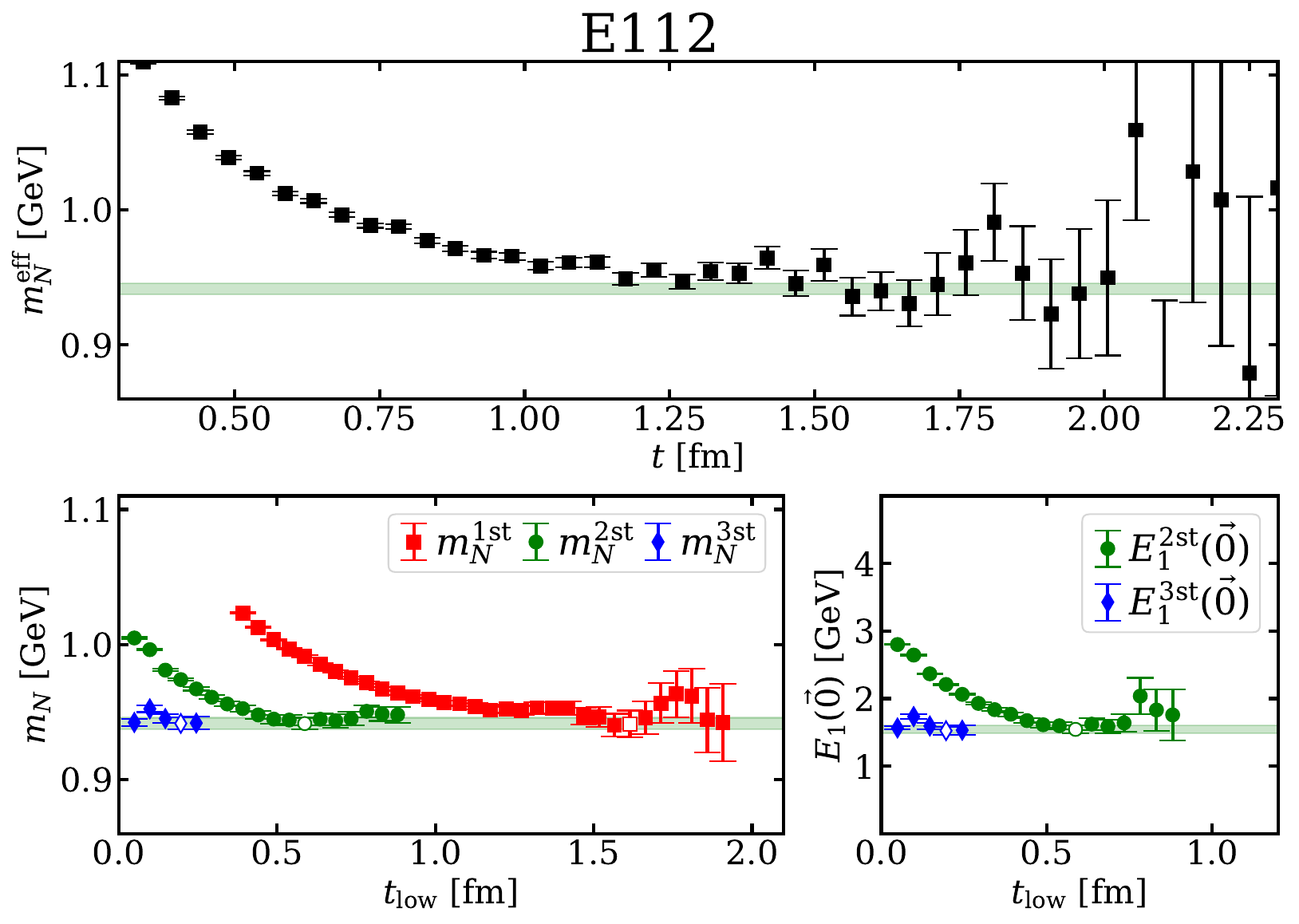}
    \caption{
        Analysis of the rest-frame nucleon two-point function for the B64 (top left), C80 (top right), D96 (bottom left), and E112 (bottom right) ensembles.
        For each ensemble, the upper panel shows the effective nucleon mass $m_N^{\rm eff}$ as a function of the source-sink separation $t$.
        The lower-left panel shows the nucleon mass $m_N$ extracted from one-state (red squares), two-state (green circles), and three-state (blue diamonds) fits as a function of the minimum fit time $t_{\rm low}$.
        The lower-right panel shows the corresponding first excited-state energy $E_1(\vec{0})$ obtained from the two-state and three-state fits.
        Open symbols indicate the selected fit results, while the green bands show the values from the selected two-state fits.
    }\label{fig:meff}
\end{figure*}

\begin{figure}[!ht]
    \centering
    \includegraphics[width=\columnwidth]{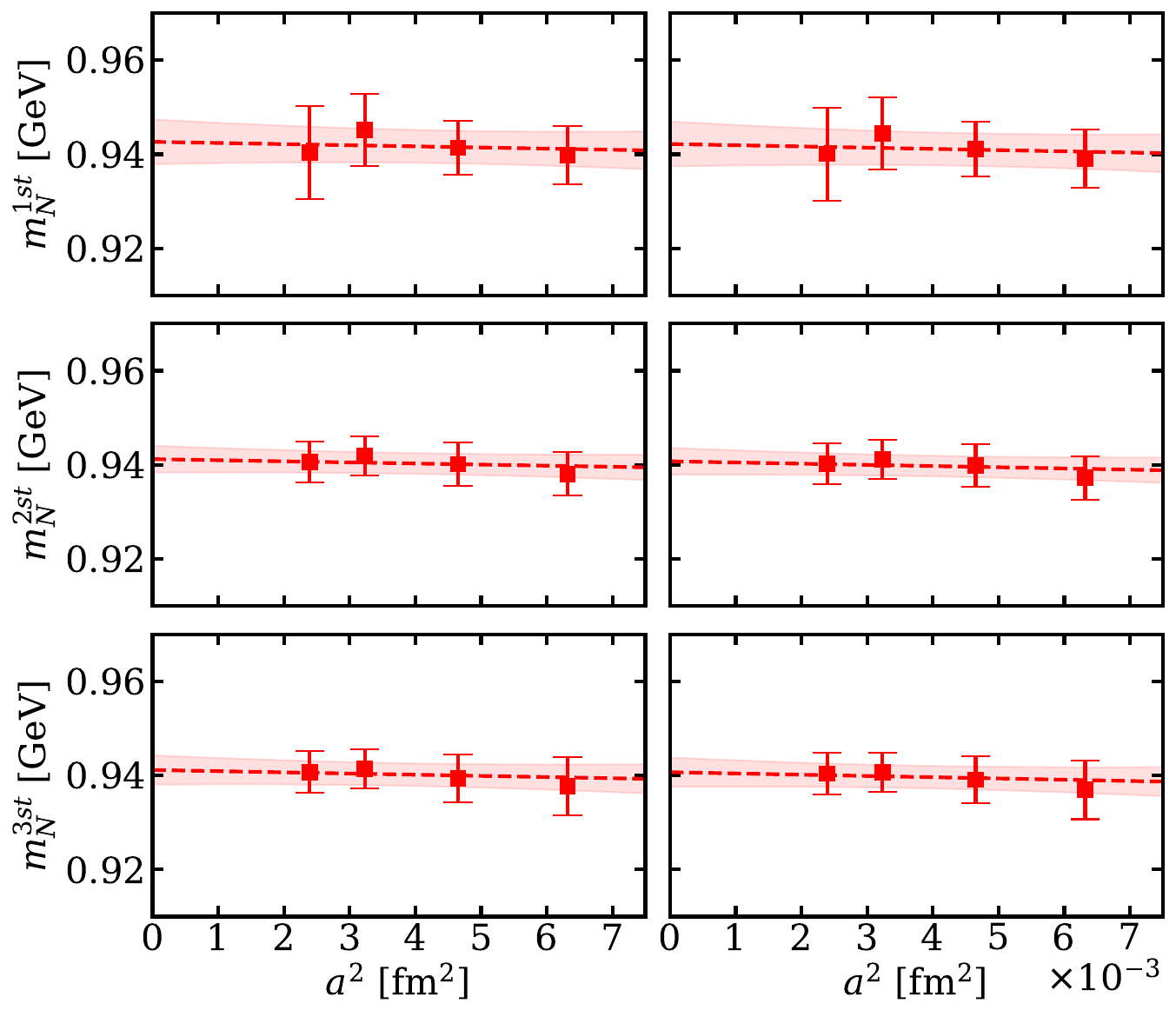}
    \caption{
        Results on the nucleon mass per ensemble and at the continuum limit. Left: we show results determined by the one-state (top), two-state (middle) and three-state (bottom) fits, corrected to the isoQCD value, $m_\pi^{\rm iso}=135\,\mathrm{MeV}$, using the sigma term determined on the same ensemble. Right: we show the results using the same fit An\"atze but using NNLO chiral perturbation theory for correcting the masses.}\label{fig:mN_ce}
\end{figure}

\begin{figure}[!ht]
    \centering
    \includegraphics[width=\columnwidth]{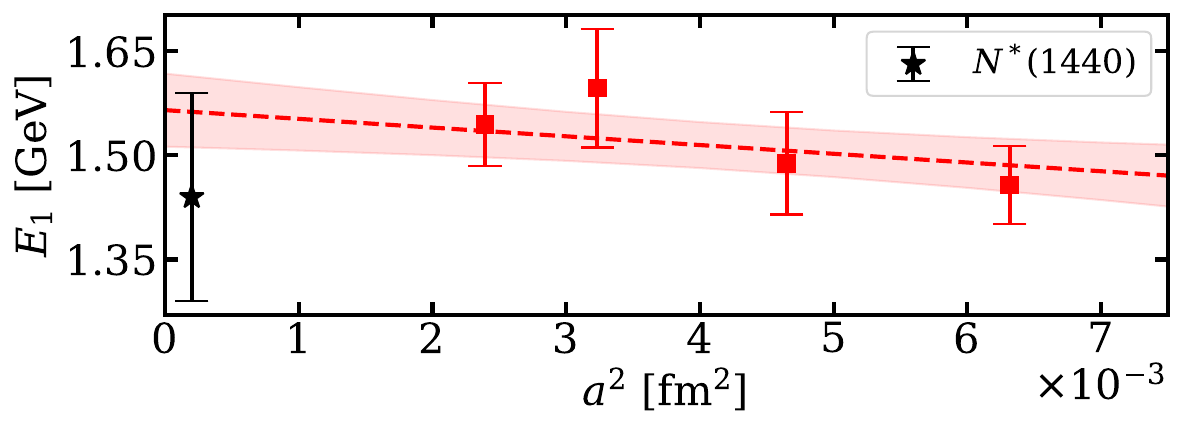}
    \caption{Results on the first excited state energy per ensemble and at the continuum limit using two-state fits. The black star symbol at $a=0$ denotes the Roper resonance $N^*(1440)$.
    }\label{fig:ce_E1}
\end{figure}

\begin{figure}[!ht]
    \centering
    \includegraphics[width=\columnwidth]{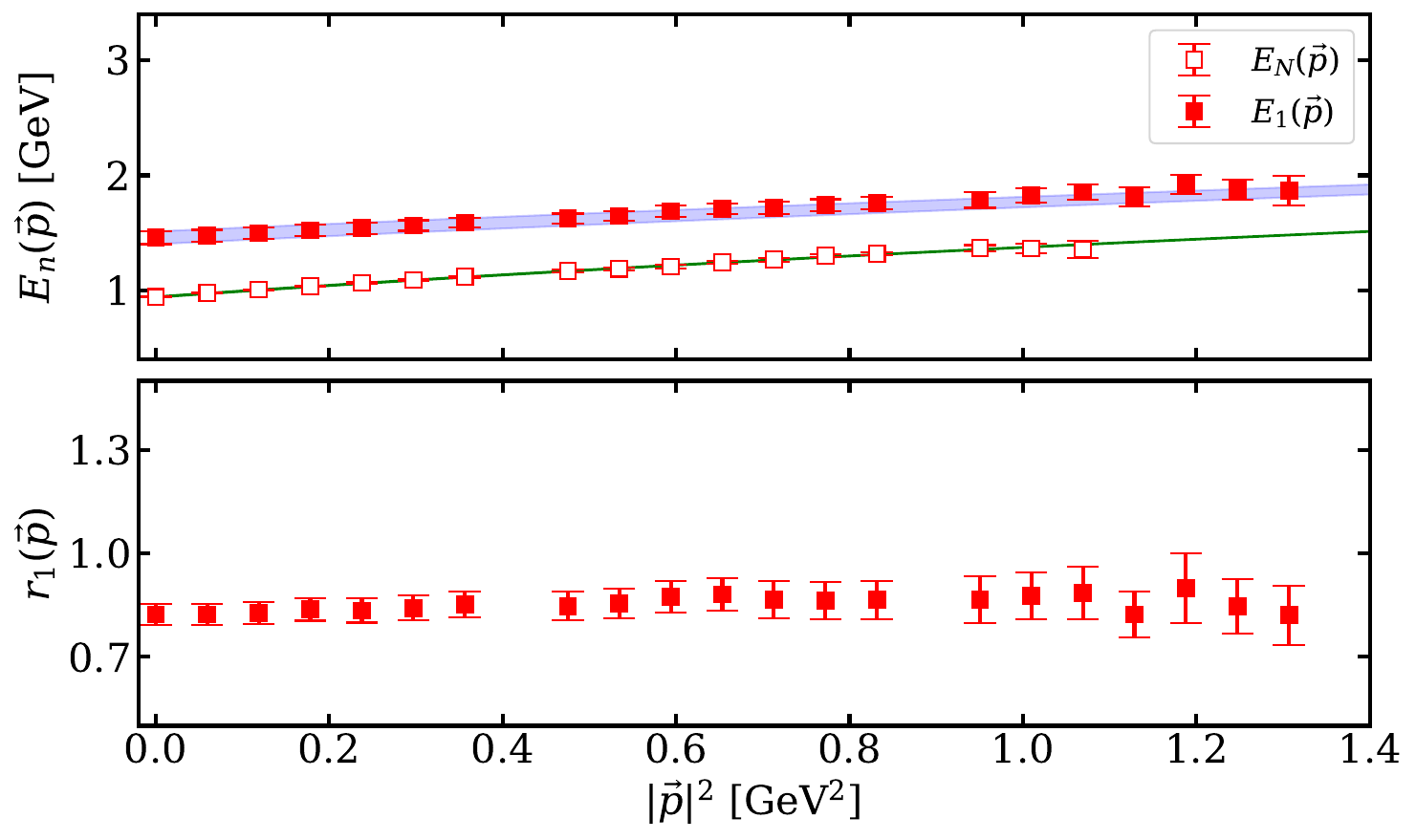}
    \caption{
    Results for the nucleon ground- and first-excited-state energies and the ratio of the overlap coefficients of the first excited to the ground state as a function of $|\vec{p}|^2$ for the B64 ensemble.
    Top: Open symbols show the ground-state energy $E_N(\vec{p})$ extracted from two-state fits without imposing the dispersion relation, while filled symbols show the first-excited-state energies $E_1(\vec{p})$ obtained from fits in which the ground-state energy is fixed using the dispersion relation. The green and blue bands show the corresponding continuum dispersion relations using the masses extracted in the rest frame.
    Bottom: The ratio $r_1(\vec{p})$ of the overlap coefficient of the first excited state to that of the ground state.
    }\label{fig:pars_2st_b}
    
\end{figure}
\begin{figure}[!ht]
    \centering
    \includegraphics[width=\columnwidth]{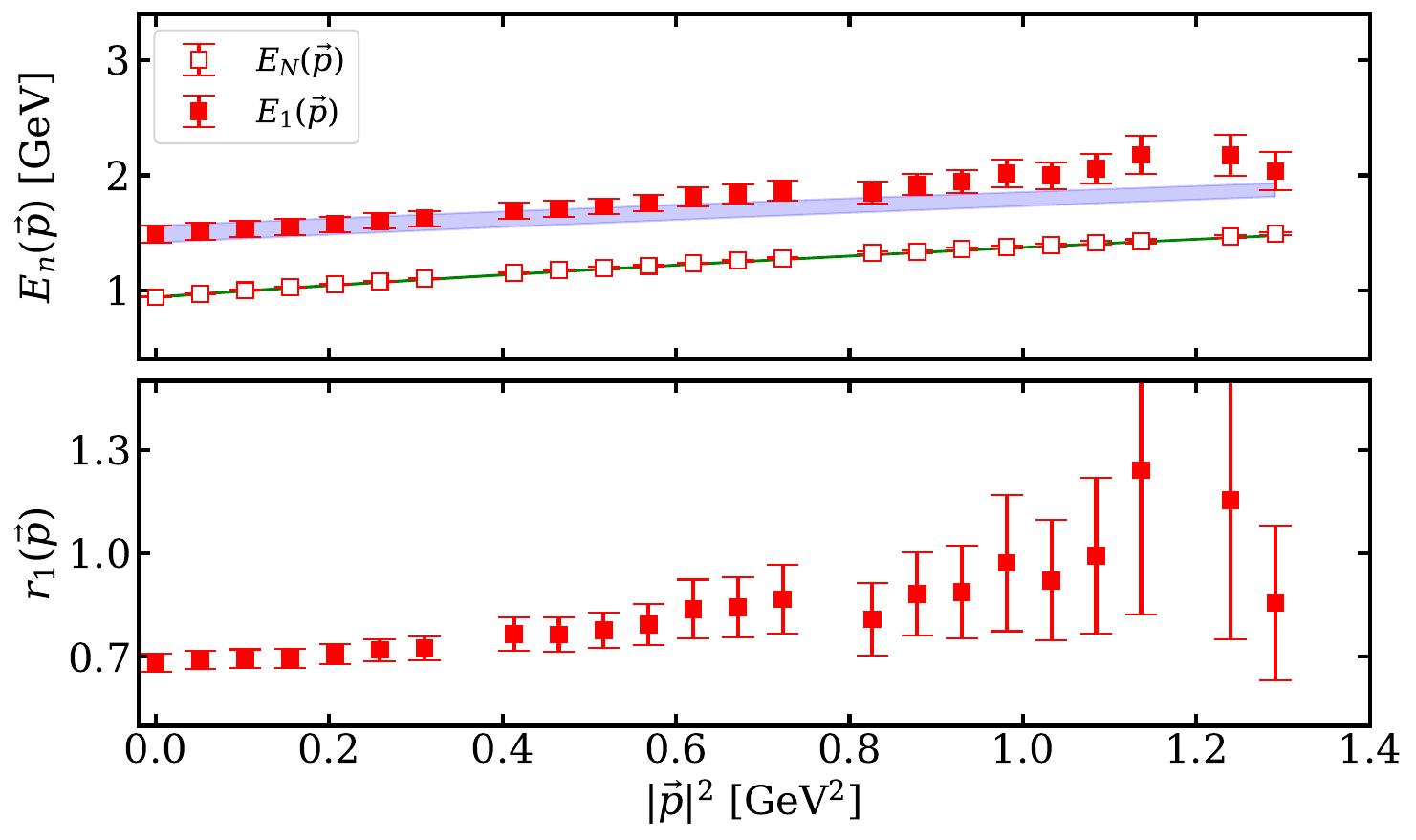}
    \caption{Results for the ground- and first-excited state energies and overlap ratio for the C80 ensemble. The notation is the same as in \cref{fig:pars_2st_b}.
    }\label{fig:pars_2st_c}
\end{figure}
\begin{figure}[!ht]
    \centering
    \includegraphics[width=\columnwidth]{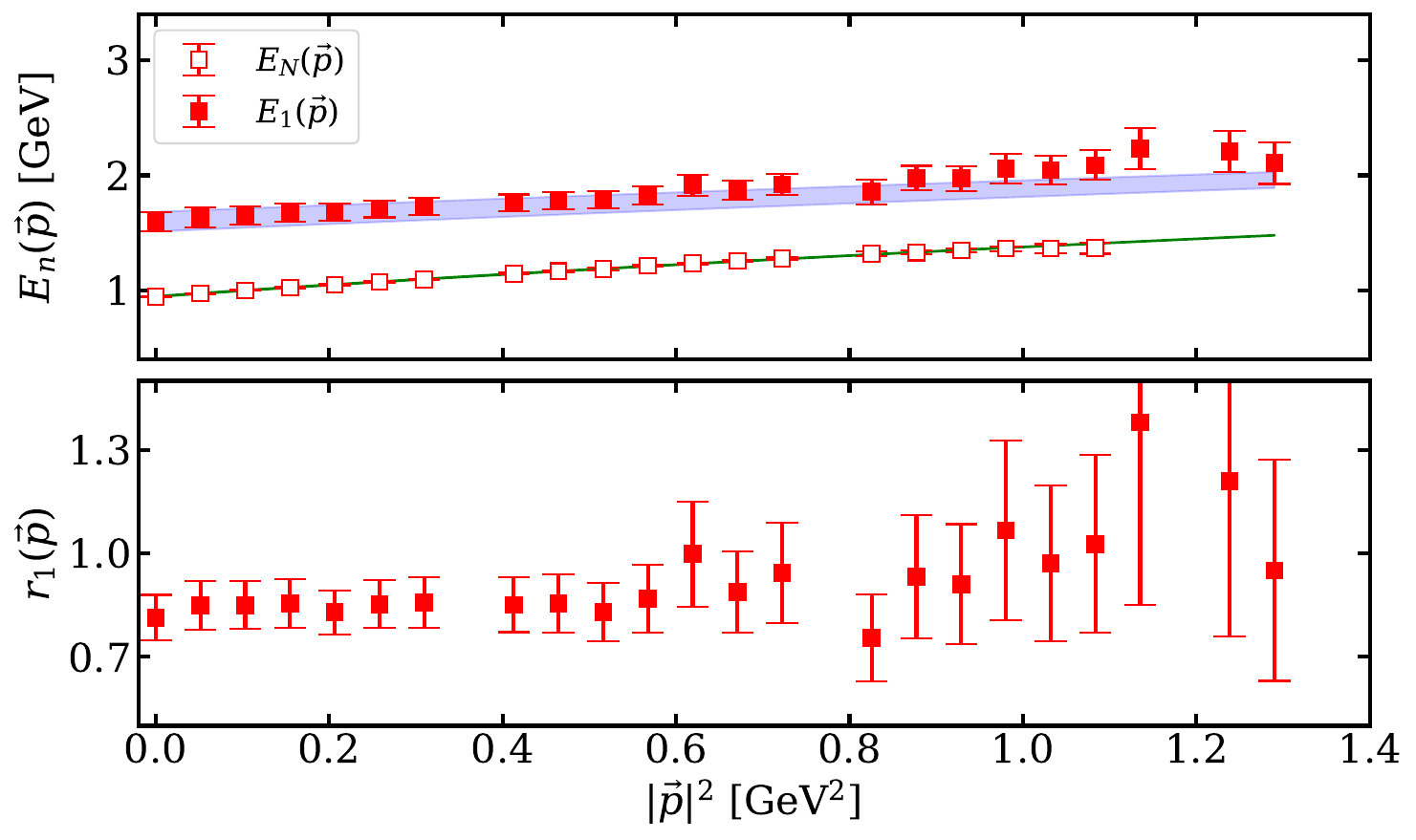}
    \caption{Results  for the ground- and first-excited state energies and overlap ratio for the D96 ensemble. The notation is the same as in \cref{fig:pars_2st_b}.
    }\label{fig:pars_2st_d}
\end{figure}
\begin{figure}[!ht]
    \centering
    \includegraphics[width=\columnwidth]{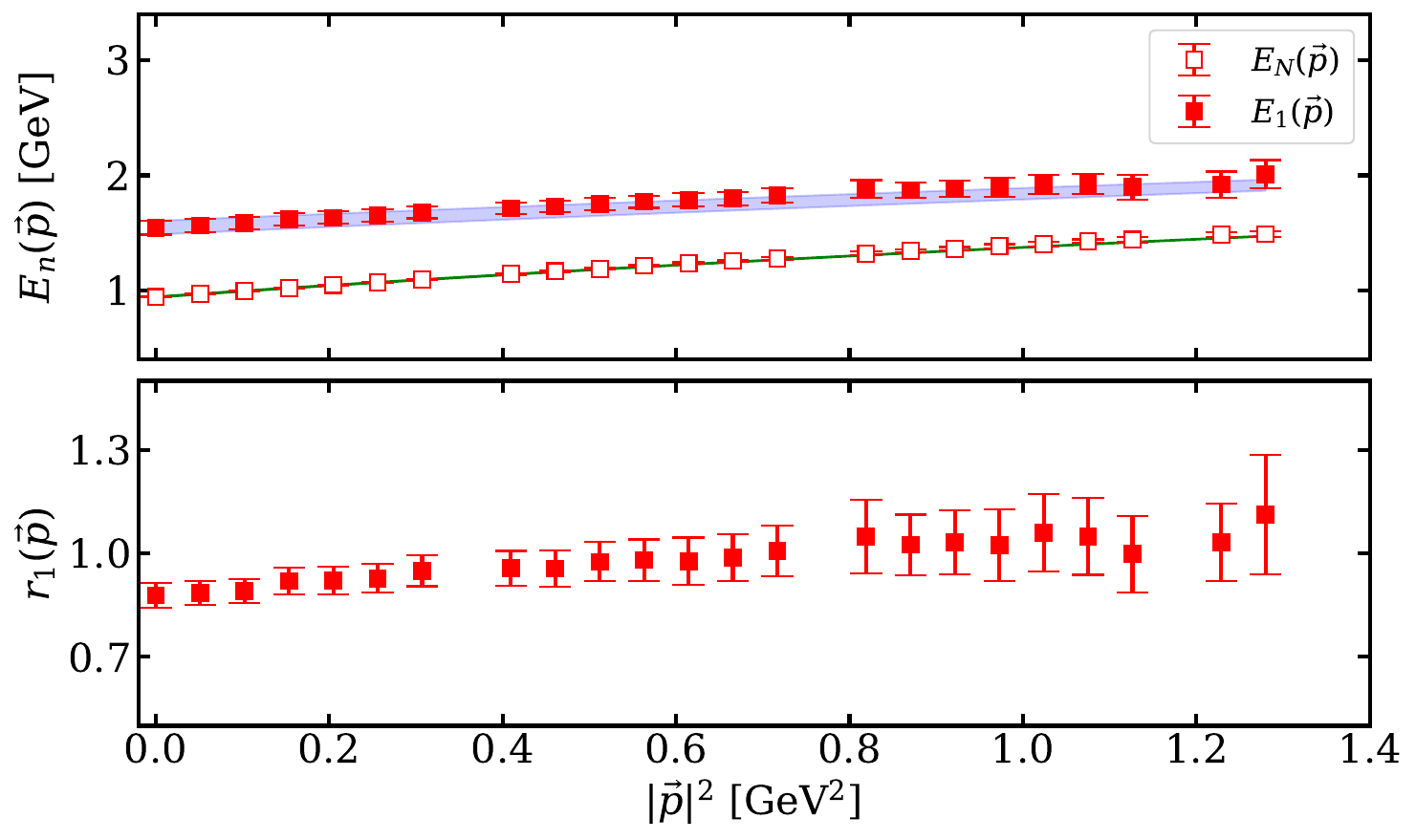}
    \caption{Results  for the ground- and first-excited state energies and overlap ratio  for the E112 ensemble. The notation is the same as in \cref{fig:pars_2st_b}.
    }\label{fig:pars_2st_e}
\end{figure}

\section{Analysis of the nucleon effective mass}\label{sec:analysis 2pt}
 The spectral decomposition of  the two-point function is given as
\begin{align}\label{eq:sd2pt}
    C_{\rm 2pt}(\vec{p},t) = \sum_n c_n(\vec{p})\, e^{-E_n(\vec{p})t}.
\end{align}

The nucleon mass is   determined by analysing  the effective mass  given by
\begin{equation}
m_{\rm eff}(\vec{p};t)=\ln[C_{\rm 2pt}(\vec{p};t)/C_{\rm 2pt}(\vec{p};t+1)].
\end{equation}
To extract the ground-state energy, we perform one-state, two-state, and three-state fits to the effective mass. 
The fit range  $t_{\rm low}\le t\le t_{\rm max}$ for these three Ans\"atze, is chosen by taking  $t_{\rm max}$ to be the largest time slice for which the relative statistical uncertainty remains below 20\%. The minimum fit time $t_{\rm low}$ is varied to investigate convergence and determine the optimal fit window.

The resulting effective masses in the  rest frame are shown in \cref{fig:meff} for all four ensembles. They are accurately determined for all ensembles up to about 1.75~fm with  a clear approach to the ground-state  as the Euclidean time separation increases. The   nucleon mass and first excited-state energies extracted using two-state and three-state fits as functions of the minimum fit time $t_{\rm low}$ show good convergence, asymptotically agreeing  between them and with the one-state fit. Based on the observed stability, we choose $t_{\rm low}\simeq 1.6\,\mathrm{fm}$, $0.6\,\mathrm{fm}$, and $0.2\,\mathrm{fm}$ for the one-state, two-state, and three-state fits, respectively. The extracted values are indicated by the open symbols in \cref{fig:meff}. 

Prior to taking the  continuum limit, we correct the nucleon masses  for the small miss-tuning of the simulation parameter of the light quark doublet to account for the deviation of  pion masses from the isoQCD value of $m_\pi^{\rm phys}=135\,\mathrm{MeV}$. For this correction, we use the pion--nucleon sigma term,
\begin{align}
    \Delta m_N = \frac{\partial m_N}{\partial m_\pi^2} \Delta m_\pi^2 = \sigma_{\pi N} \frac{\Delta m_\pi^2}{m_\pi^2},
\end{align}
computed using the same gauge ensemble, where $\Delta m_\pi^2=(m_\pi^{\rm phys})^2-m_\pi^2$. As an alternative, we also determine the correction using  NNLO chiral perturbation theory for the pion mass dependence of the nucleon mass using the result of   Ref.~\cite{Hoferichter:2015hva}:
\begin{align}
m_N &= m_0 - 4c_1 m_\pi^2 - \frac{3g_A^2 m_\pi^3}{32\pi F_\pi^2} +\mathcal{O}(m_\pi^4),
\end{align}
where $m_0$ is not needed for determining $\Delta m_N$, $c_1=-1.07(2)\times10^{-3}$ is a low-energy constant, and  for the axial charge $g_A$ and the pion decay constant $F_\pi$ we use the PDG values  $g_A=1.2756(13)$ and $F_\pi=92.1(8)$ MeV \cite{ParticleDataGroup:2026aaa}.

To take the continuum limit, we employ the superjackknife procedure of Ref.~\cite{LHPC:2010jcs}. For each fit Ansatz  and pion-mass correction scheme, we perform both constant and linear extrapolations in $a^2$. The final value in the continuum limit is obtained using the average determined via the Akaike information criterion (AIC)~\cite{Jay:2020jkz,Neil:2022joj}, where the AIC weights are applied independently to each superjackknife sample. 

The resulting continuum extrapolations are shown in \cref{fig:mN_ce}. The lattice-spacing dependence is mild and the slope is statistically consistent with zero for all fit Ans\"atze for both pion mass correction procedures. The values extracted in the continuum limit are summarized in \cref{tab:mN}. The nucleon masses obtained from the one-, two-, and three-state analyses are mutually consistent. The  corrections to the pion mass obtained using the lattice-determined sigma term and NNLO chiral perturbation theory are also consistent. To avoid reliance on external chiral perturbation theory input, we select the result obtained from the two-state analysis with the  pion mass corrections obtained using the nucleon sigma terms for the subsequent analysis.

We also perform a continuum extrapolation of the first excited state energy obtained from the two-state fits, as shown in \cref{fig:ce_E1}. For this quantity, the statistical uncertainty is at the level of about $50~{\rm MeV}$, and the expected pion mass correction is minor compared with this uncertainty. We therefore do not apply a pion mass correction to $E_1$. The extrapolated value is
$E_1=1565(53)~{\rm MeV}$, which is compatible  with the Roper resonance, $N^*(1440)$ which has  width of 150~MeV, shown as the black star in \cref{fig:ce_E1}, where the error its   half of its total decay width.
\begin{table}[!ht]
    \caption{
        Nucleon masses obtained from one-state, two-state, and three-state fits to the two-point correlation functions for the B64, C80, D96, and E112 ensembles.
        The columns $\Delta m_N^{\rm LAT}$ and $\Delta m_N^{\chi{\rm PT}}$ give the shifts applied to correct the results to the isoQCD physical-point pion mass, $m_\pi=135\,\mathrm{MeV}$, using either the lattice-determined pion--nucleon sigma term or NNLO chiral perturbation theory, respectively.
        The final two rows show the continuum-extrapolated nucleon masses obtained using the corresponding pion-mass corrections.
        The shaded entry highlights the selected continuum-extrapolated value, obtained from the two-state analysis using the lattice-determined pion-mass correction.
        All values are given in MeV.
    }\label{tab:mN}
    \centering
    \renewcommand\arraystretch{1.5}
    \begin{ruledtabular}
    \begin{tabular}{cccccc}
    & $m_N^{1st}$ & $m_N^{2st}$ & $m_N^{3st}$ & $\Delta m_N^{\rm LAT}$ & $\Delta m_N^{\rm \chi PT}$ \\
    \hline
    B64 & 943.5(6.2) & 941.8(4.6) & 941.4(6.2) & -3.7(2) & -4.5(4) \\
    C80 & 942.4(5.8) & 941.2(4.5) & 940.4(5.0) & -1.05(6) & -1.3(5) \\
    D96 & 949.7(7.7) & 946.4(4.2) & 946.0(4.1) & -4.5(4) & -5.3(4) \\
    E112 & 941(10) & 942.4(4.5) & 942.2(4.6) & -0.91(9) & -1.2(4) \\ \hline
    LAT & 942.5(4.8) & \setlength{\fboxsep}{1pt}\colorbox{gray!20}{941.7(2.9)} & 941.5(3.2) &  &  \\
    $\chi$PT & 942.0(4.8) & 941.3(3.0) & 941.0(3.2) &  &  \\
    \end{tabular}
    \end{ruledtabular}
\end{table}

The analysis is repeated for non-zero momentum, since this information is used in the three-point function analysis. For $\vec{p}\neq\vec{0}$, the two-point function is first analyzed using two-state fits with the same starting fit range as  the ones used  for the $\vec{p}=\vec{0}$ analysis, namely  we use $t_{\rm low}\simeq 0.6\,\mathrm{fm}$. The resulting ground-state energies are shown in  \cref{fig:pars_2st_b,fig:pars_2st_c,fig:pars_2st_d,fig:pars_2st_e}. We compare with the  continuum dispersion relation 
\begin{align}
E_N(\vec{p})=\sqrt{m_N^2+\vec{p}^{\,2}},
\end{align}
where $m_N$ is the nucleon mass determined from our the two-state analysis in the rest frame.
We find that the resulting energy is  fully consistent with the continuum dispersion relation.
We therefore fix the momentum dependence of the ground-state energy through the dispersion relation.

The other fit parameters entering in the effective mass fits are the first excited-state energy $E_1(\vec{p})$ and the overlap ratio $r_1(\vec{p})=c_1(\vec{p})/c_N(\vec{p})$. Fixing the ground-state through the dispersion relation we  extract  these two parameters as a function of the momentum using two-state fits with the same fit ranges. The values are shown in \cref{fig:pars_2st_b,fig:pars_2st_c,fig:pars_2st_d,fig:pars_2st_e} for the four ensembles. The first excited state energy is broadly consistent with the corresponding dispersion relation, while the overlap ratios are of order unity across the momentum range considered.

\section{Extraction of bare gravitational form factors}\label{sec:bgff}
\subsection{Procedure for extracting  ground state matrix elements}
For each ensemble,  we use  the nucleon mass and first excited energy determined from the two-state fit with the nucleon mass corrected via the sigma-term as input, namely $m_N=m_N^{2st}$ in \cref{tab:mN}, for each ensemble for the  analysis of nucleon matrix elements. We use the continuum dispersion relation for the  ground state energy, i.e. we take $E_{N}(\vp)=\sqrt{m_N^2+\vp^2}$, while for
the first excited state energy we instead use  $E_1(\vec{p})$  at the given $\vec{p}$ value as determined from the effective energy analysis.

The spectral decomposition of the three-point function is given by
\begin{align}
    &C_{\rm 3pt}^{\mu\nu}(\Gamma_\alpha,\vpp,\vp; \ts,\tins) \nonumber\\
    =& \sum_{m,n} a^{\mu\nu}_{mn}(\Gamma_\alpha,\vpp,\vp) e^{-E_N(\vpp)(\ts-\tins)} e^{-E_n(\vp)\tins}\,,
\end{align}
where $a^{\mu\nu}_{00}(\Gamma_\alpha,\vpp,\vp)=\sqrt{c_N(\vpp)c_N(\vp)}\,\Pi^{\mu\nu}(\Gamma,\vpp,\vp)$ with $\Pi^{\mu\nu}(\Gamma_\alpha,\vpp,\vp)$  the spin-projected nucleon matrix element of Eq.~(\ref{eq:me2ff}) with projector $\Gamma_\alpha$ and $c_N(\vp)$ the coefficient appearing in the spectral decomposition of the two-point function in \cref{eq:sd2pt}.

To extract the desired matrix element, we construct the ratio of three- to two-point functions
\begin{align}\label{eq:ratio}
    R^{\mu\nu}(\Gamma_\alpha,\vpp,\vp;\ts,\tins) &= \frac{C_{\rm 3 pt}^{\mu\nu}(\Gamma_\alpha,\vpp,\vp;\ts,\tins)}{\sqrt{C_{\rm 2 pt}(\vpp;\ts)C_{\rm 2 pt}(\vp;\ts)}} \nonumber\\
    &\times e^{-\left(E_{N}(\vpp)-E_{N}(\vp)\right)\,(\tins-\ts/2)} \,,
\end{align}

We opt to use the ratio $R^{\mu\nu}(\Gamma_\alpha,\vpp,\vp;\ts,\tins)$ instead of the  ratio  
\begin{align}\label{eq:ratio_std}
    & {R^\prime}^{\mu\nu}(\Gamma_\alpha,\vpp,\vp;\ts,\tins) = \frac{C_{\rm 3 pt}^{\mu\nu}(\Gamma_\alpha,\vpp,\vp;\ts,\tins)}{\sqrt{C_{\rm 2 pt}(\vpp;\ts)C_{\rm 2 pt}(\vp;\ts)}} \nonumber\\
    &\qquad \times \sqrt{\frac{C_{\rm 2 pt}(\vp;\ts-\tins)C_{\rm 2 pt}(\vpp;\tins)}{C_{\rm 2 pt}(\vpp;\ts-\tins)C_{\rm 2 pt}(\vp;\tins)}}\,.
\end{align} 
used in our previous works for the following reason:
while $R^{\prime \mu\nu}$ typically  has a smaller error due to optimizing the correlations, it exhibits larger excited state contamination arising  from using two-point functions at early time slices in the second square root. In Ref.~\cite{Alexandrou:2024tin}, $R^{\mu\nu}$ was compared with $R^{\prime \mu\nu}$. It was shown that indeed $R^{\mu\nu}$ typically has larger statistical errors but its time dependence is milder as compared to that of  $R^{\prime \mu\nu}$ leading to better control of ground state extraction.
In the asymptotic limit $t_{\rm ins}\Delta E(\vec{p})\gg 1$ and $(\ts-\tins)\Delta E(\vec{p})\gg1$,
\begin{align}
    R^{\mu\nu}(\Gamma_\alpha,\vpp,\vp;\ts,\tins)\xrightarrow[(\ts-\tins)\Delta E(\vec{p})\gg1]{\tins\Delta E(\vec{p})\gg 1} \Pi^{\mu\nu}(\Gamma_\alpha,\vpp,\vp) \,,
\end{align}
where $\Delta E(\vec{p})$ is the energy gap between the ground state and first excited state.

The gravitational form factors
$A_{20}(Q^2)$, $B_{20}(Q^2)$, and $C_{20}(Q^2)$ all contribute  to the nucleon matrix element according to \cref{eq:me2ff} and must therefore be disentangled. This is achieved by constructing an overconstrained system of equations using all available combinations of momentum transfer, Lorentz indices, and spin projectors for a given value of $Q^2$. The system is written at the level of the ratio in Eq.~(\ref{eq:ratio}) as
\begin{align}
\mathcal{R}_i(\ts,\tins)=\sum_{j=1}^{3} G_{ij}F_j(Q^2;\ts,\tins),
\end{align}
where $\mathcal{R}_i$ denotes the ratio extracted for a given choice of momentum transfer, Lorentz indices, and projector, $G_{ij}$ are known kinematical coefficients, and
\begin{align}
F(Q^2;\ts,\tins)=\begin{pmatrix}
A_{20}(Q^2;\ts,\tins)\\
B_{20}(Q^2;\ts,\tins)\\
C_{20}(Q^2;\ts,\tins)
\end{pmatrix}.
\end{align}
 The coefficients $G_{ij}$ for the case $\vpp=\vec{0}$ are provided in Appendix \ref{app:gffs}, \cref{eq:gff1,eq:gff2,eq:gff3,eq:gff4} for the $\mu=\nu$ case, and \cref{eq:gff5,eq:gff6,eq:gff7,eq:gff8} for the $\mu\neq\nu$ case. For the analysis using a boosted frame opted for disconnected contributions, we give the relevant coefficients in what follows. 

To take into account for the statistical errors of the ratios, we define the  quantities
\begin{align}
\widetilde{\mathcal{R}}_i(\ts,\tins)=\frac{\mathcal{R}_i(\ts,\tins)}{w_i(\ts,\tins)},\qquad
\widetilde{G}_{ij}=\frac{G_{ij}}{w_i(\ts,\tins)},
\end{align}
where $w_i(\ts,\tins)$ is the statistical jackknife error of $\widetilde{\mathcal{R}}_i(\ts,\tins)$. As explained in Ref.~\cite{Alexandrou:2020sml}, the singular value decomposition (SVD) solution is equivalent to minimizing the corresponding $\chi^2$ function. The system is solved through the decomposition
\begin{align}
\widetilde{G}=U\Sigma V^\dagger,
\end{align}
yielding the pseudoinverse solution
\begin{align}
F=V\Sigma^{-1}U^\dagger \widetilde{\mathcal{R}}.
\end{align}

In the forward limit when $\vp=\vpp$ and $Q^2=0$, only $A_{20}(0)=\braket{x}$ contributes. It can therefore be extracted directly from
\begin{align}
\Pi^{44}(\Gamma_0;\vec{0},\vec{0})&=-\frac{3m_N}{4}\,\braket{x}, \label{eq:pi44} \\ 
\Pi^{4i}(\Gamma_0;\vec{p},\vec{p})&=i p_i \,\braket{x}. \label{eq:pi4i}
\end{align}

\begin{figure}[h!]
    \centering
    \includegraphics[width=\columnwidth]{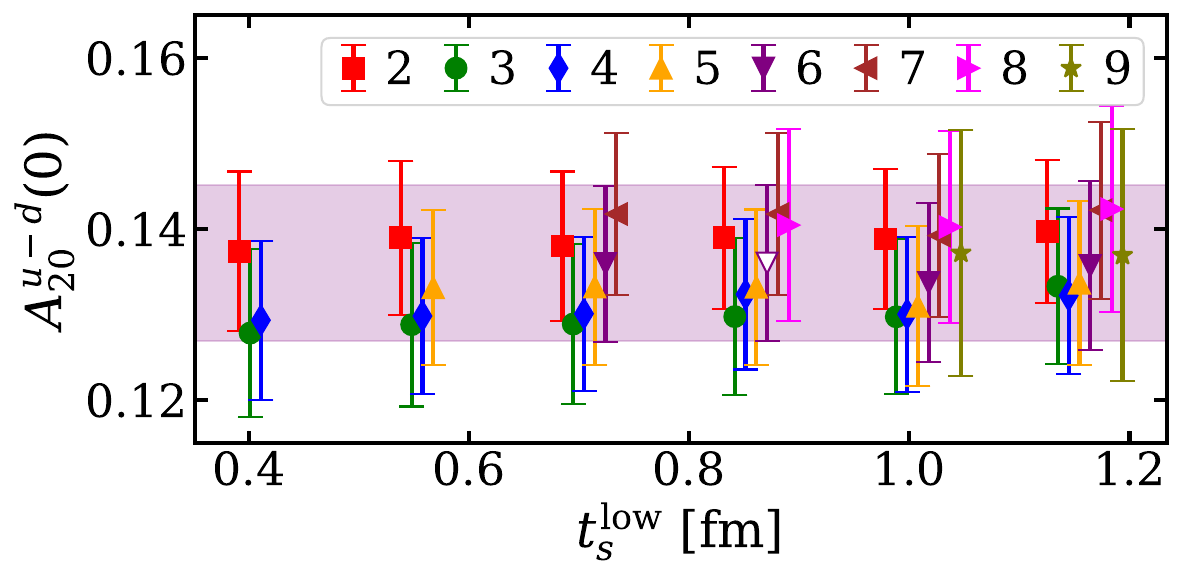}
    \caption{
    Stability of the extracted bare form factor $A_{20}^{u-d}(0)$ for the E112 ensemble under variations of the two-state fit ranges.
    The results are shown as a function of the minimum source-sink separation $t_s^{\rm low}$ for different insertion-time cuts $t_{\rm ins}^{\rm cut}/a$, indicated by different colors and symbols.
    The horizontal band shows the selected fit result.
    The open symbol denotes the selected fit, obtained with $t_s^{\rm low}\simeq0.8~{\rm fm}$ and $t_{\rm ins}^{\rm cut}\simeq0.3~{\rm fm}$.
    }\label{fig:A20_0_jm_conn_tins}
\end{figure}

\begin{figure}[h!]
    \centering
    \includegraphics[width=\columnwidth]{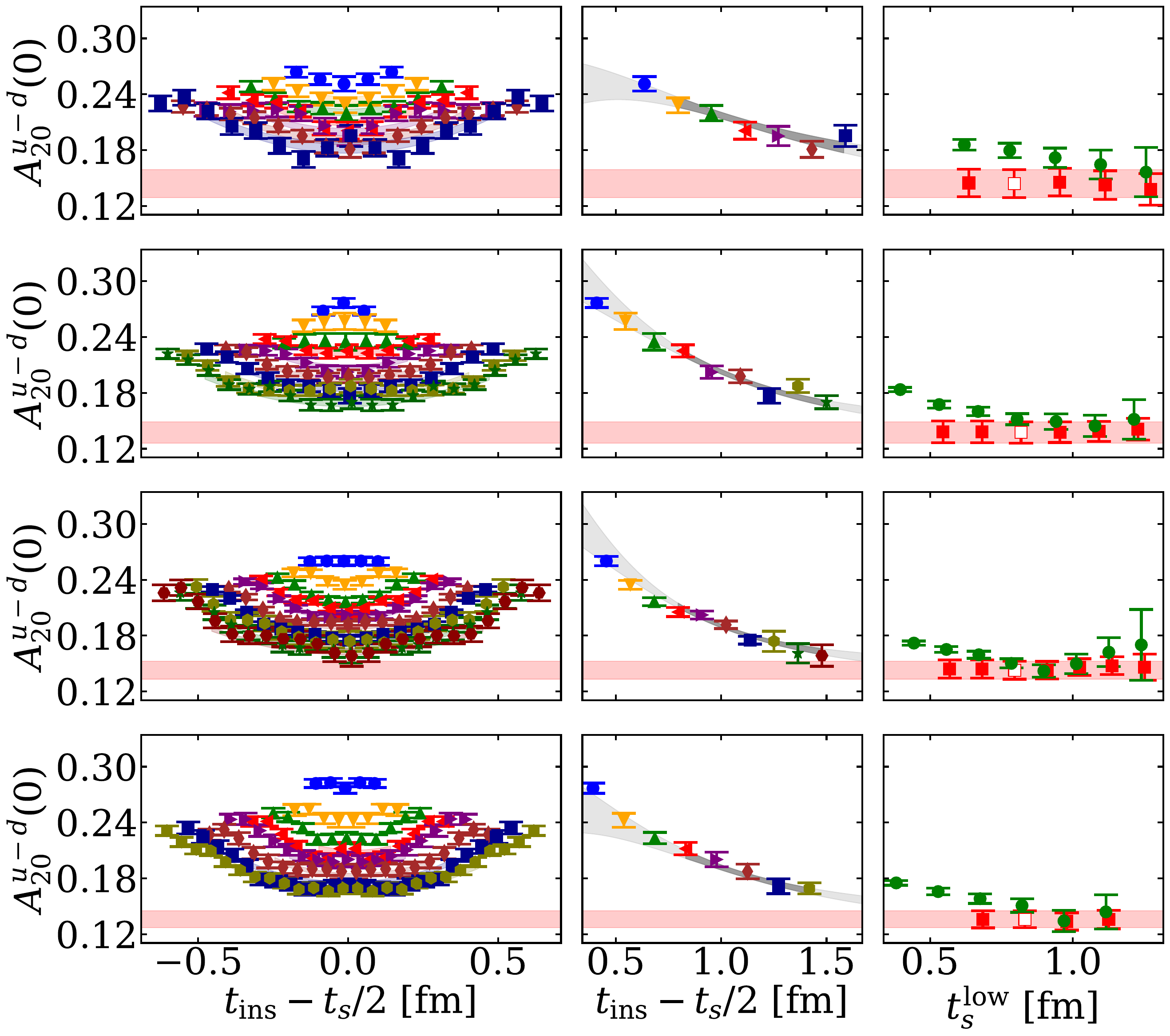}
    \caption{
        Results for the bare isovector ${A}_{20}^{u-d}(0)$ extracted 
         using \cref{eq:pi44} in the nucleon rest frame.
        From top to bottom  we show results  for the B64, C80, D96, and E112 ensembles.
        Left panels show the ratio as a function of $t_{\rm ins}-t_s/2$ for different source-sink separations $t_s$ shown by points with different colors and symbols.
       The  middle panels show the ratio at the midpoint as a function of $t_s$ using the same symbols for each time separation $t_s$ as in the left panels. For the midpoint, we use the ratio at $t_{\rm ins}=t_s/2$ when $t_s/a$ is even and the average of the ratio between $(t_s-a)/2$ and $(t_s+a)/2$ when $t_s/a$ is odd. 
        The right panels show the values extracted from two-state fits (red squares) as a function of the smallest value of $t_s$, $t_s^{\rm low}$, used in fit. The open symbols indicate the selected fit. The horizontal bands correspond to the chosen fit result. In the left and middle panels, the additional bands show the prediction of the selected two-state fit. Results from the summation method (green circles) are also shown for comparison.
    }\label{fig:A20_0_jm_conn}
\end{figure}

\begin{figure}[h!]
    \centering
    \includegraphics[width=\columnwidth]{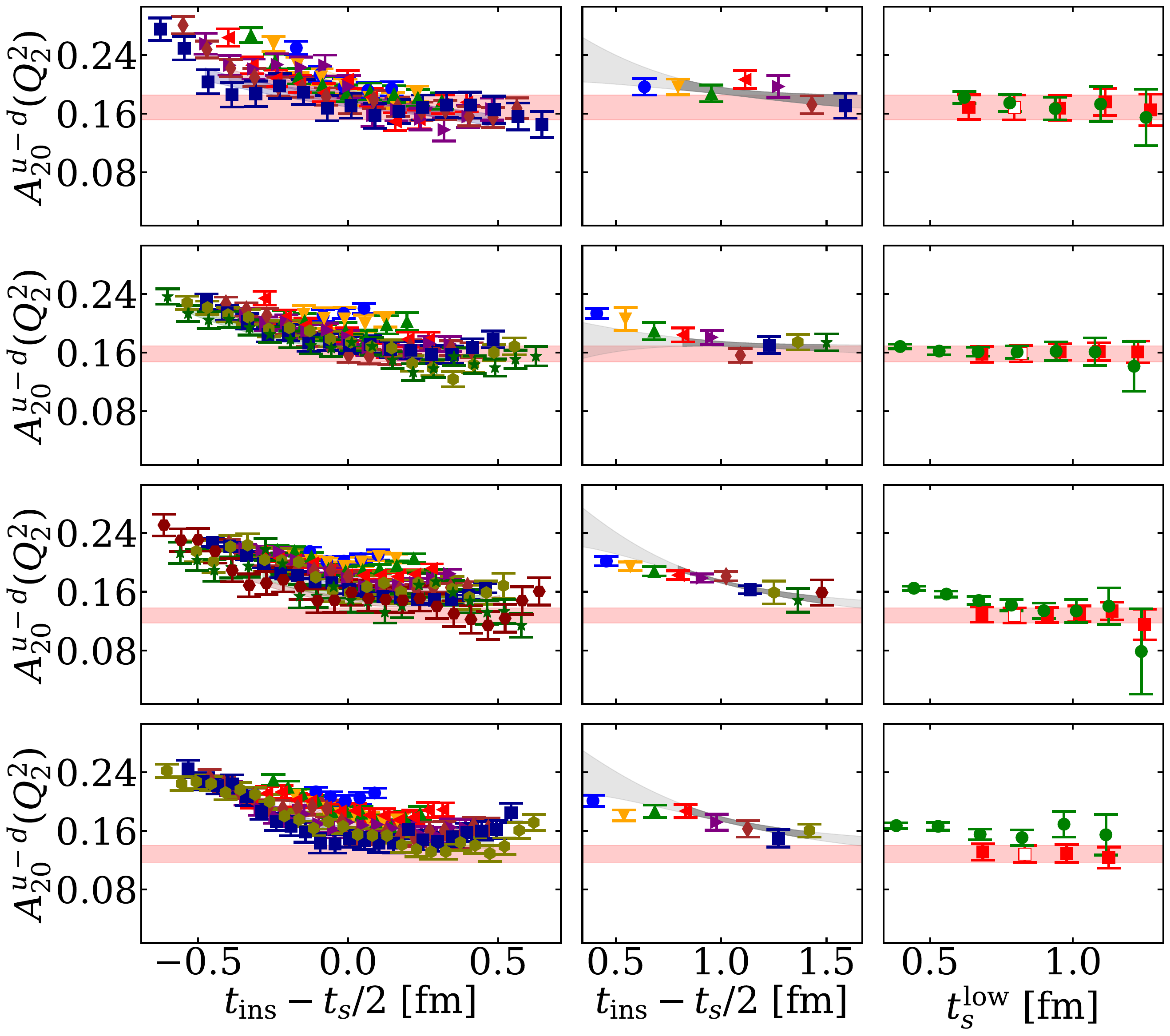}
    \caption{
        Results for the bare ${A}_{20}^{u-d}(Q_2^2)$ for $Q_2^2\approx0.1$ GeV$^2$ that corresponds to two units of momentum transfer. The notation is the same as in \cref{fig:A20_0_jm_conn}.
    }\label{fig:A20_1_jm_conn}
\end{figure}

\begin{figure}[h!]
    \centering
    \includegraphics[width=\columnwidth]{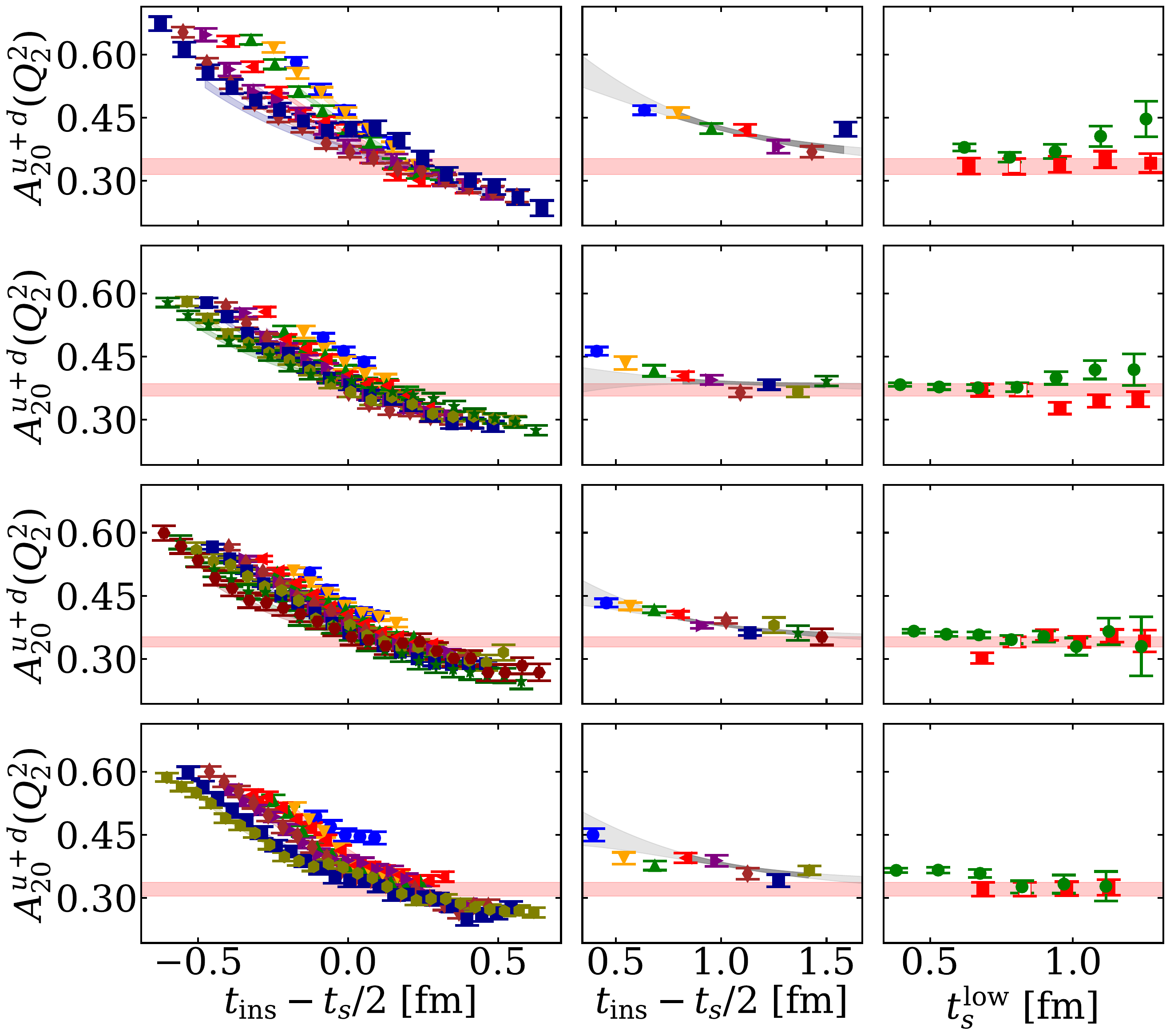}
    \caption{
        Results for the  connected contribution to the isoscalar  bare ${A}_{20}^{u+d}(Q_2^2)$ for $Q_2^2\approx0.1$ GeV$^2$ that corresponds to two units of momentum transfer. The notation is the same as in \cref{fig:A20_0_jm_conn}.
    }\label{fig:A20_2_jp_conn}
\end{figure}

\begin{figure}[h!]
    \centering
    \includegraphics[width=\columnwidth]{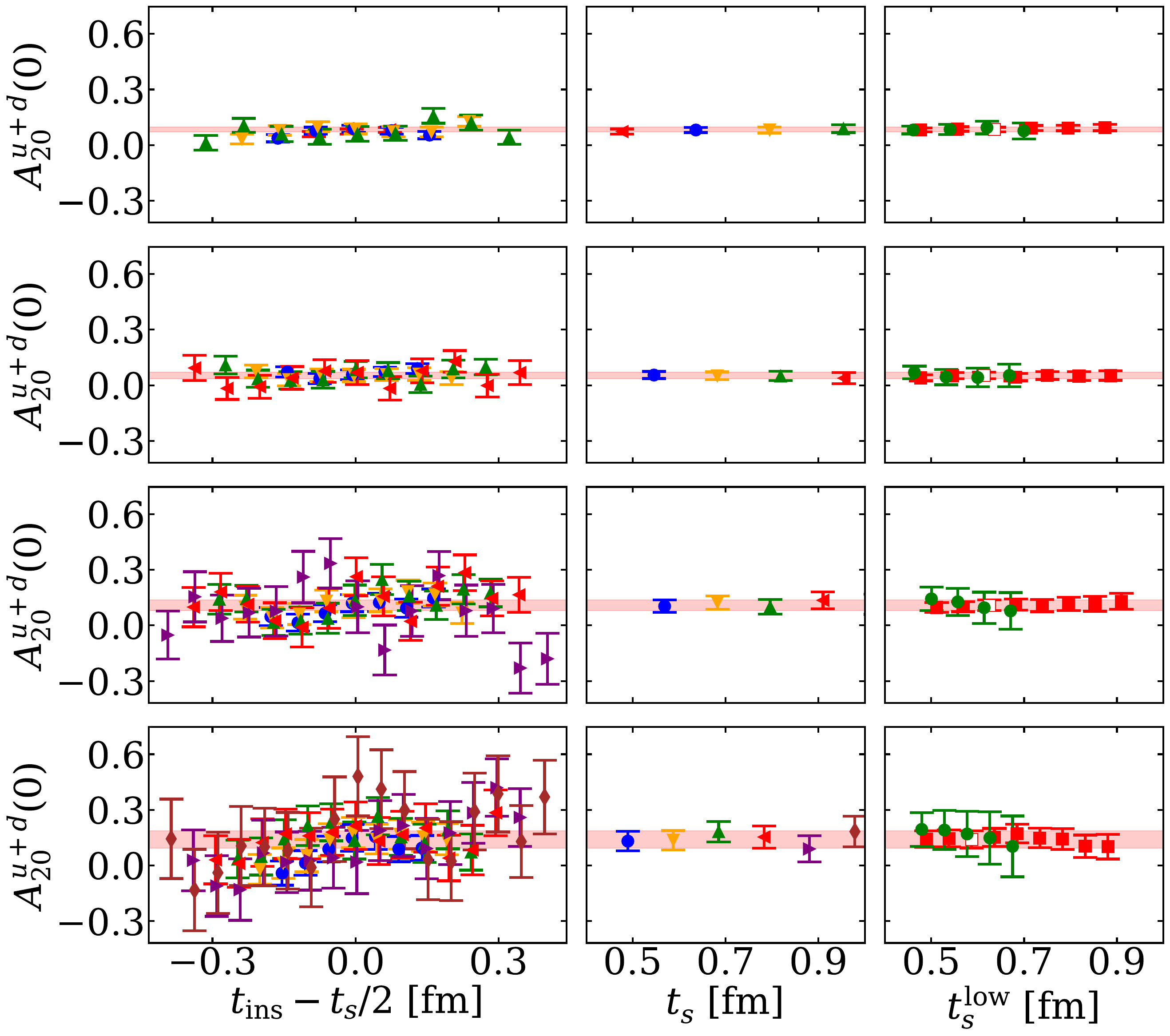}
    \caption{
        Results for the disconnected contribution to the bare isoscalar ${A}_{20}^{u+d}(0)$ for $\mu\neq\nu$.
        The middle panels show the values extracted from constant fits to each source-sink separation $t_s$ individually as a function of $t_s$. 
        The right panels show the results from constant fits (red squares) including all source-sink separations satisfying $t_s\geq t_s^{\rm low}$, where the open symbols indicate the selected fit. The band corresponds to the chosen fit. Results from the summation method (green circles) are also shown for comparison. The rest of the notation is the same as in 
        \cref{fig:A20_0_jm_conn}. 
    }\label{fig:A20_2_jp_disc}
\end{figure}

\begin{figure}[h!]
    \centering
    \includegraphics[width=\columnwidth]{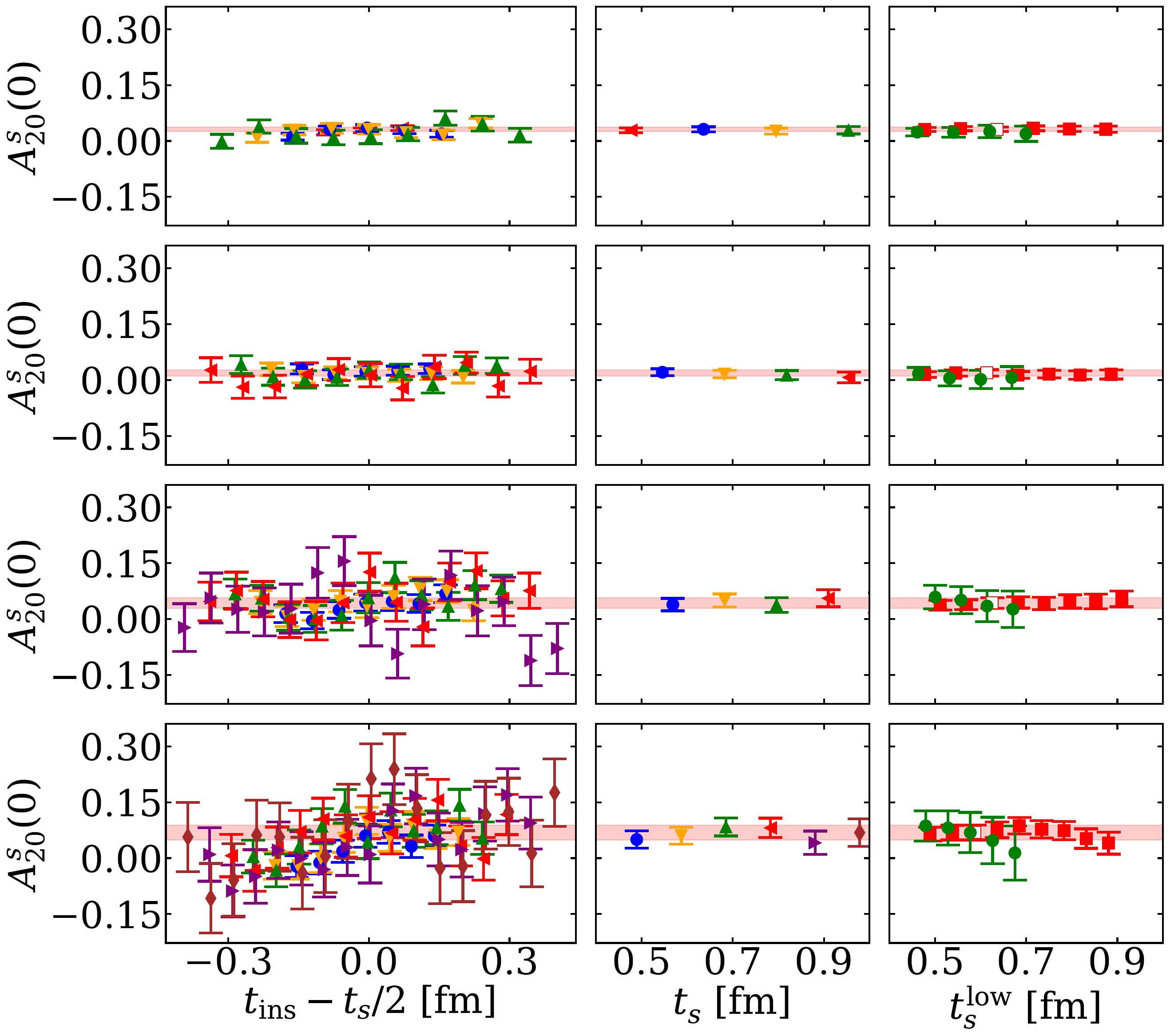}
    \caption{
        Results for the bare ${A}_{20}^{s}(0)$.
        The notation is the same as in 
        \cref{fig:A20_2_jp_disc}. 
    }\label{fig:A20_2_js_disc}
\end{figure}

\begin{figure}[h!]
    \centering
    \includegraphics[width=\columnwidth]{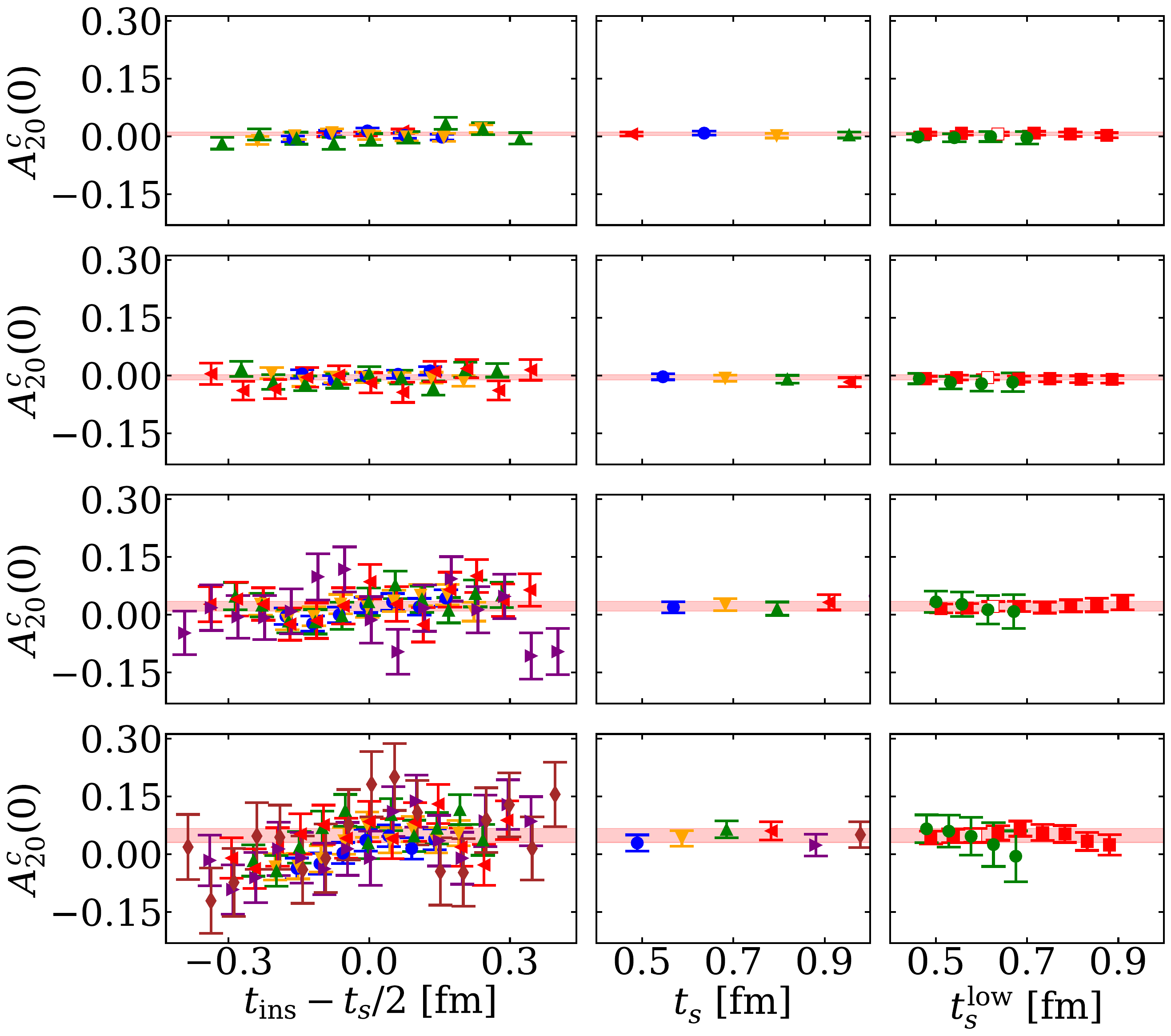}
    \caption{
        Results for the bare ${A}_{20}^{c}(0)$.
        The notation is the same as in 
        \cref{fig:A20_2_jp_disc}. 
    }\label{fig:A20_2_jc_disc}
\end{figure}

\begin{figure}[h!]
    \centering
    \includegraphics[width=\columnwidth]{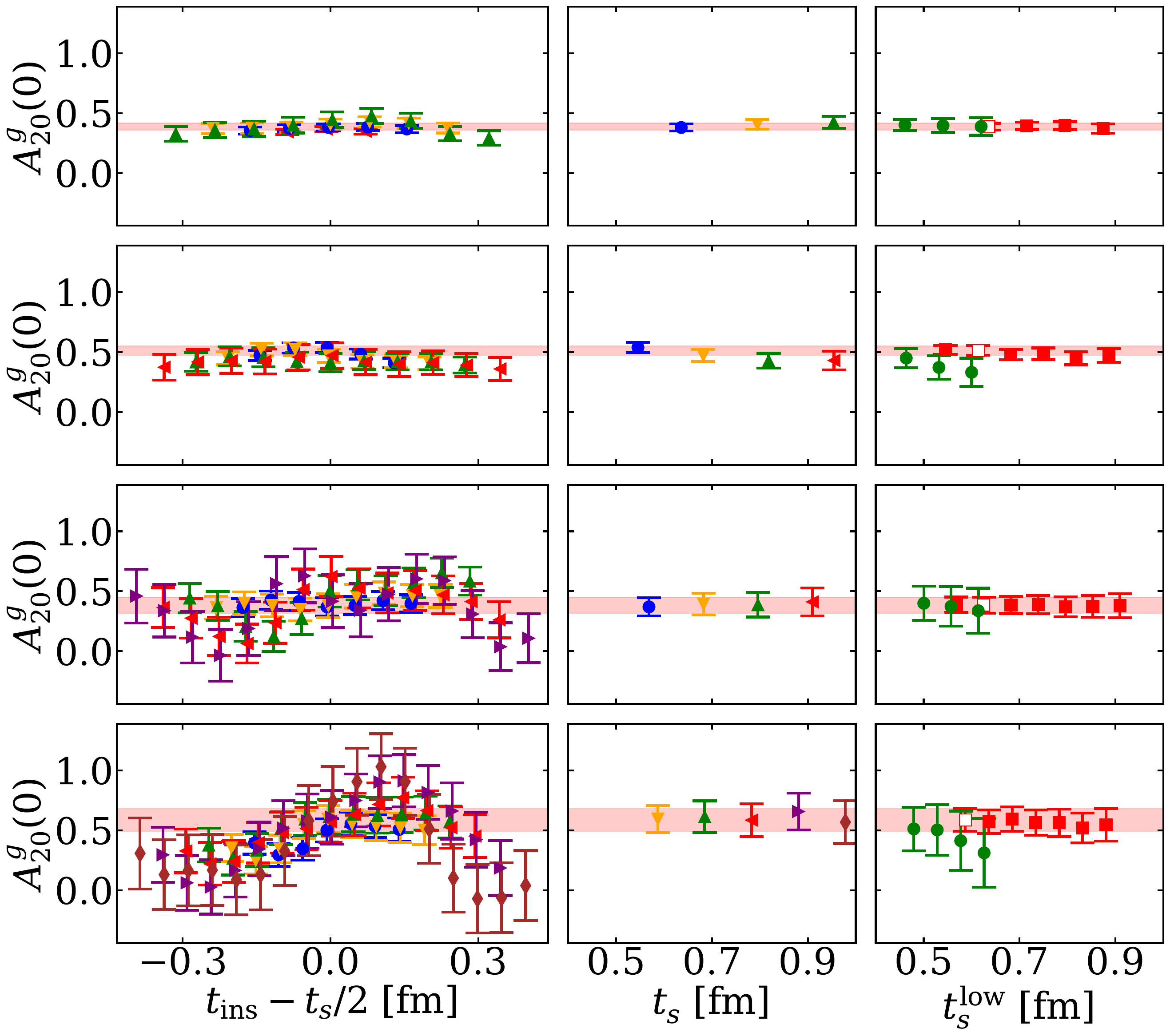}
    \caption{
        Results for the bare gluon ${A}_{20}^{g}(0)$ with stout smearing $n_s=10$.
        The notation is the same as in 
        \cref{fig:A20_2_jp_disc}. 
    }\label{fig:A20_2_jg_stout10}
\end{figure}

\subsection{Determination of  $A_{20}(Q^2)$ }

The momentum fraction $\braket{x}=A_{20}(0)$ is directly accessible from the forward-limit matrix elements in \cref{eq:pi44,eq:pi4i}. For connected contributions, only data with $\vpp=\vec{0}$ are available, so only \cref{eq:pi44} can be used. For disconnected contributions, for both quarks and gluons, the matrix element in \cref{eq:pi44} is significantly noisier than that in \cref{eq:pi4i}, as also observed in Ref.~\cite{Alexandrou:2020sml}. Therefore, for all disconnected contributions we only employ a boosted frame, namely \cref{eq:pi4i}.

For the  extraction of the isovector $A_{20}^{u-d}(0)$, we fit the ratio of \cref{eq:ratio} keeping terms up to the first excited state (two-state fits), where the ground state and first  excited state energies are fixed  from the corresponding two-state fits to the nucleon effective mass. 
Since in the forward-limit the ratio is expected to be symmetric under $t_{\rm ins}\leftrightarrow t_s-t_{\rm ins}$, we symmetrize it around the midpoint $t_s/2$. We perform different fits  within the range  $t_{\rm ins}^{\rm cut}\leq t_{\rm ins}\leq t_s/2$, eliminating time slices from the sink  to identify convergence to a time-independent value. We find that leaving $t_{\rm ins}^{\rm cut}\approx0.3~{\rm fm}$ from the sink is an appropriate starting range. We also vary the source-sink time separations satisfying $t_s\geq t_s^{\rm low}$. We find that   $t_s^{\rm low}\approx0.8~{\rm fm}$ is a suitable value for all ensembles. These values are chosen from examining convergence to a time-independent value. A representative example of the convergence dependence is shown in \cref{fig:A20_0_jm_conn_tins} for $A_{20}^{u-d}(0)$ computed using the
E112 ensemble. The selected values of $t_{\rm ins}^{\rm cut}$ and $t_s^{\rm low}$ are in agreement with subsequent values at larger values.  

For comparison and a cross-check, we also analyze the results using the summation method. Namely we perform a linear fit in $\ts$ to the following summed ratio
\begin{align}
R^{\mu\nu}_{\rm sum}
(\Gamma_\rho,\vec{p}^{\,\prime},\vec{p};t_s)
&=
\sum_{t_{\rm ins}=2a}^{t_s-2a}
R^{\mu\nu}(\Gamma_\rho,\vec{p}^{\,\prime},\vec{p};t_s,t_{\rm ins})
\nonumber\\
&=
c+\Pi^{\mu\nu}(\Gamma_\rho,\vec{p}^{\,\prime},\vec{p})\,t_s
+\mathcal{O}\left(e^{-\Delta E t_s}\right).
\label{eq:summed_ratio}
\end{align}
In \cref{fig:A20_0_jm_conn}, we show the bare ratio of \cref{eq:ratio}
 from which the bare isovector  $A^{u-d}_{20}(0)$ is extracted, as well as, the resulting values from the two-state  and summation method fits.  As can be seen, the selected values from the two-state fits show a good convergence and are in agreement with the values from summation method.

Since we have data for non-zero momentum transfer in the lab frame, we also extract the GFFs for non-zero $Q^2$, using the expressions  given in  Appendix \ref{app:gffs}. Here both $\mu=\nu$ and $\mu\ne\nu$ can be used. We opt to show only examples  for the $\mu=\nu$ case since the results for the $\mu\ne\nu$ are very similar.     In \cref{fig:A20_1_jm_conn}, we show the case corresponding to $Q^2\approx0.1~{\rm GeV}^2$, obtained from two units of momentum transfer. In this case, the ratio is no longer symmetric, and we therefore allow for asymmetric elimination of time slices from the source and sink and fit in the time range $t_{\rm ins}^{\rm cut,a}\leq t_{\rm ins}\leq t_s-t_{\rm ins}^{\rm cut,b}$ with $t_{\rm ins}^{\rm cut,a}+t_{\rm ins}^{\rm cut,b}=2\,t_{\rm ins}^{\rm cut}\approx0.6~{\rm fm}$. We chose $t_s^{\rm low}$   to be the same as that used in our analysis for $Q^2=0$. We perform fits for all possible choices of $(t_{\rm ins}^{\rm cut,a},t_{\rm ins}^{\rm cut,b})$ satisfying the above condition and select the value with the largest probability.

We perform a similar  analysis for the connected contribution to the isoscalar ratio. Since the disconnected contributions are extracted only from the $\mu\neq\nu$ matrix elements, we consistently use only the $\mu\neq\nu$ case also for the connected  contribution in the SVD analysis. However, the connected three-point function is only computed in the lab frame, and thus one cannot use \cref{eq:pi4i} to determine the $A^{u+d}_{20}(0)$ directly.  We instead compute $A^{u+d}_{20}(Q^2)$ for finite $Q^2$ values and extrapolated  to $Q^2=0$.

When using only the $\mu\neq\nu$ matrix elements, one unit of momentum transfer does not provide a sufficient number of equations to determine the GFFs, and therefore, we start from two units of momentum transfer. In \cref{fig:A20_2_jp_conn}, we show the extracted values corresponding to $Q^2\approx0.1~{\rm GeV}^2$. The fit ranges are chosen to be the same as for the isovector case, since we observe good convergence to the ground state with these parameters.

For the disconnected contributions, we employ only \cref{eq:pi4i}. In \cref{fig:A20_2_jp_disc,fig:A20_2_js_disc,fig:A20_2_jc_disc,fig:A20_2_jg_stout10}, we show the  ratios  for the disconnected $u+d$, $s$, $c$, and gluon contributions, respectively. 
Since we do not observe  significant excited state contributions within our statistical error, especially for the ensembles B64 and C80 where the statistical errors are the smallest,  we use one-state fits.  Fitting  to a constant at each   $t_s$ we can probe the dependence on $t_s$. We use $t_{\rm ins}^{\rm cut}\approx0.2~{\rm fm}$ for the quark-disconnected contributions and $t_{\rm ins}^{\rm cut}\approx0.3~{\rm fm}$ for the gluon contributions.
As can be seen, the values extracted within $0.5~{\rm fm}<t_s<0.9~{\rm fm}$  using constant fits  are fully consistent. Therefore, we perform a final constant fit to all values of $t_s$ starting from $t_s^{\rm low}\approx0.6~{\rm fm}$ and  using the same $t_{\rm ins}^{\rm cut}$. As can be seen in  \cref{fig:A20_2_jp_disc,fig:A20_2_js_disc,fig:A20_2_jc_disc,fig:A20_2_jg_stout10}, the values are stable as we vary $t_s^{\rm low}$ and in agreement with the values  obtained from fits at each $t_s$. 
As a cross-check, we also perform fits using the summation method, which are statistically nosier but compatible with the constant fit results.

\begin{figure}[!ht]
    \centering
    \includegraphics[width=\columnwidth]{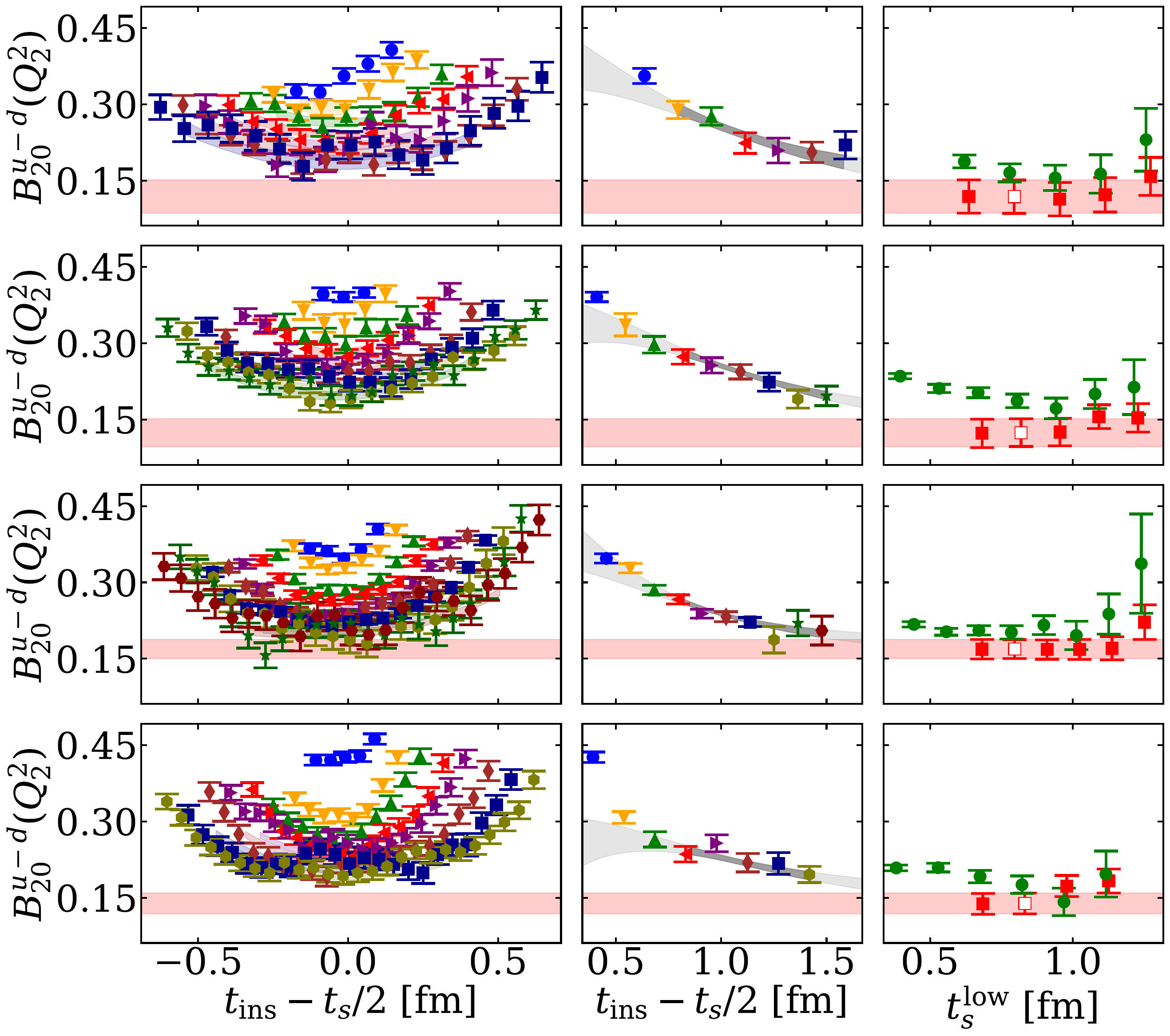}
    \caption{
        Results for $B_{20}^{u-d}(Q_2^2)$ for $Q_2^2\approx0.1$ GeV$^2$ that corresponds to two units of momentum transfer. The notation is the same as in \cref{fig:A20_0_jm_conn}.
    }\label{fig:B20_1_jm_conn}
\end{figure}

\begin{figure}[!ht]
    \centering
    \includegraphics[width=\columnwidth]{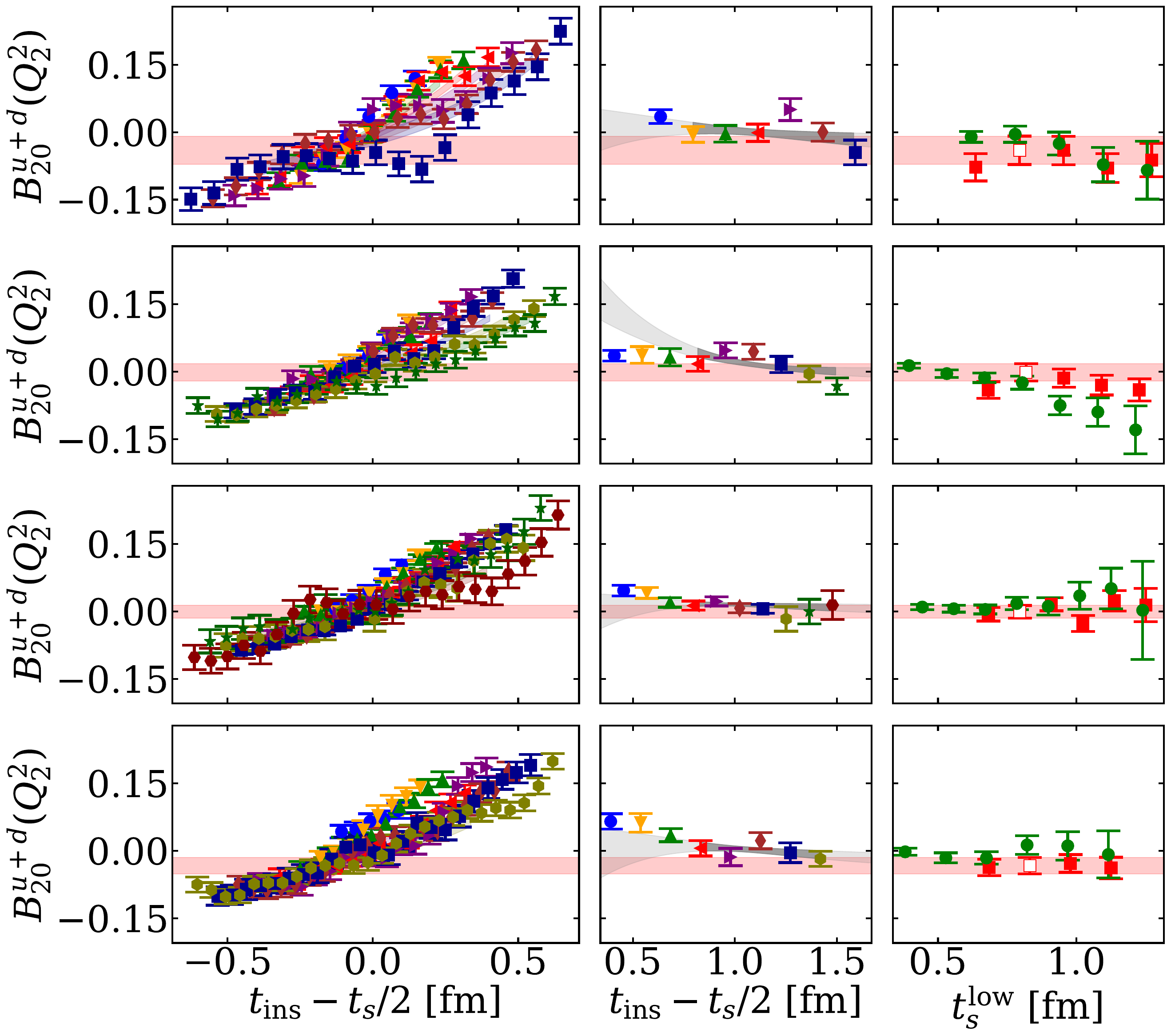}
    \caption{
        Results for ${B}_{20}^{u+d}(Q_2^2)$ (connected) for $Q_2^2\approx0.1$ GeV$^2$ that corresponds to two units of momentum transfer. The notation is the same as in \cref{fig:B20_2_jp_conn}.
    }\label{fig:B20_2_jp_conn}
\end{figure}

\begin{figure}[!ht]
    \centering
    \includegraphics[width=\columnwidth]{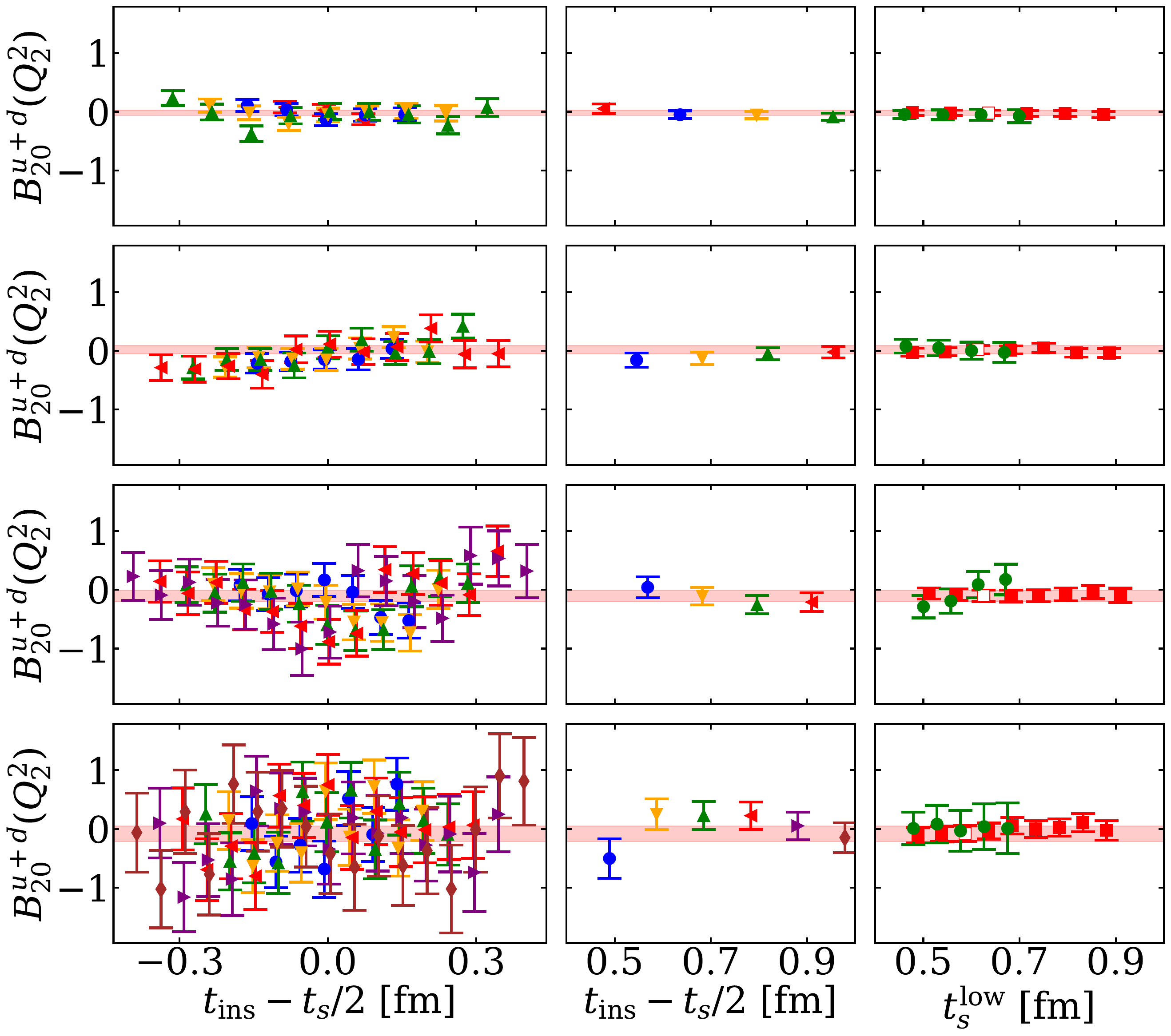}
    \caption{
        Results for ${B}_{20}^{u+d}(Q_2^2)$ (disconnected) for $Q_2^2\approx0.1$ GeV$^2$ that corresponds to two units of momentum transfer. The notation is the same as in \cref{fig:A20_2_jp_disc}.
    }\label{fig:B20_2_jp_disc}
\end{figure}

\begin{figure}[!ht]
    \centering
    \includegraphics[width=\columnwidth]{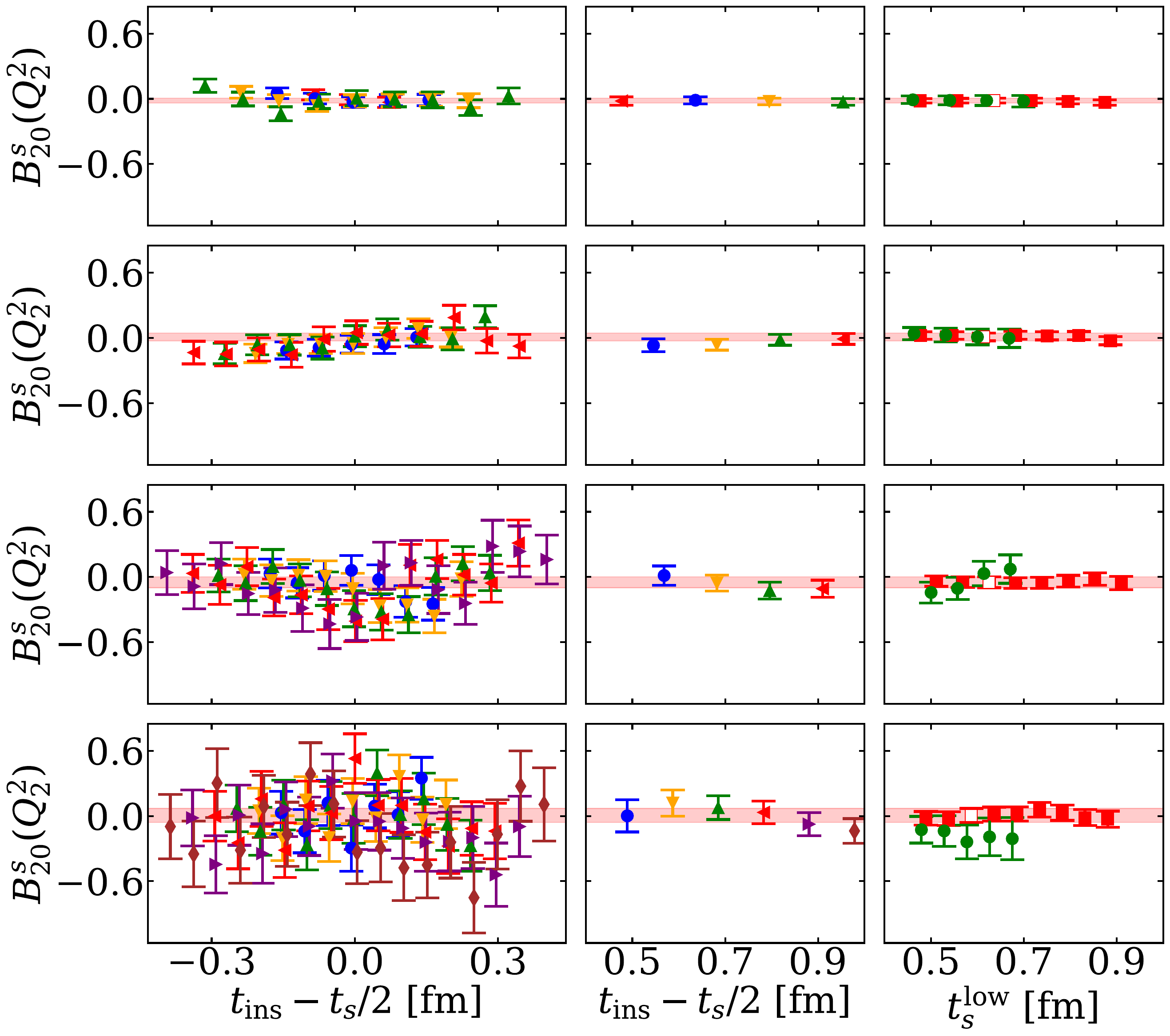}
    \caption{
        Results for ${B}_{20}^{s}(Q_2^2)$ for $Q_2^2\approx0.1$ GeV$^2$ that corresponds to two units of momentum transfer. The notation is the same as in \cref{fig:A20_2_jp_disc}.
    }\label{fig:B20_2_js_disc}
\end{figure}

\begin{figure}[!ht]
    \centering
    \includegraphics[width=\columnwidth]{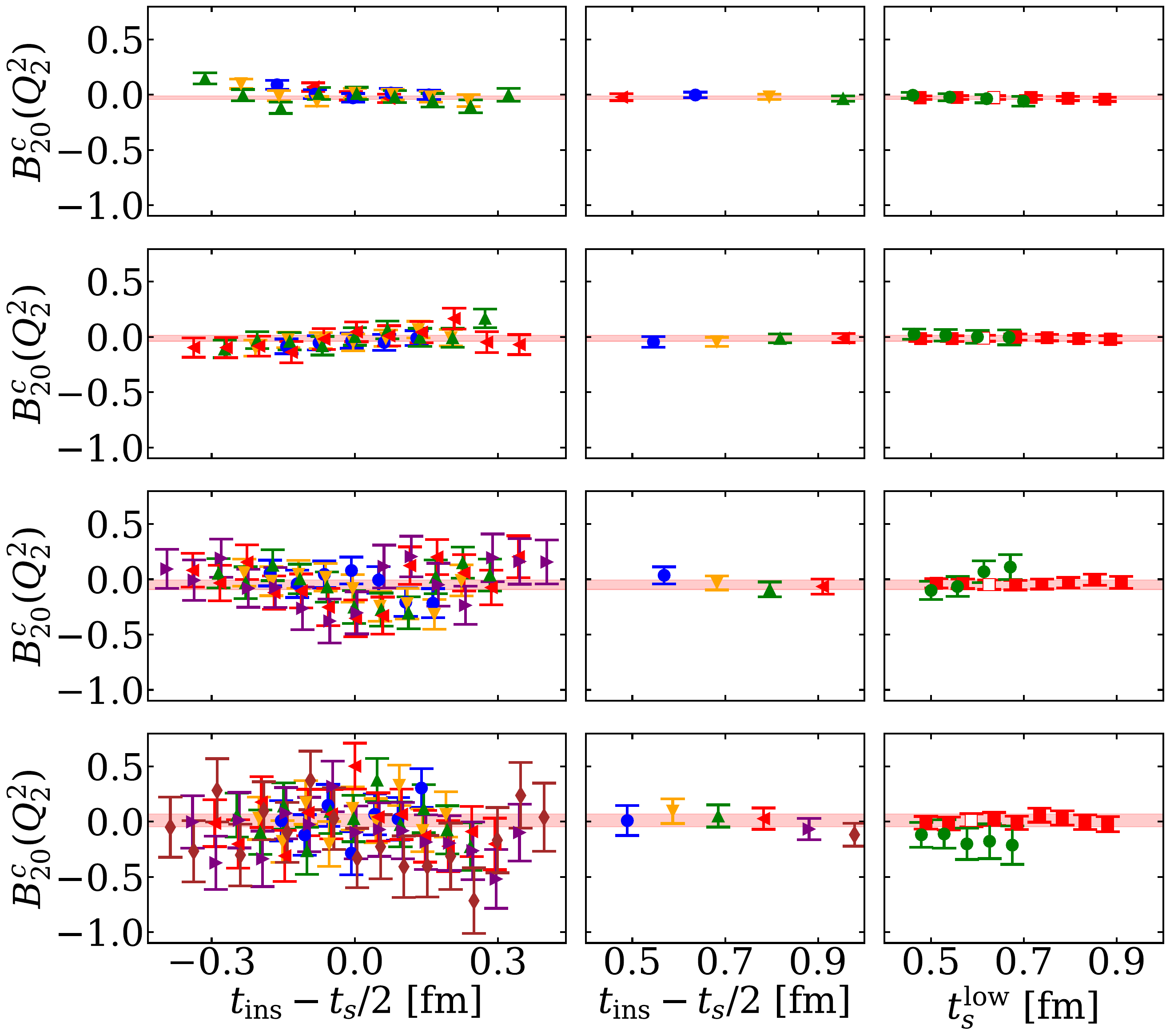}
    \caption{
        Results for ${B}_{20}^{c}(Q_2^2)$ for $Q_2^2\approx0.1$ GeV$^2$ that corresponds to two units of momentum transfer. The notation is the same as in \cref{fig:A20_2_jp_disc}.
    }\label{fig:B20_2_jc_disc}
\end{figure}

\begin{figure}[!ht]
    \centering
    \includegraphics[width=\columnwidth]{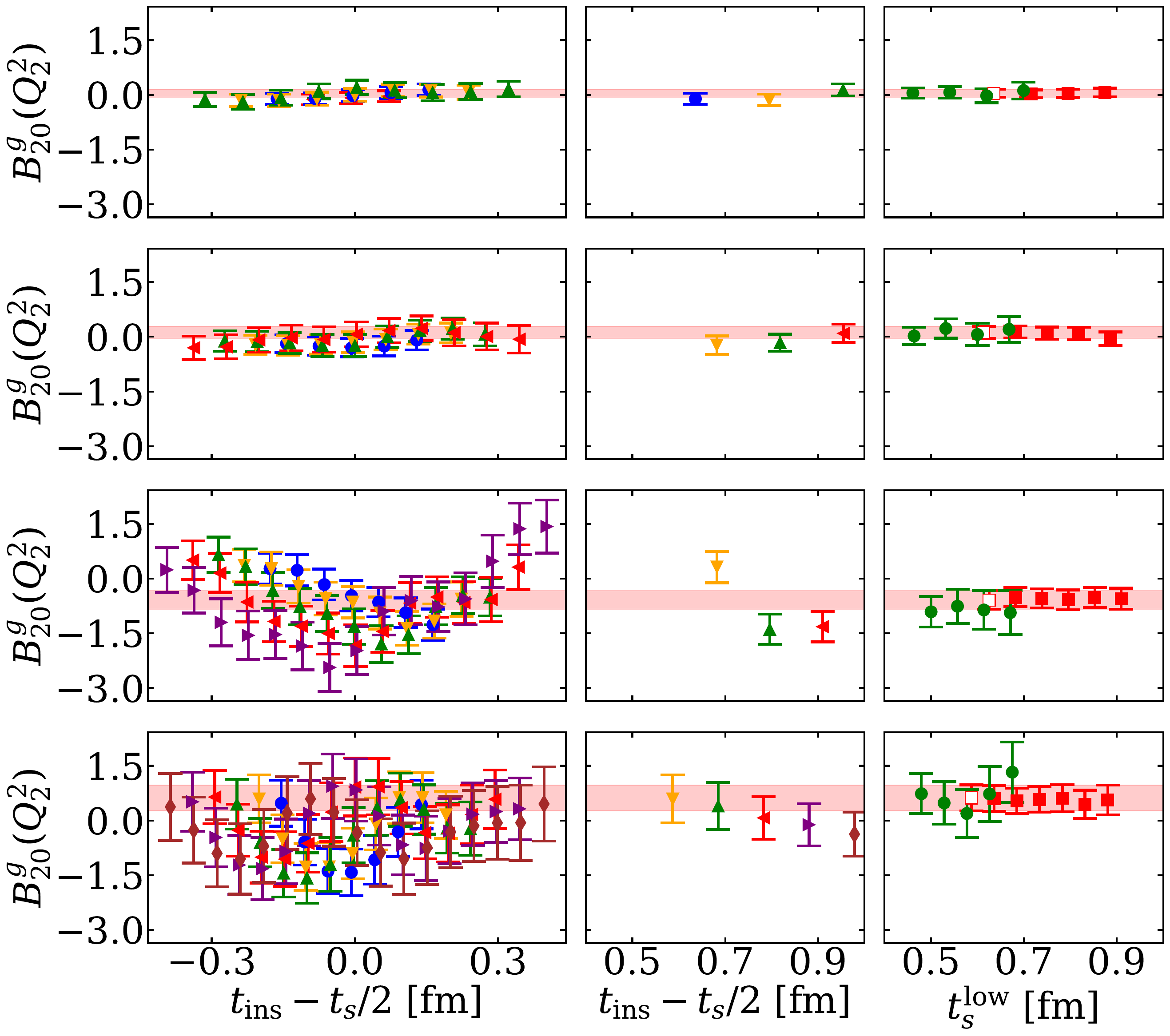}
    \caption{
        Results for ${B}_{20}^{g}(Q_2^2)$ with $n_s=10$ for $Q_2^2\approx0.1$ GeV$^2$ that corresponds to two units of momentum transfer. The notation is the same as in \cref{fig:A20_2_jp_disc}.
    }\label{fig:B20_2_jg_stout10}
\end{figure}

\subsection{Determination of $B_{20}(Q^2)$}
Unlike $A_{20}(Q^2)$, the GFF $B_{20}(Q^2)$ is not directly accessible at $Q^2=0$ and therefore requires an extrapolation from finite momentum transfer. Like for the case of the isovector $A^{u-d}_{20}(Q^2)$, one can use both the $\mu=\nu$ and $\mu\neq\nu$ EMT  and, for our final results, the analysis is carried out for both cases. To illustrate our analysis,   we only show examples for the $\mu\neq\nu$ case here since again the results are similar. 
In \cref{fig:B20_1_jm_conn}, we show the extraction of $B_{20}^{u-d}(Q^2)$ at $Q^2\approx0.1~{\rm GeV}^2$, corresponding to two units of momentum transfer. The extraction is performed using two-state fits, following the same strategy as for the connected $A_{20}(Q^2)$ analysis. In particular, we employ the same fit ranges, namely $t_s^{\rm low}\approx0.8~{\rm fm}$ and $t_{\rm ins}^{\rm cut,a}+t_{\rm ins}^{\rm cut,b}\approx0.6~{\rm fm}$.

For the connected contribution to the isoscalar ${B}_{20}^{u+d}(Q^2)$, as well as for all disconnected contributions, we  use the $\mu\neq\nu$ case matrix elements for the same reason as for the determination of these quantities for $A_{20}(Q^2)$. 
In \cref{fig:B20_2_jp_conn}, we show the results  for the connected  ${B}_{20}^{u+d}(Q^2)$ at $Q^2\approx0.1~{\rm GeV}^2$, corresponding to two units of momentum transfer. The excited state effects are treated using the same two-state fit procedure and fit ranges as in the isovector case.

The disconnected quark and gluon ratios are shown in
\cref{fig:B20_2_jp_disc,fig:B20_2_js_disc,fig:B20_2_jc_disc,fig:B20_2_jg_stout10} for $Q^2=0.1~{\rm GeV}^2$. As in the case of the corresponding ratios for the extraction of  $A_{20}(Q^2)$, no significant excited state contamination is observed.  Therefore, we  perform a similar analysis as for the disconnected contributions to $A_{20}(Q^2)$. Namely, we perform constant fits using $t_s^{\rm low}$ and $t_{\rm ins}^{\rm cut,a}+t_{\rm ins}^{\rm cut,b}=2\,t_{\rm ins}^{\rm cut}$ that are consistent with the choices adopted in the disconnected $A_{20}$ analysis and make a cross-check with the summation method.

\begin{figure*}[h!]
    \centering
    \includegraphics[width=\textwidth]{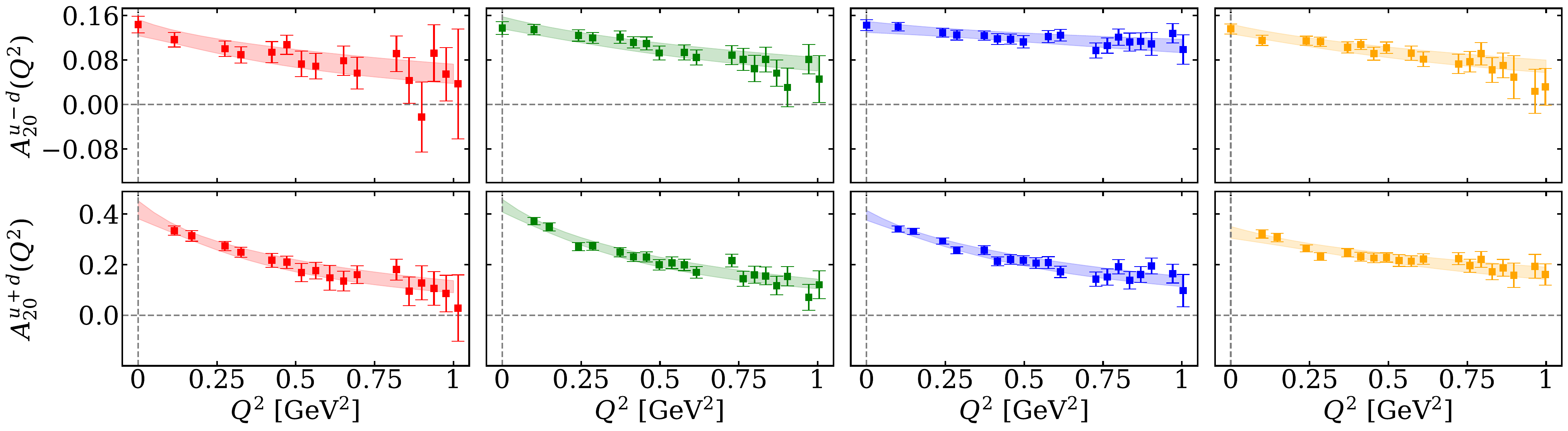}
    \caption{
        Results for  bare ${A}^{u-d}_{20}(Q^2)$ for $\mu=\nu $ (top row) and connected ${A}^{u+d}_{20}(Q^2)$ for $\mu\ne\nu$ (bottom row) as functions of $Q^2$ for the B64, C80, D96, and E112 ensembles (from left to right). The bands correspond to the dipole fits used to obtain the form factors at $Q^2=0$.
    }\label{fig:A20_Q2_conn}
\end{figure*}

\begin{figure*}[!ht]
    \centering
    \includegraphics[width=\textwidth]{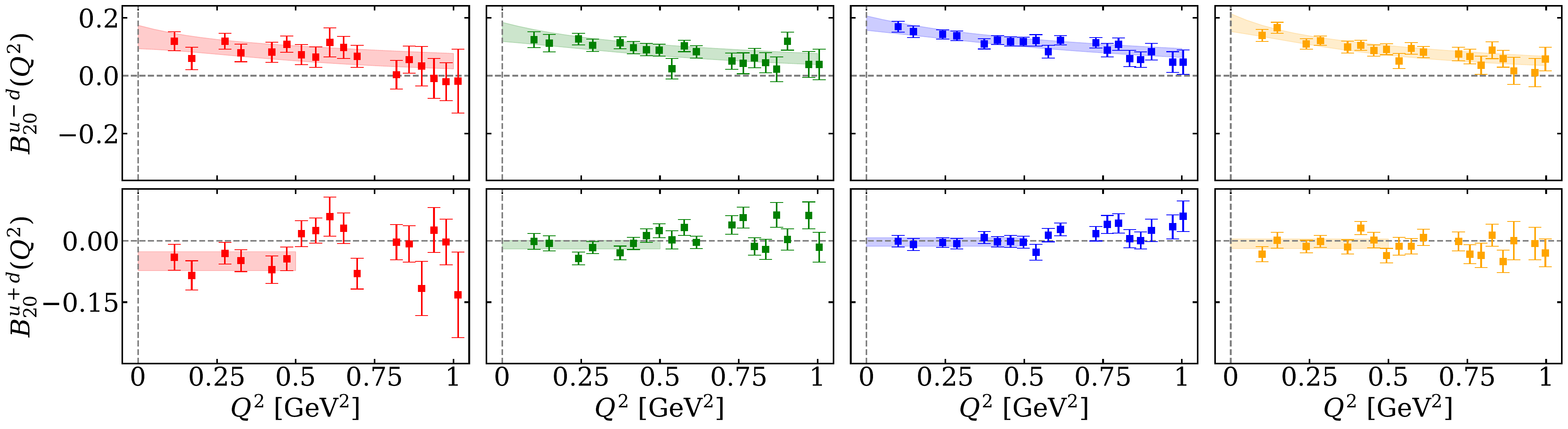}
    \caption{
        Results for connected bare ${B}^{u-d}_{20}(Q^2)$ for $\mu\ne\nu$ (top row) and ${B}^{u+d}_{20}(Q^2)$ for $\mu\ne\nu$ (bottom row). The notation is the same as in \cref{fig:A20_Q2_conn}. $\tilde{B}^{u+d}_{20}(Q^2)$ is fitted with a constant model up to $Q^2 = 0.5$ GeV$^2$.
    }\label{fig:B20_Q2_conn}
\end{figure*}

\begin{figure*}[!ht]
    \centering
    \includegraphics[width=\textwidth]{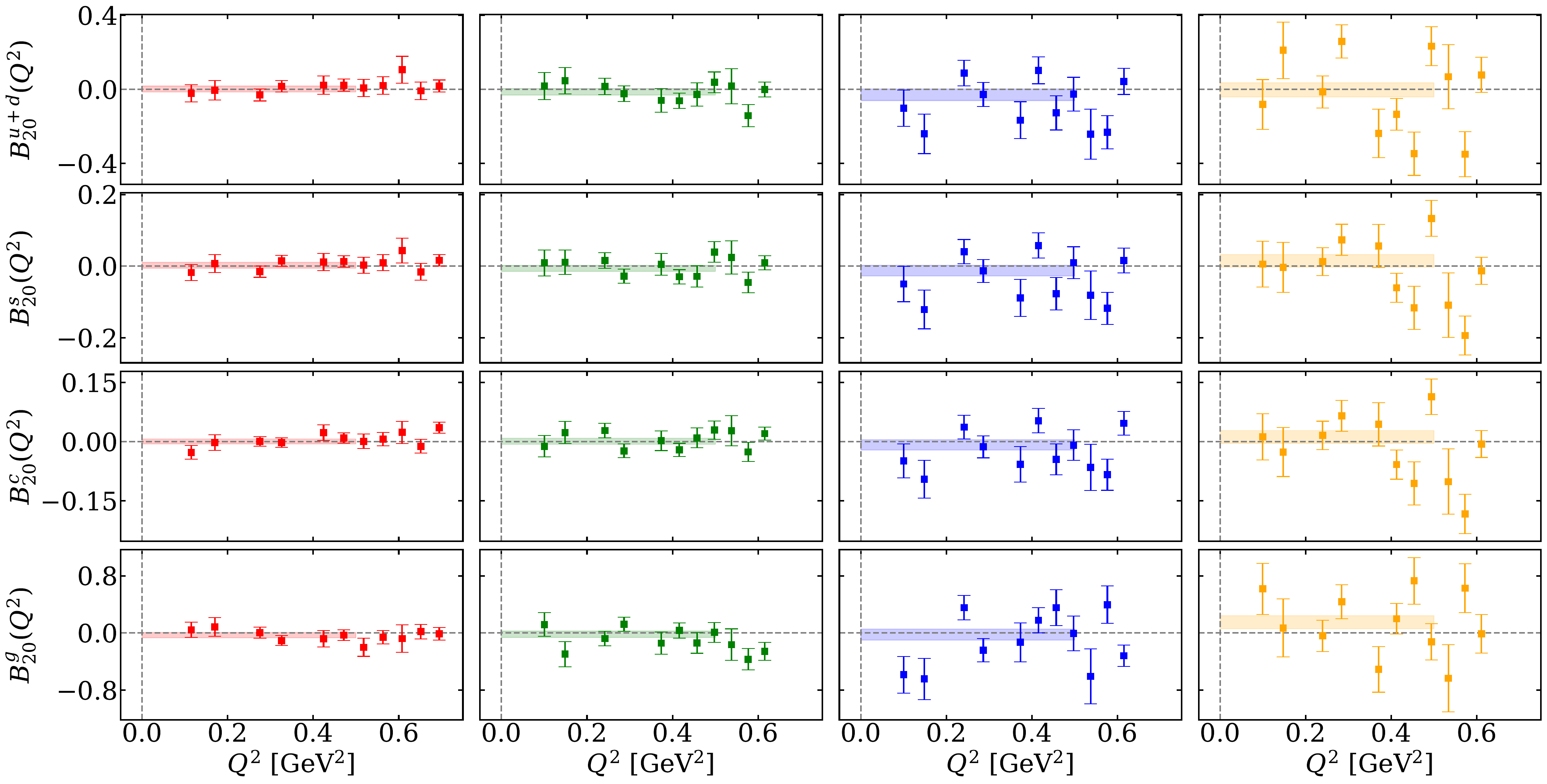}
    \caption{
        Results for disconnected bare ${B}^{q,g}_{20}(Q^2)$ for $\mu\ne\nu$. The notation is the same as in \cref{fig:A20_Q2_conn}. They are all fitted with a constant model up to $Q^2 = 0.5$ GeV$^2$.
    }\label{fig:B20_Q2_disc}
\end{figure*}

\subsection{Determination of $A_{20}(0)$ and $B_{20}(0)$}
For the isovector flavor combination $u-d$, one can increase statistics by using both $\mu=\nu$ and $\mu\ne\nu$ matrix elements. 
Here we opt to illustrate the $Q^2$-dependence using only  the $\mu=\nu$ case for $A^{u-d}(Q^2)$, where  the value determined directly at $Q^2=0$ is also available, and the $\mu \ne\nu$ case  for  $B^{u-d}(Q^2)$. This allows us to show bare quantities as done for the isoscalar combination $u+d$ and all disconnected contributions, where  only the $\mu\neq\nu$ ratios are used.

Performing the analysis for momentum transfers up to $Q^2\approx1~{\rm GeV}^2$, we obtain $A_{20}^{u-d}(Q^2)$ and connected $A_{20}^{u+d}(Q^2)$ for all ensembles. We fit the $Q^2$-dependence using a dipole Ansatz
\begin{align}
f(Q^2)=\frac{g}{(1+Q^2/m^2)^2},
\end{align}
from which we determined $A^{u-d}_{20}(0)$ and the connected $A^{u+d}_{20}(0)$.
We show the results in \cref{fig:A20_Q2_conn}, where for the isovector case,  we also include the value of $A_{20}(0)$ extracted directly at $Q^2=0$ in the fit but still allowing  $g$ to be a fit parameter. As can be seen, the dipole fit provides a good description of the data. This determination makes best use of all the data on the connected contributions. For the disconnected, we opt to use the value extracted directly at $Q^2=0$ since there we can have the $\mu\ne\nu$ case in the boosted frame.

We repeat a similar  analysis  for the isovector  connected  $B^{u-d}_{20}(Q^2)$ using a dipole Ansatz.  The results are shown in Fig.~\ref{fig:B20_Q2_conn}, showing again that the dipole fit provides a good description of  the  extrapolation to $Q^2=0$. For the isoscalar case, $B^{u+d}_{20}(Q^2)$, on the other hand, the   connected  contributions show no significant $Q^2$-dependence and and thus we perform a constant  up to $Q^2\approx0.5~{\rm GeV}^2$ to extrapolate to the $Q^2=0$ values, as shown in Fig.~\ref{fig:B20_Q2_conn}. Similarly,  no statistically significant $Q^2$-dependence is observed for all disconnected contributions, as shown in Fig.~\ref{fig:B20_Q2_disc}. We therefore employ a constant fit  up to $Q^2\approx0.5~{\rm GeV}^2$ to determine the $Q^2=0$ values for all disconnected contributions to $B_{20}(Q^2)$.

\section{Renormalization}\label{sec:renormalization}

\subsection{Theoretical Setup}
The matrix elements of $T^{\mu\nu}_{q}$ and 
$T^{\mu\nu}_{g}$ are renormalized nonperturbatively 
in the RI$'$-MOM scheme~\cite{Martinelli:1994ty} and converted 
perturbatively to the $\overline{\rm MS}$ scheme 
at the scale $\bar{\mu} = 2$ GeV. We consider the quark flavor non-singlet ($ns$) combinations  $T^{\mu\nu;ns}_{q} \in \{T^{\mu\nu}_{u-d}$, $T^{\mu\nu}_{u+d-2s}$, $T^{\mu\nu}_{u+d+s-3c}\}$, as well as the flavor-singlet ($s$) combination  $T^{\mu\nu;s}_{q} \equiv T^{\mu\nu}_{u+d+s+c}$, which have definite renormalization properties. Renormalized matrix elements of the individual quark flavors are obtained from appropriate linear combinations of the renormalized singlet and nonsinglet components. The nonsinglet operators are renormalized multiplicatively with a common renormalization factor $Z_{qq}$,
\begin{equation}
    (T^{\mu\nu;ns}_{q})^{\mathrm{R}} = Z_{qq} \, T^{\mu\nu;ns}_{q}.
\end{equation}
In contrast, the quark singlet and gluon EMT components mix under renormalization~\cite{Caracciolo:1991cp}. Also, according to global symmetries, they can mix with gauge-noninvariant operators, which are BRST-variations of some operators or are vanished by the equations of motion. In the continuum and for a minimal renormalization scheme, the mixing set can be reduced to three operators, as discussed in Ref.~\cite{Panagopoulos:2020qcn}\footnote{From the calculation of Ref.~\cite{Panagopoulos:2020qcn} in dimensional regularization, it is concluded that finite mixing effects from other gauge-noninvariant operators are also absent in the RI$'$-MOM scheme, at least up to two loops, when Landau gauge is employed.}. Then, the renormalization of the quark singlet and gluon EMT components, up to possible finite mixing effects on the lattice, is given by 
\begin{equation}
  \begin{pmatrix}
    (T^{\mu\nu;s}_{q})^{\mathrm{R}}\\
    (T^{\mu\nu}_{g})^{\mathrm{R}}\\
    (T^{\mu\nu}_{c})^{\mathrm{R}}
  \end{pmatrix}
  =
  \begin{pmatrix}
    Z_{qq}^s & Z_{qg} & Z_{qc}\\
    Z_{gq} & Z_{gg} & Z_{gc}\\
    Z_{cq} & Z_{cg} & Z_{cc}\\
  \end{pmatrix}
  \begin{pmatrix}
    T^{\mu\nu;s}_{q} \\
    T^{\mu\nu}_{g} \\
    T^{\mu\nu}_{c}
  \end{pmatrix},
  \label{Eq:3x3_mixing_matrix}
\end{equation}
where $T^{\mu\nu}_{c}$ is a gauge-noninvariant operator defined in Eq.\,(127) of Ref.~\cite{Panagopoulos:2020qcn}. This operator does not enter physical matrix elements. In this case, the mixing matrix is reduced to $2\times2$,
\begin{equation}
  \begin{pmatrix}
    (T^{\mu\nu;s}_{q})^{\mathrm{R}}\\
    (T^{\mu\nu}_{g})^{\mathrm{R}}
  \end{pmatrix}
  =
  \begin{pmatrix}
    Z_{qq}^s & Z_{qg}\\
    Z_{gq} & Z_{gg} \\
  \end{pmatrix}
  \begin{pmatrix}
    T^{\mu\nu;s}_{q} \\
    T^{\mu\nu}_{g}
  \end{pmatrix}.
  \label{Eq:2x2_mixing_matrix}
\end{equation}
However, when considering the RI$'$-MOM scheme, which involves  Green's functions with elementary external fields, the mixing with the gauge-noninvariant operator cannot be ignored {\it a priori}. In previous nonperturbative renormalization studies, $T^{\mu\nu}_{c}$ was always neglected and a $2\times 2$ mixing matrix was adopted. A complication in including such operator is that it depends on ghost fields, and thus, the nonperturbative calculation of its vertex functions is not feasible by compact lattice simulations. However, as explained below, in determining the $2\times2$ block of the mixing matrix in Eq.~\eqref{Eq:2x2_mixing_matrix} that survives in the calculation of the matrix elements, the vertex functions of $T^{\mu\nu}_{c}$ enter only in the very final step. 
Thus, in the current work, we keep the $3\times3$ form of the mixing matrix in all intermediate steps of the analysis, and in the last step we truncate the vertex functions of $T^{\mu\nu}_{c}$ to their tree-level values. Our approach can suppress ultra-violet (UV)-divergent mixing effects in the determination of the RI$'$-MOM renormalization coefficients, stemming from gauge-nonvariant operators.

 Due to the breaking of rotational symmetry on the lattice, operators with $\mu=\nu$ and $\mu \neq \nu$ belong to different irreducible representations of the hypercubic group $H(4)$, and thus, they are renormalized with a different renomalization factor or mixing matrix. In our work, we consider both $\mu=\nu$ and $\mu \neq \nu$ cases for the nonsinglet operators, whereas for the singlet operators we restrict ourselves to $\mu \neq \nu$, which provides an improved signal-to-noise ratio.

The RI$'$-MOM scheme is defined for vertex functions of the operators under study with external offshell quark or gluon states in the Landau gauge,
\begin{align}
    &\Lambda^{\mu\nu}_{i} (p) \equiv \frac{a^{12}}{V} \sum_{x,y,z} e^{-i p (x-y)} \, \left\langle q (x) \, T^{\mu\nu}_i (z) \, \bar{q} (y) \right\rangle_{\rm amp.}, \\
    &V^{\mu\nu\rho\sigma}_{i} (p) \equiv \frac{a^{12}}{4 V} \sum_{x,y,z} e^{-i p (x-y)} e^{-i p (\hat{\rho}-\hat{\sigma})/2} \, \times\nonumber \\
    & \qquad \qquad \qquad \langle {\rm Tr}[A^{\rho} (x + \frac{\hat{\rho}}{2}) A^{\sigma} (y + \frac{\hat{\sigma}}{2})] \, T^{\mu\nu}_i (z) \rangle,
\end{align}
where $i \in \{q,g,c\}$ and 
\begin{eqnarray}
A^{\mu} (x + \frac{\hat{\mu}}{2}) &=& \frac{1}{2i g_0} \Big[(U_\mu (x) - U^\dagger_\mu (x)) - \nonumber \\
   && \ \frac{1}{3} {\rm Tr} (U_\mu (x) - U^\dagger_\mu (x)) \openone \Big] + \mathcal{O} (a^2). \,   
\end{eqnarray}
The fermionic vertex functions $\Lambda^{\mu\nu}_{i} (p)$ are taken to be amputated, while the gluonic vertex functions $V^{\mu\nu\rho\sigma}_{i} (p)$ are kept nonamputated. The reason is that the gluon propagator, which enters the process of amputation, is not invertible in Landau gauge. Although one can overcome this issue by setting some components of the momentum $p$ to zero~\cite{Panagopoulos:2020qcn}, this choice does not allow the use of democratic momenta, which have smaller rotational breaking effects. Thus, we do not follow this approach in this work. For the nonsinglet combinations, which are multiplicatively renormalizable, we need only the fermionic vertex functions. Given that all three nonsinglet combinations share the same renormalization factor with the nonsinglet flavor nondiagonal operator $T^{\mu\nu}_{\bar{u}d}$, we employ this operator for the extraction of $Z_{qq}$.

In the continuum limit, $\Lambda^{\mu\nu}_{i} (p)$ and $V^{\mu\nu\rho\sigma}_{i} (p)$ are decomposed into a number of independent structures allowed by Lorentz symmetry~\cite{Panagopoulos:2020qcn}. Some of the structures arise only at higher perturbative order, while at tree level the vertex functions take, in Euclidean space, the following form 
\begin{eqnarray}
[\Lambda^{\mu\nu}_{q} (p)]^{\rm tree} &=& i \gamma^{\{\mu} p^{\nu\}}, \\
{[V^{\mu\nu\rho\sigma}_{g} (p)]}^{\rm tree} &=& \frac{1}{p^2} \Bigg[(\delta^{\mu\rho} \delta^{\nu\sigma} + \delta^{\mu\sigma} \delta^{\nu\rho}) - \nonumber \\
&& \hspace{-2cm}(\delta^{\mu\rho} \frac{p^{\nu} p ^{\sigma}}{p^2} + \delta^{\mu\sigma} \frac{p^{\nu} p ^{\rho}}{p^2} + \delta^{\nu\rho} \frac{p^{\mu} p ^{\sigma}}{p^2} + \delta^{\nu\sigma} \frac{p^{\mu} p ^{\rho}}{p^2}) - \nonumber \\
&& \ \delta^{\mu\nu}  (\delta^{\rho\sigma} - \frac{p^{\rho} p^{\sigma}}{p^2}) +2\delta^{\rho\sigma} \frac{p^{\mu} p^{\nu}}{p^2} \Bigg], \\
{[V^{\mu\nu\rho\sigma}_{c} (p)]}^{\rm tree} &=& \frac{1}{p^2} \Bigg[-2 \,(\delta^{\mu\rho} \delta^{\nu\sigma} + \delta^{\mu\sigma} \delta^{\nu\rho}) + \nonumber \\
&& \hspace{-2cm}2\,(\delta^{\mu\rho} \frac{p^{\nu} p ^{\sigma}}{p^2} + \delta^{\mu\sigma} \frac{p^{\nu} p ^{\rho}}{p^2} + \delta^{\nu\rho} \frac{p^{\mu} p ^{\sigma}}{p^2} + \delta^{\nu\sigma} \frac{p^{\mu} p ^{\rho}}{p^2}) + \nonumber \\
&& \delta^{\mu\nu}  (\delta^{\rho\sigma} - \frac{p^{\rho} p^{\sigma}}{p^2}) -4\,\frac{p^{\mu} p^{\nu} p^{\rho} p^{\sigma}}{{(p^2)}^2} \Bigg], \\
{[\Lambda^{\mu\nu}_{g} (p)]}^{\rm tree} &=& {[\Lambda^{\mu\nu}_{c} (p)]}^{\rm tree} = {[V^{\mu\nu\rho\sigma}_{q} (p)]}^{\rm tree} = 0.
\end{eqnarray}
In perturbation theory, the renormalization conditions are typically defined in terms of the part of each vertex function that is proportional to its tree-level structure. In nonperturbative studies, these parts can be isolated by applying suitable projectors. In the case of singlet operators, where mixing is present,  isolating these terms can be done by projecting to specific sub-structures that are protected from mixing. In particular, the terms $\delta^{\rho\sigma} p^{\mu} p^{\nu}/{(p^2)}^2$ in ${[V^{\mu\nu\rho\sigma}_{g} (p)]}^{\rm tree}$ and $p^{\mu} p^{\nu} p^{\rho} p^{\sigma}/{(p^2)}^3$ in ${[V^{\mu\nu\rho\sigma}_{c} (p)]}^{\rm tree}$ are the only sub-structures of gluonic vertex functions that are not shared among the tree-level vertex functions of the mixing operators. Forming the conditions through the projection of these sub-structures allows us to determine each element of the inverse of the mixing matrix in terms of one ``projected'' vertex function for each element, as given in   Eqs.\eqref{Ziq}--\eqref{Zic}. We stress that when considering democratic momenta, which have nonzero components, it is more convenient to isolate the combinations $2\delta^{\rho\sigma} p^{\mu} p^{\nu}/{(p^2)}^2-(\delta^{\mu\rho} p^\nu p^\sigma + \delta^{\mu\sigma} p^\nu p^\rho + \delta^{\nu\rho} p^\mu p^\sigma + \delta^{\nu\sigma} p^\mu p^\rho)/{(p^2)}^2$ and $2 p^{\mu} p^{\nu} p^{\rho} p^{\sigma}/{(p^2)}^3 -(\delta^{\mu\rho} p^\nu p^\sigma + \delta^{\mu\sigma} p^\nu p^\rho + \delta^{\nu\rho} p^\mu p^\sigma + \delta^{\nu\sigma} p^\mu p^\rho)/{(p^2)}^2$, appearing in the tree-level structures ${[V^{\mu\nu\rho\sigma}_{g} (p)]}^{\rm tree}$ and ${[V^{\mu\nu\rho\sigma}_{c} (p)]}^{\rm tree}$, respectively. These combinations are also protected from mixing as long as $\delta^{\rho\sigma} p^{\mu} p^{\nu}/{(p^2)}^2$ and $p^{\mu} p^{\nu} p^{\rho} p^{\sigma}/{(p^2)}^3$ do not vanish simultaneously.

Based on these considerations, we impose the following conditions, \footnote{It is understood that $\Lambda^{\mu\nu;ns}_{q}$ denotes $\Lambda^{\mu\nu}_{q}$ using the nonsinglet operator $T^{\mu\nu;ns}_{q}$. Similarly, in Eqs. (\ref{Ziq}--\ref{Zic}), the singlet operator $T^{\mu\nu;s}_{q}$ is employed when $i=q$.}
\begin{eqnarray}
{(Z_{qq}^{\mu \neq \nu})}^{{\rm RI}'} {(Z_q^{{\rm RI}'})}^{-1} \frac{1}{72} \sum_{\mu<\nu} {\rm Tr} [\Lambda_{qq}^{\mu\nu;ns} \, \mathcal{P}_q^{\mu\neq\nu}] \Big|_{p^2 = \mu_0^2} \!\!\! &=& \!\!1, \quad \ \ \\
{(Z_{qq}^{\mu = \nu})}^{{\rm RI}'} {(Z_q^{{\rm RI}'})}^{-1} \frac{1}{48} \sum_{\mu} {\rm Tr} [\Lambda_{qq}^{\mu\mu;ns} \, \mathcal{P}_q^{\mu=\nu}] \Big|_{p^2 = \mu_0^2} \!\!\! &=& \!\!1, \quad \ \
\end{eqnarray}
\begin{eqnarray}
{(\overline{Z}_{iq}^{\mu \neq \nu})}^{{\rm RI}'} \!\! &=& \!\! {(Z_q^{{\rm RI}'})}^{-1} \frac{1}{72} \sum_{\mu<\nu} {\rm Tr} [\Lambda_{i}^{\mu\nu} \, \mathcal{P}_q^{\mu\neq\nu}] \Big|_{p^2 = \mu_0^2}, \quad \label{Ziq} \\ 
{(\overline{Z}_{ig}^{\mu \neq \nu})}^{{\rm RI}'} \!\! &=& \!\! Z_g^{{\rm RI}'} \frac{1}{6} \sum_{\mu<\nu} \sum_\rho V_{i}^{\mu\nu\rho\rho} \, \mathcal{P}_g^{\mu\neq\nu,\rho} \Big|_{p^2 = \mu_0^2}, \\
{(\overline{Z}_{ic}^{\mu \neq \nu})}^{{\rm RI}'} \!\! &=& \!\! Z_g^{{\rm RI}'} \frac{1}{6} \sum_{\mu<\nu} \sum_\rho V_{i}^{\mu\nu\rho\rho} \, \widetilde{\mathcal{P}}_g^{\mu\neq\nu,\rho} \Big|_{p^2 = \mu_0^2}, \quad \label{Zic}
\end{eqnarray}
where $i \in \{q,g,c\}$,  $\overline{Z}$ is the matrix to be inverted only  at the final step,
\begin{equation}
  \overline{Z} = \begin{pmatrix}
    Z_{qq}^s & Z_{qg} & Z_{qc}\\
    Z_{gq} & Z_{gg} & Z_{gc}\\
    Z_{cq} & Z_{cg} & Z_{cc}\\
  \end{pmatrix}^{-1},  
\end{equation}
and 
\begin{eqnarray}
   \mathcal{P}_q^{\mu\neq\nu} &=&  -i\,\frac{2 (\slashed{\tilde{p}}^\mu + \slashed{\tilde{p}}^\nu) - \slashed{\tilde{p}}}{\tilde{p}^\mu \tilde{p}^\nu}, \\
   \mathcal{P}_q^{\mu=\nu} &=&  -\frac{4i\gamma_\mu}{3\tilde{p}^\mu}, \\
   \mathcal{P}_g^{\mu\neq\nu,\rho} &=& \frac{{(\tilde{p}^2)}^2}{4 \tilde{p}^\mu \tilde{p}^\nu} \left\{ 
   \begin{array}{cc}
       \frac{\tilde{p}^2-{(\tilde{p}^{\mu})}^2-{(\tilde{p}^{\nu})}^2}{2\tilde{p}^2-{(\tilde{p}^{\mu})}^2-{(\tilde{p}^{\nu})}^2},  & \rho \in \{\mu,\nu\} \\
       1, & \rho \notin \{\mu,\nu\}
   \end{array}
   \right., \\
   \widetilde{\mathcal{P}}_g^{\mu\neq\nu,\rho} &=& \frac{{(\tilde{p}^2)}^2}{4 \tilde{p}^\mu \tilde{p}^\nu} \left\{ 
   \begin{array}{cc}
       \frac{\tilde{p}^2}{2\tilde{p}^2-{(\tilde{p}^{\mu})}^2-{(\tilde{p}^{\nu})}^2},  & \rho \in \{\mu,\nu\} \\
       0, & \rho \notin \{\mu,\nu\}
   \end{array}
   \right..
\end{eqnarray}
$\tilde{p}^\mu \equiv \sin(a p^\mu)$, $\slashed{\tilde{p}}^\mu \equiv \gamma^\mu \tilde{p}^\mu$ (no sum), $\slashed{\tilde{p}} = \sum_\mu \slashed{\tilde{p}}^\mu$ and $\tilde{p}^\mu\neq 0$, $\tilde{p}^\nu \neq 0$. $\mu_0$ represents the RI$'$-MOM scale. $Z_q^{{\rm RI}'}$ and $Z_g^{{\rm RI}'}$are the renormalization factors of the quark and gluon fields, respectively, defined by~\cite{ExtendedTwistedMass:2021gbo}
\begin{eqnarray}
     Z_q^{{\rm RI}'} &=& \frac{1}{12} {\rm Tr} \left[S^{-1} (p) \cdot \sum_{\mu} \frac{-i \ \gamma^\mu}{4 {\tilde{p}}^\mu}\right] \Big|_{p^2 = \mu_0^2}, \\
     {(Z_g^{{\rm RI}'})}^{-1} &=& \left[\sum_\mu D^{\mu\mu} (p) \,\frac{\hat{p}^2}{3}\right] \Big|_{p^2 = \mu_0^2},
\end{eqnarray}
where $S(p)$ and $D^{\mu\nu} (p)$ are the quark and gluon propagators, respectively, in the momentum space,
\begin{eqnarray}
    S (p) &=& \frac{a^{8}}{V} \sum_{x,y} e^{-i p (x-y)} \langle q (x) \, \bar{q} (y) \rangle, \\
    D^{\mu\nu} (p) &=& \frac{a^{8}}{4 V} \sum_{x,y} e^{-i p (x-y)} e^{-i p (\hat{\mu} - \hat{\nu})/2} \times \nonumber \\
    && \quad \langle {\rm Tr}[A^{\mu} (x + \frac{\hat{\mu}}{2}) A^{\nu} (y + \frac{\hat{\nu}}{2})] \rangle,
\end{eqnarray}
and $\hat{p} \equiv 4 \sum_\mu \sin^2(ap^\mu/2)$. Note that we avoid using $V_{i}^{\mu\nu\rho\sigma}$ with $\rho \neq \sigma$ as they are more noisy. 

The RI$'$-MOM renormalization factors $Z_{qq}^{\mu\neq\nu}$, $Z_{qq}^{\mu=\nu}$ and mixing matrix $\bar{Z}^{\mu\neq\nu}$ are converted to the $\overline{\rm MS}$ scheme at the reference scale $\bar{\mu} = 2$ GeV through
\begin{eqnarray}
    {(Z_{qq}^{\mu\neq\nu})}^{\overline{\rm MS}} (\bar{\mu}) &=& C_{qq}^{\overline{\rm MS},{\rm RI}'} (\bar{\mu},\mu_0) \, {(Z_{qq}^{\mu\neq\nu})}^{{\rm RI}'} (\mu_0), \label{Cqq} \\
    {(Z_{qq}^{\mu=\nu})}^{\overline{\rm MS}} (\bar{\mu}) &=& C_{qq}^{\overline{\rm MS},{\rm RI}'} (\bar{\mu},\mu_0) \, {(Z_{qq}^{\mu=\nu})}^{{\rm RI}'} (\mu_0), \\
    {(\overline{Z}_{ij}^{\mu\neq\nu})}^{\overline{\rm MS}} (\bar{\mu}) &=& \, {(\overline{Z}_{ik}^{\mu\neq\nu})}^{{\rm RI}'} (\mu_0)\overline{C}_{kj}^{{\rm RI}',\overline{\rm MS}} (\bar{\mu},\mu_0) , \label{Cbar_ij}
\end{eqnarray}
where $i,j \in \{q,g,c\}$ and $k$ is summed over $q,g,c$. The nonsinglet matching coefficient $C_{qq}^{\overline{\rm MS},{\rm RI}'}$ has been calculated in Ref.~\cite{Gracey:2003mr} up to three loops and the singlet matching coefficients $\overline{C}_{kj}^{{\rm RI}',\overline{\rm MS}}$ in Ref.~\cite{Panagopoulos:2020qcn} up to two loops. An alternative method splits the matching procedure into the following three steps: i) the renormalization factors and mixing matrix are evolved to an intermediate scale $\mu_0'\gg \bar{\mu}$; ii) they are converted to $\overline{\rm MS}$ at the same scale $\mu_0'$; and  iii) they are evolved from $\mu_0'$ to $\bar{\mu}$ scale. Even though the two methods should agree to all perturbative orders, the truncation of the perturbative series can lead to small differences. Given that the evolution functions between two scales are typically more convergent objects than the conversion factors between two schemes, we adopt the second method in this work, in which the conversion is performed only at a large scale $\mu_0'$, where a better convergence is expected. The perturbative input needed for this method is provided in Appendix~\ref{app:evolution}.
    
\subsection{Determination of the renormalization factors and mixing coefficients}

The vertex functions and propagators are calculated using gauge ensembles with \Nf{4} mass-degenerate twisted-mass clover-improved fermions at the same four lattice spacings as the \Nf{2}{1}{1} ensembles analyzed in this work. These \Nf{4} gauge ensemble are simulated by ETMC for our renormalization program. For the singlet operators, we focus on the calculation of the first two rows of the  mixing matrix $\overline{Z}$, in which vertex functions of the gauge-noninvariant operator $T^{\mu\nu}_c$ do not enter. Only in the end we will take the  inverse to extract the matrix elements of $Z$. The computation contains both connected and disconnected diagrams involving quark and gluon loops. For the connected parts, we employ the momentum-source method~\cite{Gockeler:1998ye}, which gives high statistical accuracy. For the nonsinglet operators, in which disconnected diagrams are absent, it is sufficient to use  30 configurations. For the singlet operators, where disconnected quark and gluon loops enter the calculation, high statistics are required to reduce the substantial gauge noise. Thus, we employ gauge ensembles of $\mathcal{O}(10^4)$ configurations generated specifically for the computation of EMT renormalization. Given that RI$'$-MOM renormalization factors do not significantly depend on the lattice size, the ensembles are simulated at small lattice volumes. The parameters of these ensembles are given in Table ~\ref{tab:Nf4ensembles}. The E gauge  ensemble at the smallest lattice spacing has been recently produced and is used in this work for the first time. 
\begin{table}[!h]
\caption{\Nf{4} gauge ensembles and their parameters: $\beta = (2 N/g^2)$, in the third, the  lattice volume ($(L/a)^3 \times (T/a)$), twisted-mass parameter $(a \mu_{\rm sea})$, hopping parameter ($\kappa$) and clover coefficient ($c_{\rm SW}$). In the last column, the number of configurations $n_{\rm confs}$ are given.}
  \centering
\begin{tabular} {ccccccc}
  \hline
  \hline
  \noalign{\vskip 5pt}
   Ensemble & $\beta$ & $(\frac{L}{a})^3 \times (\frac{T}{a})$ & $a \mu_{\rm sea}$ & $\kappa$ & $c_{\rm SW}$ & $n_{\rm confs}$ \\[1ex]
   \hline
   B  & 1.778 & $12^3 \times 24$ & 0.0060 & 0.1393050 & 1.6900 & 30,978 \\
   C  & 1.836 & $24^3 \times 48$ & 0.0050 & 0.1386735 & 1.6452 & 25,002 \\
   D  & 1.900 & $24^3 \times 48$ & 0.0040 & 0.1379346 & 1.6112 & 10,000 \\ 
   E & 1.960 & $24^3 \times 48$ & 0.0035 & 0.137386 & 1.5792 & 12,917 \\
   \hline
\end{tabular}
    \label{tab:Nf4ensembles}
    \end{table}
\begin{figure*}
\centering
\includegraphics[width=0.49\textwidth]{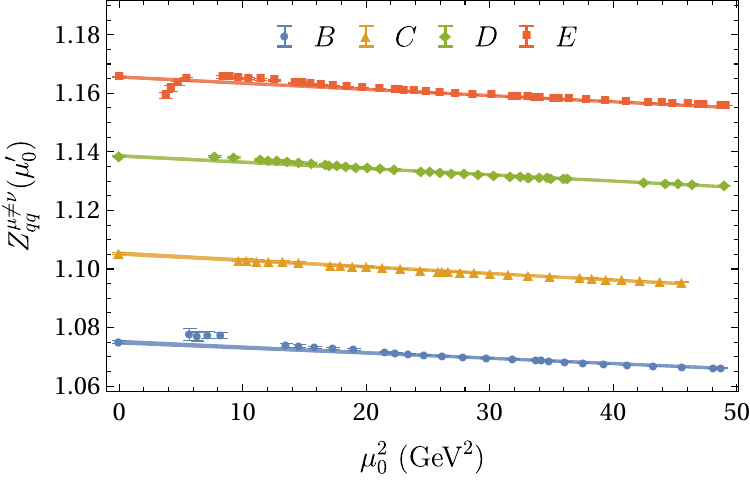} \,
\includegraphics[width=0.49\textwidth]{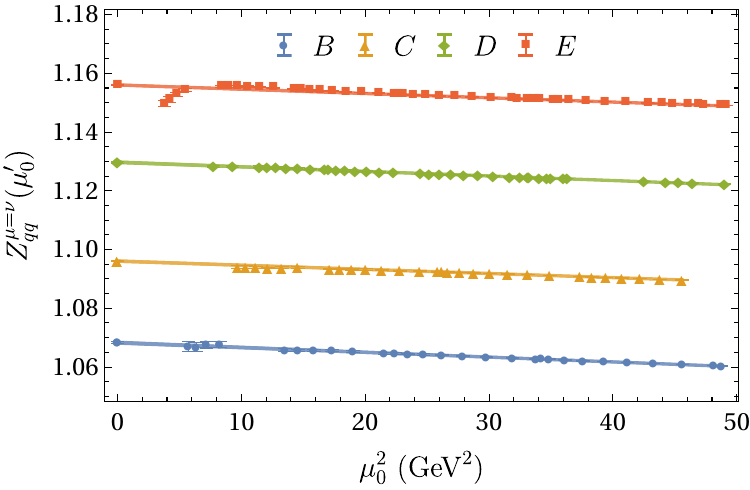} \\
\includegraphics[width=0.49\textwidth]{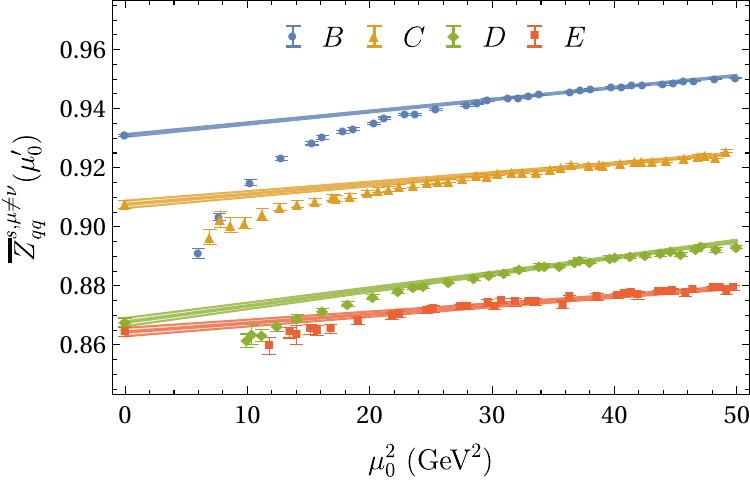} \,
\includegraphics[width=0.49\textwidth]{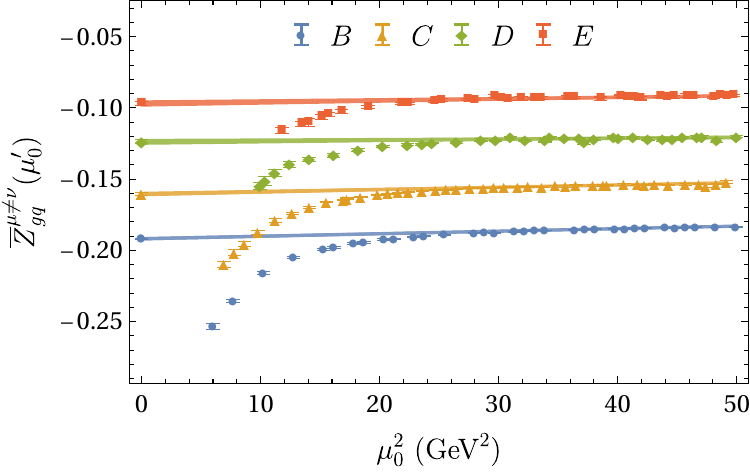} \\
\includegraphics[width=0.49\textwidth]{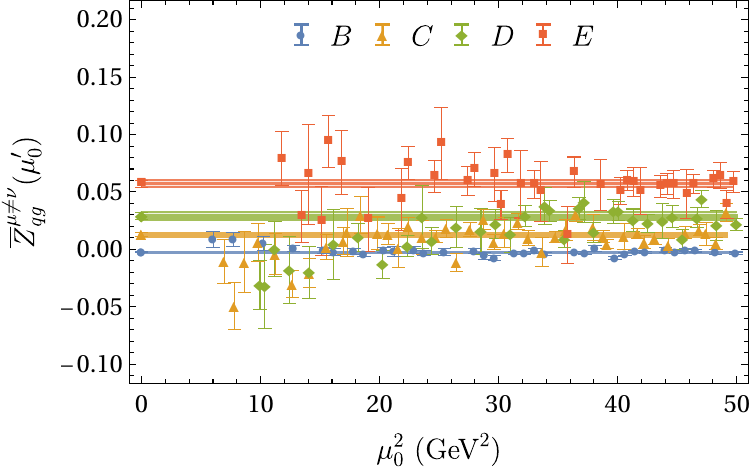} \,
\includegraphics[width=0.49\textwidth]{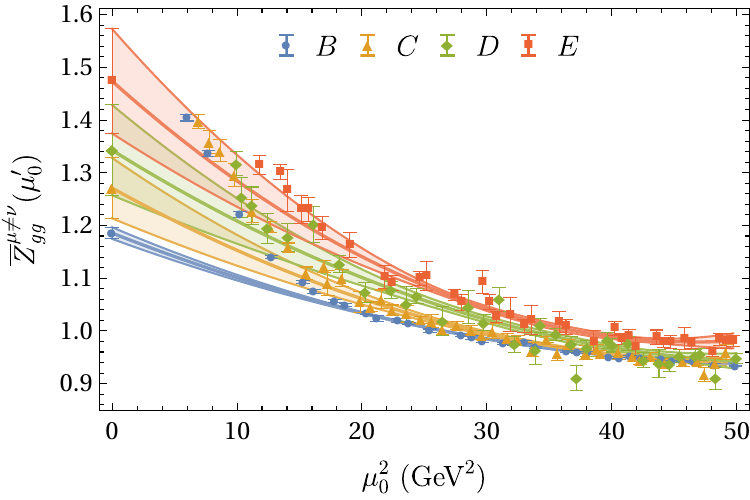} \\
\includegraphics[width=0.49\textwidth]{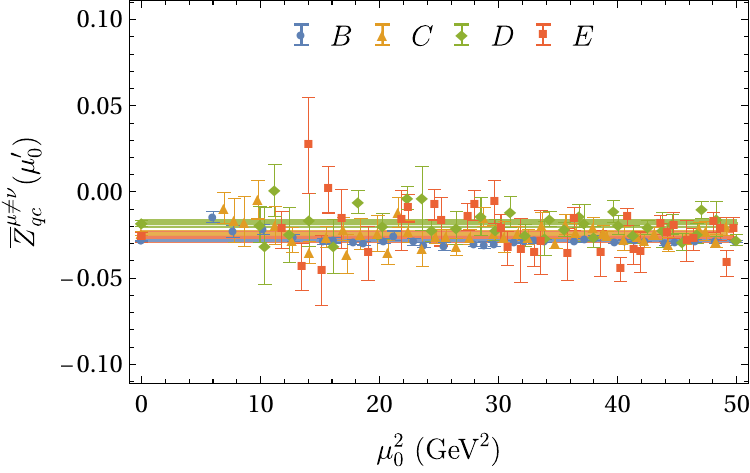} \,
\includegraphics[width=0.49\textwidth]{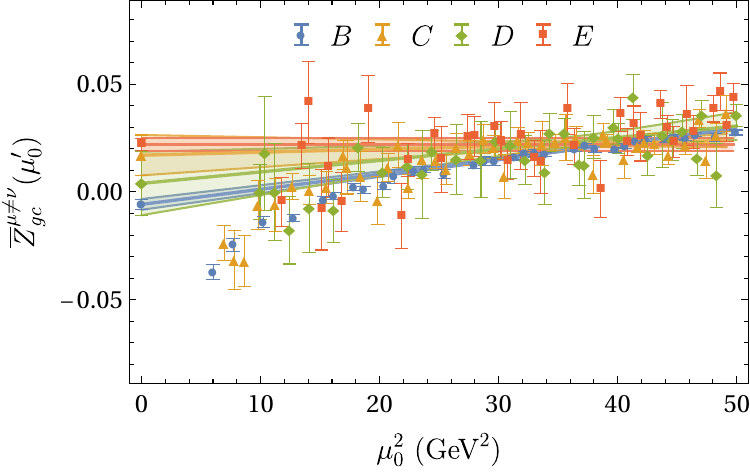} 
    \caption{Results for the renormalization factors $Z_{qq}^{\mu\neq\nu}$ (top, left), $Z_{qq}^{\mu=\nu}$ (top,right) and mixing-matrix coefficients $\overline{Z}^{\mu\ne\nu}_{ij}$, $i,j \in \{q,g,c\}$ given in the 2nd, 3rd and 4th rows, respectively, in the RI$'$-MOM scheme at the reference scale of ${\mu_0'}^2 = 21$ GeV$^2$. Blue circles, yellow triangles, green diamonds and red squares show, respectively, results for the B, C, D and E gauge ensembles, accompanied by the same-colored band showing the AIC-average value. Constant, linear or quadratic fits in $a^2 \mu_0^2$ are employed, as explained in the text. }
    \label{fig:Zfactors}
\end{figure*}

For the determination of the nonsinglet renormalization factors, the chiral limit is obtained by using, for each lattice spacing, four ensembles generated at different values of the twisted-mass parameter $\mu_{\rm sea}$. The dependence on $\mu_{\rm sea}$ is found to be mild, in agreement with our previous studies~\cite{Alexandrou:2020sml,ExtendedTwistedMass:2024kjf}. This residual dependence is removed by performing a linear fit in $\mu_{\rm sea}$. For the singlet operators, however, computing both the connected and disconnected vertex functions for four values of $\mu_{\rm sea}$ and for all number of configurations is computationally expensive. Since no significant dependence on $\mu_{\rm sea}$ is observed in the connected contribution, where statistical errors are very small, we restrict the computation for the singlet operators to the single value of $\mu_{\rm sea}$ per lattice spacing given in Table~\ref{tab:Nf4ensembles}.

To minimize rotational $O(4)$ breaking lattice effects, we consider momenta close to the body-diagonal direction (``near democratic'') by imposing $\sum_{\mu} (p^\mu)^4/[\sum_\mu (p^\mu)^2]^2 < 0.3$. Additionally, we improve our nonperturbative values by subtracting lattice artifacts from all vertex functions and propagators involved. The artifacts are computed in lattice perturbation theory to all orders in the lattice spacing. For the fermionic vertex functions $\Lambda^{\mu\nu}_i$ and quark propagator $S$, the artifacts are calculated to one loop. The gluonic vertex functions $V^{\mu\nu\rho\sigma}_i$ and the gluon propagator $D^{\mu\nu}$ show nonzero artifacts even at  tree level. The calculation of one-loop discretization effects in gluonic vertex functions is very demanding because of the large number of terms arising in the relevant Feynman diagrams, mainly coming from the use of Symanzik-improved gluon actions and the use of stout smearing in the gluon EMT operator. An exception is $V^{\mu\nu\rho\sigma}_q$, which gives a controllable number of terms, and thus we were able to complete the one-loop calculation. Therefore, the subtraction of artifacts from the gluonic vertex functions and the gluon propagator is restricted to  tree level, except for $V^{\mu\nu\rho\sigma}_q$. The perturbative calculations and subtraction formulae are described in Ref.~\cite{ExtendedTwistedMass:2024kjf}. The additional cases considered here follow this framework closely.

\begin{table}
\begin{center}
 \caption{Renormalization factors and mixing coefficients in the $\overline{\rm MS}$ scheme at 2 GeV for the B, C, D and E ensembles. The numbers in parentheses are the total errors, obtained by adding the statistical and systematic uncertainties in quadrature. The systematic uncertainty is determined from the AIC procedure.}
  \begin{tabular}{lcccc}
  \hline
  \hline
\ & \ B & \ C & \ D & \ E \\
\hline
\ $Z_{qq}^{\mu=\nu}$ & \ 1.1161(4) \ & \ 1.1450(3) \ & \ 1.1802(3)\ & \ 1.2078(4) \\
\ $Z_{qq}^{\mu \neq \nu}$ & \ 1.1241(7) \ & \ 1.1541(5) \ & \ 1.1896(4)\ & \ 1.2174(4) \\
\ $Z_{qq}^{{\rm s}, \mu \neq \nu}$ & \ 1.1284(34) \ & \ 1.1574(48) \ & \ 1.2110(53)\ & \ 1.2160(62) \\
\ $Z_{qg}^{\mu \neq \nu}$ & \ -0.0440(11) \ & \ -0.0536(34) \ & \ -0.0679(50) \ & \ -0.0807(70) \\
\ $Z_{gq}^{\mu \neq \nu}$ & \ \ 0.0822(37) \ & \ \ 0.0410(76) \ & \ \ 0.0070(69) \ & \ -0.0299(66) \\
\ $Z_{gg}^{\mu \neq \nu}$ & \ 0.873(22) \ & \ 0.795(46) \ & \ 0.781(50) \ & \ 0.665(55) \\
\hline
  \end{tabular}
  \label{tab:Zfactors}
  \end{center}
\end{table}

After chiral extrapolation of nonsinglet vertex functions and subtraction of lattice artifacts, the renormalization factors and the mixing matrix are evolved to the intermediate reference scale $(\mu_0')^2=21$ GeV$^2$, using Eqs. (\ref{evol_nomix} -- \ref{evol_mix}). Note that the evolution of each row of the inverse mixing matrix $\overline{Z}$ is done independently from the other rows of $\overline{Z}$; thus, the third row of $\overline{Z}$, which has not calculated, is not needed in this step. 

To eliminate any residual dependence on the initial scale resulting from discretization effects, we perform  a polynomial fit in $a^2 \mu_0^2$.  In particular, a linear fit is sufficient for $Z_{qq}$, $\overline{Z}_{qq}^s$, $\overline{Z}_{gq}$, and $\overline{Z}_{gc}$, while a quadratic fit is needed for $\overline{Z}_{gg}$.  For the more noisy mixing coefficients $\overline{Z}_{qg}$ and $\overline{Z}_{qc}$, a constant fit is adopted. We employ several fit ranges within $20 \leq \mu_0^2 \leq 50$ GeV$^2$. The extrapolated values at $\mu_0 = 0$ from all fits are combined using model averaging based on  AIC~\cite{Jay:2020jkz}. Momenta with $\mu_0^2 < 20$ GeV$^2$ are excluded from the analysis, as they suffer from significant hadronic contamination and the perturbative evolution is not reliable in this low-momentum region. Fig.~\ref{fig:Zfactors} displays the extrapolation to $\mu_0 \rightarrow 0$ for all renormalization factors and mixing coefficients. We point out that the off-diagonal elements $\overline{Z}_{qc}$ and $\overline{Z}_{gc}$, neglected in previous studies, make a small but nonzero contribution.

After the elimination of the $\mu_0$-dependence, we convert to the $\overline{\rm MS}$ scheme and evolve to the reference scale $\bar{\mu} = 2$ GeV, using Eqs. (\ref{evol_nomix} -- \ref{evol_mix}). As explained above, the conversion and evolution of each row of the inverse mixing matrix $\overline{Z}$ is done independently from the other rows of $\overline{Z}$; thus, the third row of $\overline{Z}$ is still not needed in this stage. The final step is to invert the matrix $\overline{Z}$. This involves all elements of $\overline{Z}$  and thus the third row of $\overline{Z}$ is needed. This is extracted from the vertex functions of $T^{\mu\nu}_{c}$. By truncating the vertex functions to their tree-level values and using the relevant matching coefficients to the same perturbative order, the third row of $\overline{Z}$ becomes trivial: $\overline{Z}_{cj} = \delta_{cj}$. Then, we invert $\overline{Z}$ to take the final values of the $2\times2$ block of Eq.~\eqref{Eq:2x2_mixing_matrix}.

The final values for the renormalization factors and mixing coefficients needed for the renormalization of the matrix elements are summarized in Table~\ref{tab:Zfactors}. We provide results using 10 steps of stout smearing, since this is what was used in our final analysis.  
When combining the renormalization functions with the matrix elements, we propagate the errors as follows: For each renormalization function, we generate a set of random samples from a Gaussian distribution with mean and standard deviation set to its central value and error, as given in Table~\ref{tab:Zfactors}. The number of random samples is chosen to match the number of gauge-field
configurations of the physical ensembles used for the matrix elements. We then construct jackknife samples of the renormalization functions from these random samples and combine them with the jackknife samples of the bare matrix elements.
\
\begin{figure*}[]
    \centering
    \includegraphics[width=\textwidth]{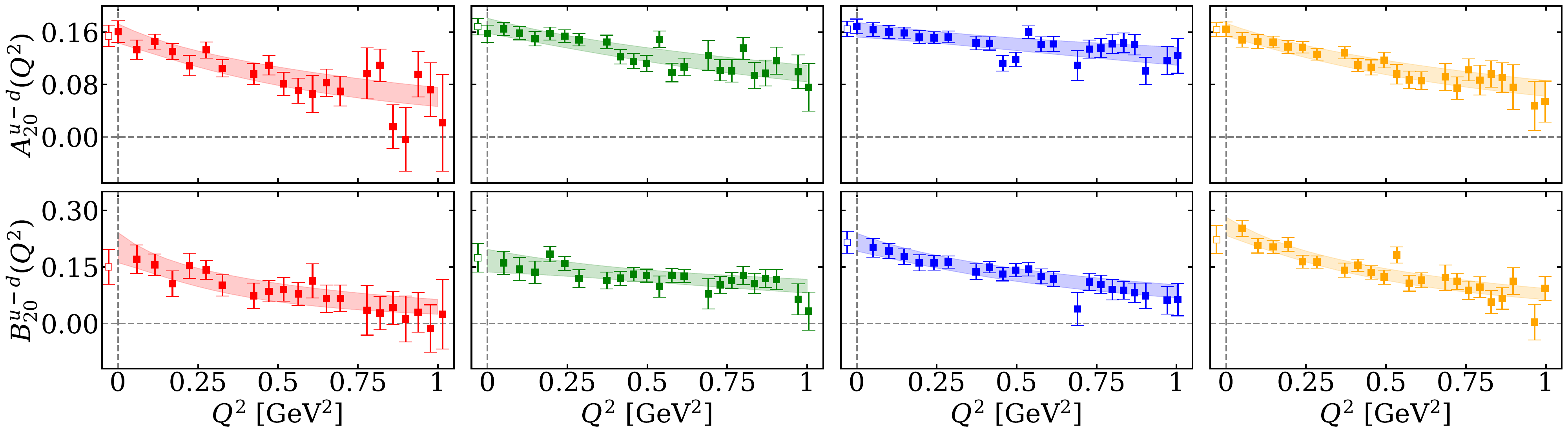}
    \caption{
        Results for renormalized ${A}^{u-d}_{20}(Q^2)$ and ${B}^{u-d}_{20}(Q^2)$ in the $\overline{\rm MS}$ at 2~GeV using both $\mu=\nu$ and $\mu\ne\nu$ matrix elements in the SVD analysis. The notation is the same as in \cref{fig:A20_Q2_conn}.
        The open symbols at $a=0$ are the results (after renormalization) obtained in \cref{fig:A20_Q2_conn} for ${A}^{u-d}_{20}(Q^2)$ using only $\mu=\nu$ matrix elements, and in \cref{fig:B20_Q2_conn} for ${B}^{u-d}_{20}(Q^2)$ using only $\mu\ne\nu$ matrix elments.
    }\label{fig:B20_Q2_both}
\end{figure*}

\section{Continuum extrapolation}\label{sec:continuum}

The isovector GFFs can be extracted using both $\mu=\nu$ and $\mu\ne\nu$ to increase statistics. One can either do the SVD independently for the two cases  or use both in the SVD after renormalizing them multiplicatively. For our final analysis, we carry out a combined SVD and  perform that excited state analysis on the resulting combined SVD results, following the same procedure as described in Sec.~\ref{sec:bgff}. The fit on the $Q^2$-dependence also follows the approach in Sec.~\ref{sec:bgff}. The renormalized results are shown in Fig.~\ref{fig:B20_Q2_both}, where we also show the  $Q^2=0$ values  using only $\mu=\nu$ for $A_{20}(Q^2)$ and $\mu\ne\nu$ for $B_{20}(Q^2)$ illustrated in \cref{sec:bgff}. Comparing to the two results at $Q^2=0$, one can see that, overall, the errors are reduced, as expected. Thus, for the forward limit of the isovector GFFs we use the results from the combined SVD analysis from now on. For all other cases we use the ratios for $\mu\ne \nu$ and thus the analysis is as demonstrated  in Sec.~\ref{sec:bgff}.
\begin{figure}[h!]
    \centering
    \includegraphics[width=\columnwidth]{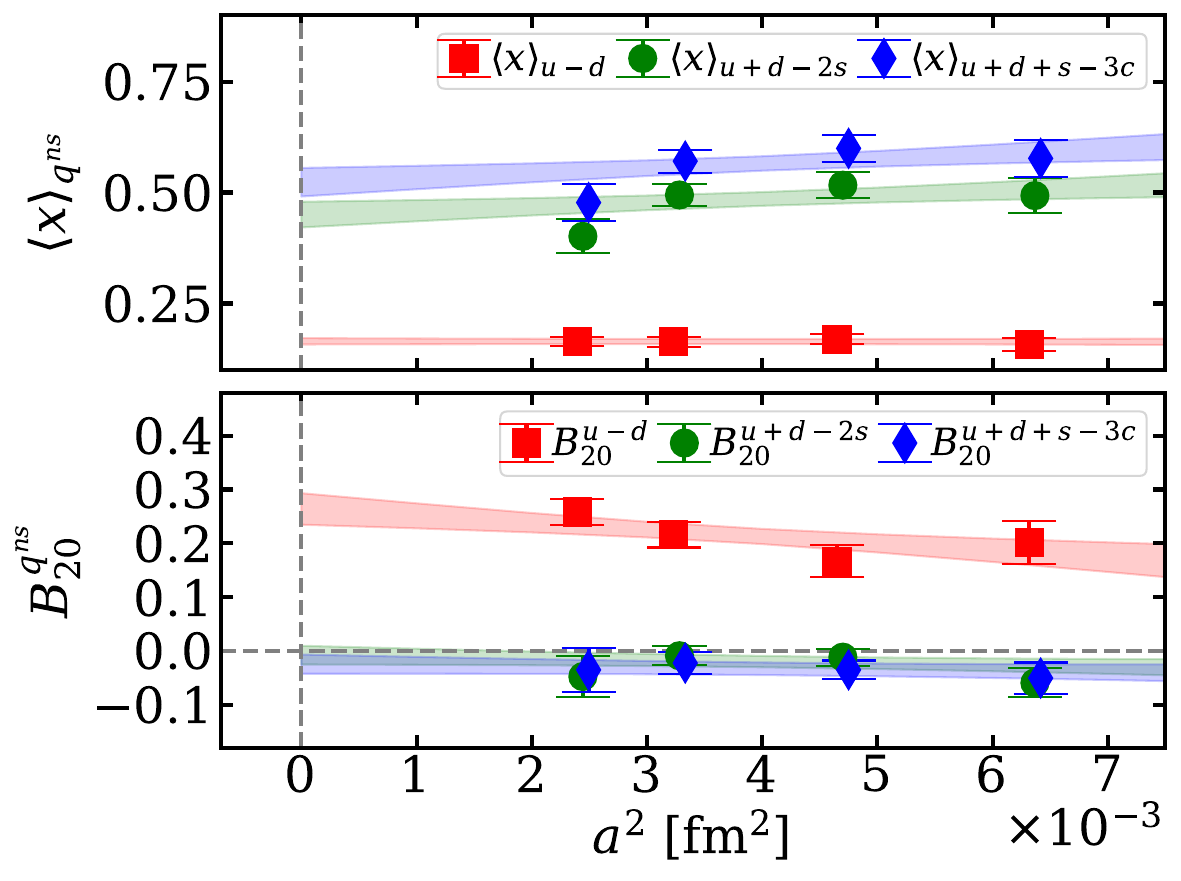}
    \caption{
    Continuum extrapolation of the renormalized  quark momentum fraction $\braket{x}_{u-d}$ (red squares), $\braket{x}_{u+d-2s}$ (green circles), $\braket{x}_{u+d+s-3c}$ (blue diamonds) (top),  and bottom, with the same notation,  $B_{20}^{u-d}(0)$, $B_{20}^{u+d-2s}(0)$ and $B_{20}^{u+d+s-3c}(0)$.
    The bands show the AIC model averaged continuum extrapolations using constant and linear fits in $a^2$. All results are given in the $\overline{\rm MS}$ at 2~GeV.
    }\label{fig:ce_AB20_v123}
\end{figure}

Within the twisted mass fermion formulation and after properly performing  the renormalization, the leading lattice discretization effects are of ${\cal{O}}(a^2$). We therefore perform linear extrapolations in  $a^2$. Because in most cases the data show no significant $a^2$-dependence within our statistical errors, we include also a constant fit and take the model average of the constant and linear fits using the Akaike information criterion (AIC).

The non-singlets, $u-d$, $u+d-2s$ and $u+d+s-3c$ renormalize multiplicatively and do not mix with the gluon. In \cref{fig:ce_AB20_v123}, we show the renormalized values of $\braket{x}_{q^{ns}}$ and $B^{q^{ns}}_{20}(0)$ in the $\overline{\rm MS}$ at 2~GeV  for the four ensembles, as well as the continuum extrapolation.

\begin{figure}[h!]
    \centering
    \includegraphics[width=\columnwidth]{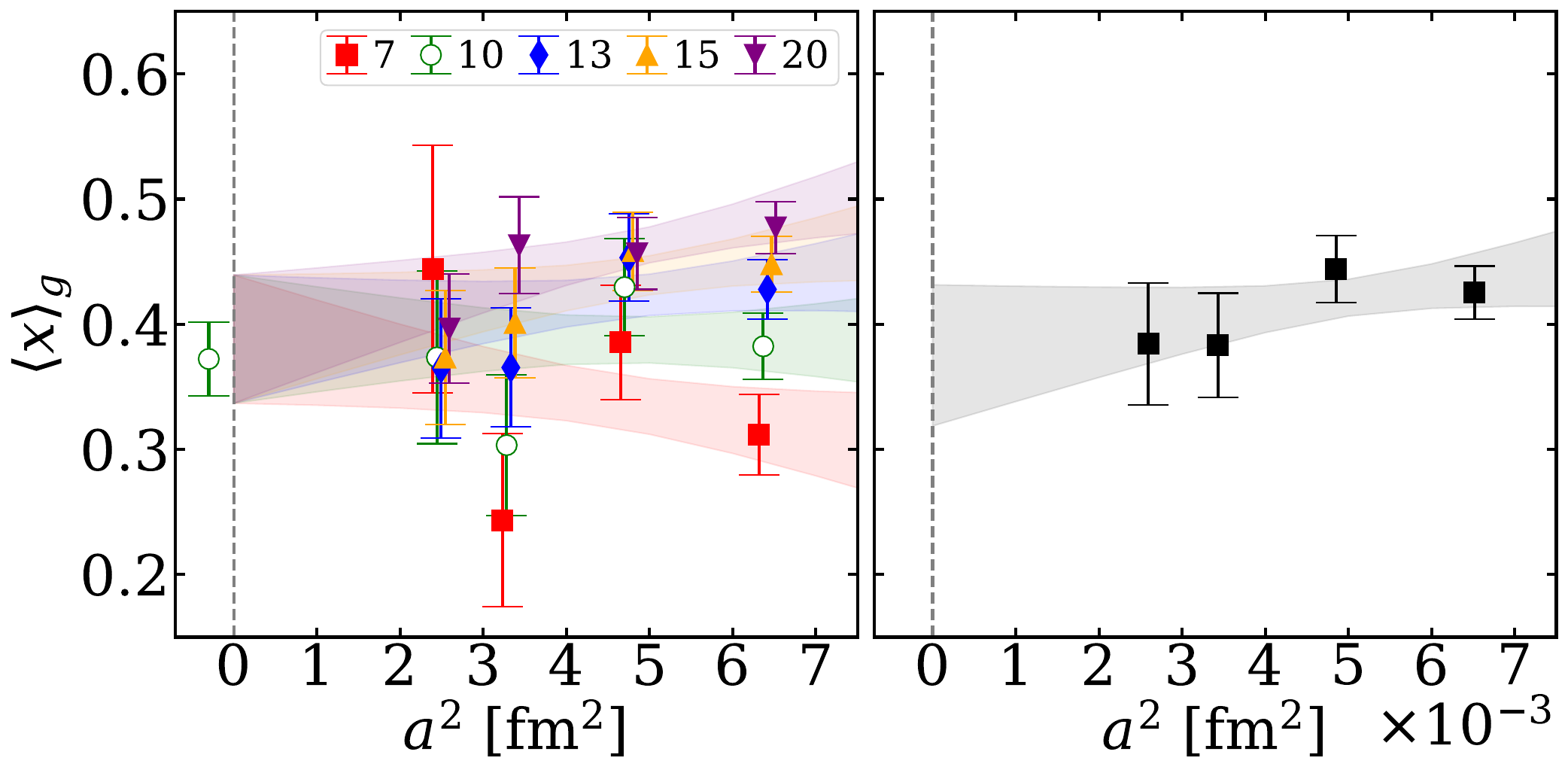}\\
     \includegraphics[width=\columnwidth]{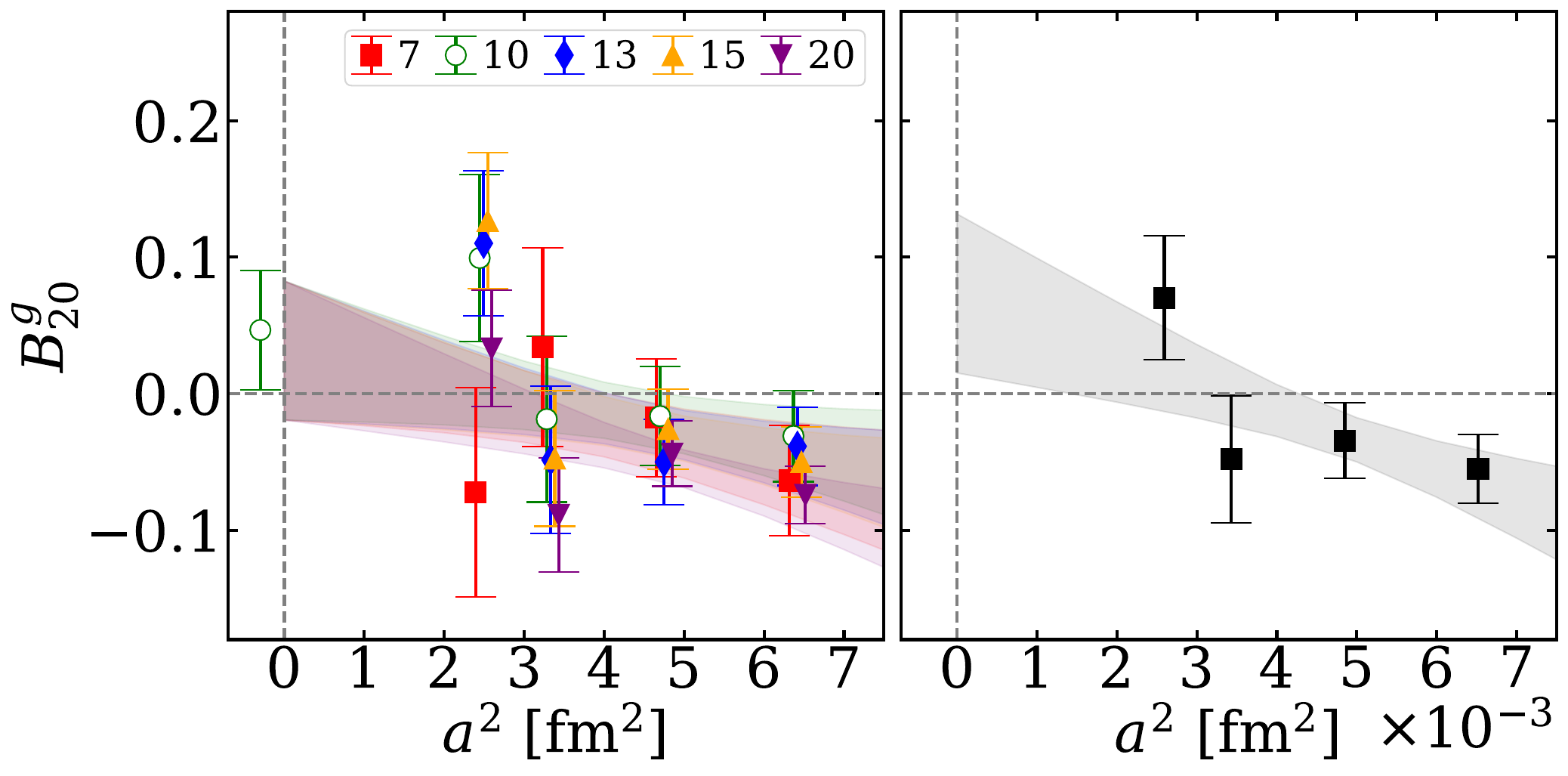}
    \caption{
    Continuum extrapolation of the renormalized gluon momentum fraction $\braket{x}_g$ (top) and  $B_{20}^{g}$ (bottom) in the $\overline{\rm MS}$ at 2~GeV.
    Left: Renormalized results for the gluon momentum fraction $\langle x\rangle_g$ versus $a^2$ for  stout-smearing steps $n_s=7,10,13,15,20 $, shown with the red squares, green circles, blue diamonds, yellow upper-pointing triangles and purple down-pointing triangles, respectively.   The corresponding colored bands show the resulting fits when  using \cref{eq:joint_fit_xg}. The value at $a^2=0$, shown with the green open circle, is  the value obtained for $n_s=10$ from the AIC average of  constant and linear in $a^2$ extrapolations. 
    Right: Results after a weighted average of the values at each   $n_s$ for each gauge ensemble  (black squares). The grey band shows the resulting  continuum limit when extrapolating linearly in $a^2$ .}
    \label{fig:jointLinear_avgx}
\end{figure}

 For the gluonic contributions,  to reduce gauge  noise, we apply stout smearing. After renormalization using vertex functions with the  same stout smearing the results should be independent of the stout smearing within uncertainties in the continuum limit. We show results for the different stout  smearings  for  the  renormalized gluon momentum fraction in Fig.~\ref{fig:jointLinear_avgx}.  
 One procedure for taking the continuum limit  is to perform a joint extrapolation using all
ensembles and all values of $n_s$ simultaneously by employing
\begin{align}
\braket{x}_g^{\,n_s}(a) = \braket{x}_g + c_{n_s} a^2 ,
\label{eq:joint_fit_xg}
\end{align}
where the slope parameters, $c_{n_s}$, are different for different values of $n_s$ while the
continuum-limit value $\braket{x}_g$ is constrained to be the same  for all $n_s$. The resulting fit is shown in Fig.~\ref{fig:jointLinear_avgx}, and describes well the data. 
\begin{figure}[h!]
    \centering
    \includegraphics[width=\columnwidth]{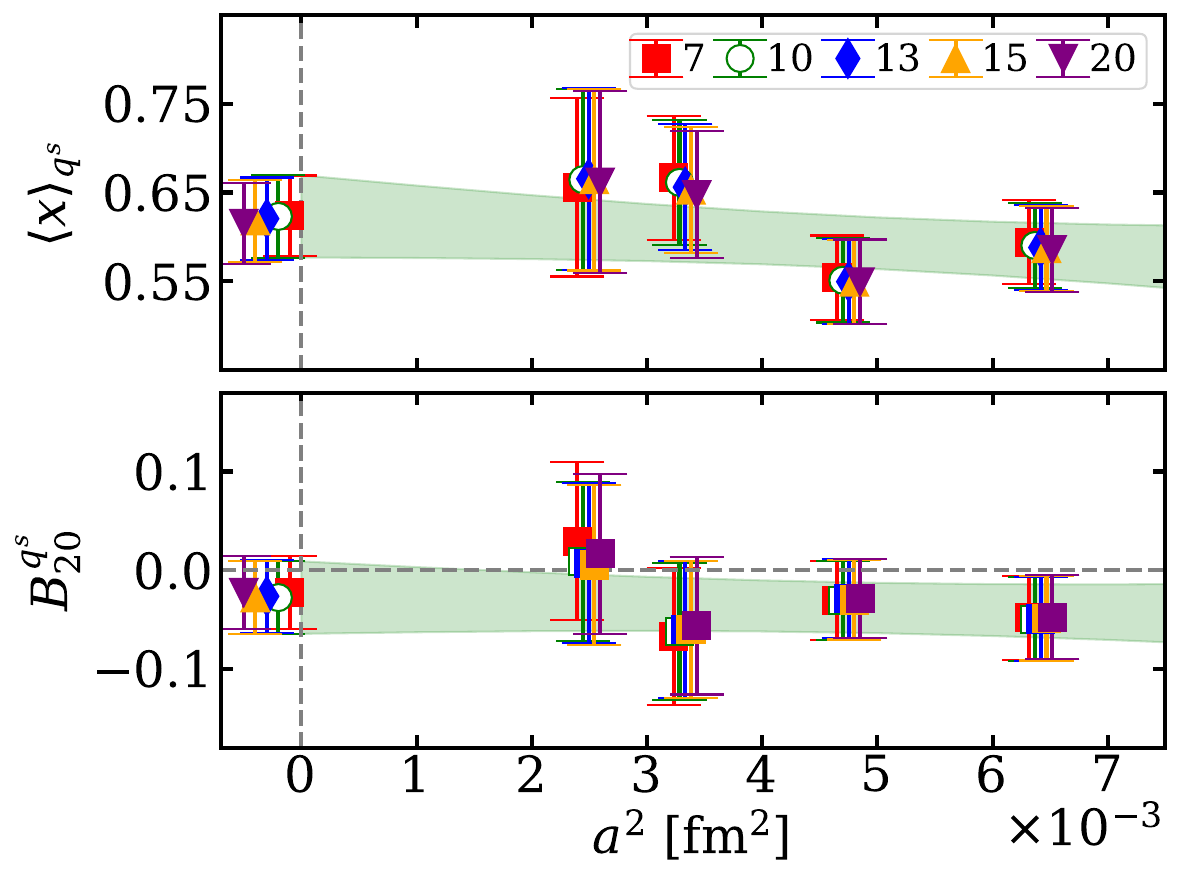}
    \caption{Continuum extrapolation of the renormalized singlet quark momentum fraction $\braket{x}_{q}$ and GFF $B_{20}^q(0)$ for different numbers of stout-smearing $n_s=7,10,13,15,$ and $20$, using the notation of \cref{fig:jointLinear_avgx}.
    For each fixed $n_s$, the continuum limit values, shifted horizontally for better visibility,  are the  AIC  average of constant and linear  in $a^2$ fits. 
  }
    \label{fig:jointLinear_quark}
\end{figure}

\begin{figure*}
    \centering
    \includegraphics[width=\textwidth]{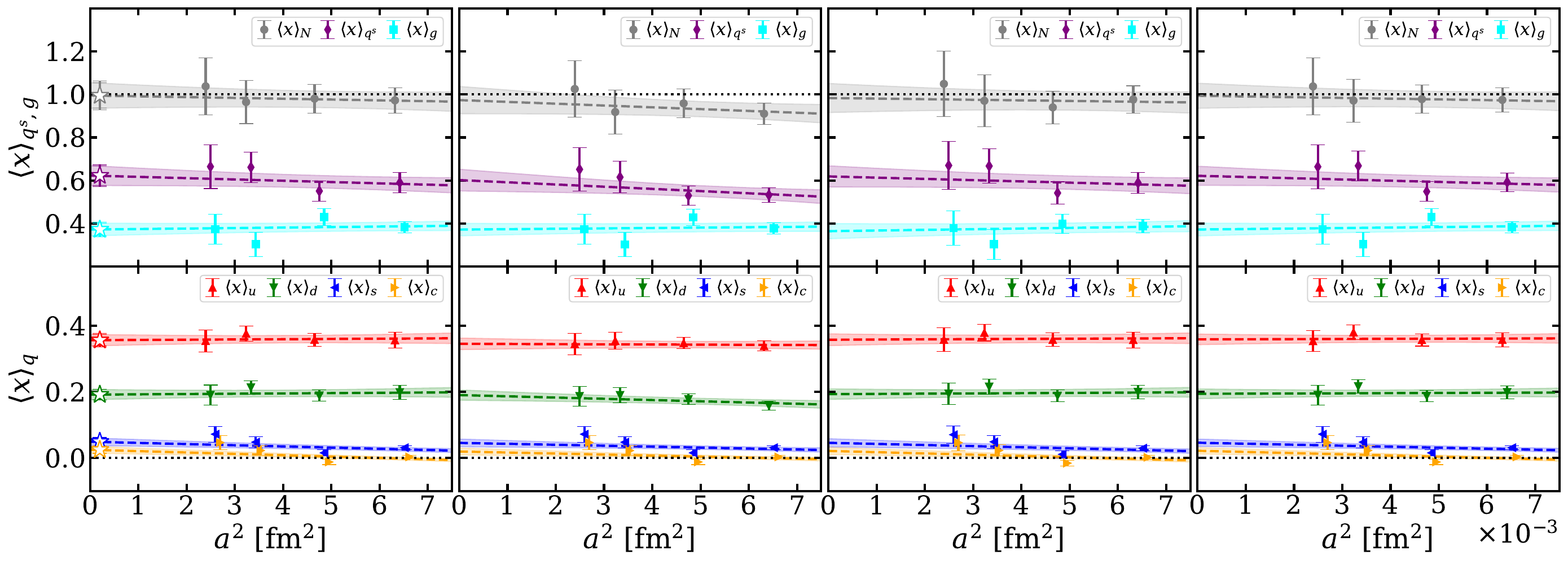}
    \caption{
    Continuum extrapolation of the nucleon momentum fractions $\braket{x}_{q,g}$ in the $\overline{\rm MS}$ at 2~GeV and the total nucleon $\braket{x_N}$. The first column shows the  analysis adopted as described in Secs.~\ref{sec:bgff} and \ref{sec:continuum}, with the stars showing the values at $a=0$, where the inner error bars show the statistical error and  the outer  the total uncertainty after adding the systematic errors in   quadrature. The second, third, and fourth columns correspond to variations of the analysis from which the systematic error is obtained, as explained in Sec.~\ref{sec:systematics}. Namely, in the second and third columns, we show respectively, the results obtained when the summation method is used for the connected contributions instead of two-state fits and when different constant fits are used for the disconnected contributions. In the last column we show the results obtained when changing the upper range of  the $Q^2$  fits of the GFFs. 
    The upper panels show the total nucleon (gray), total quark (purple) and  gluon (cyan) momentum fractions; the lower panels show the flavor decomposition into up (red), down (green), strange (blue), and charm (yellow) quark contributions.
    }\label{fig:a2dep_A20_compare}
\end{figure*}

Alternatively, one can  first take the weighted average of results for different  $n_s$, $\braket{x}_g^{n_s}$, for each ensemble  and then extrapolate 
linearly in $a^2$. This procedure yields a result at the continuum limit which is  compatible with the one from the joint fit of \cref{eq:joint_fit_xg}. 
Performing  the continuum extrapolation of the results for $n_s=10$ using a constant and a linear in $a^2$ fit and combining according to AIC also yields a result that is compatible with the one from \cref{eq:joint_fit_xg}, as shown   in \cref{fig:jointLinear_avgx}. 
Therefore, we choose $n_s=10$, motivated by that fact that the gauge noise is sufficiently reduced and the smearing is still kept at a mild level.

\begin{figure*}
    \centering
    \includegraphics[width=\textwidth]{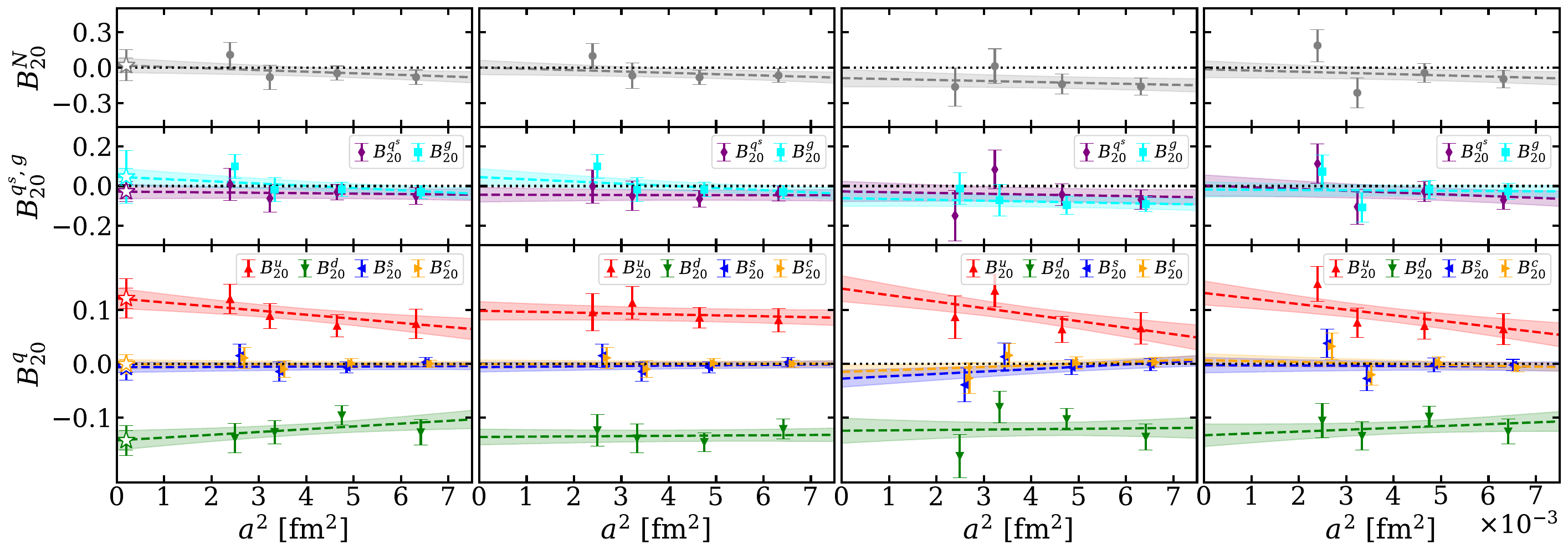}
    \caption{Continuum extrapolation of $B^{N}_{20}(0)$ (top), $ B^{q^s}_{20}(0)$ and $B^{g}_{20}(0)$ (middle) and $ B^{q}_{20}(0)$ bottom.  The notation is the same as in \cref{fig:a2dep_A20_compare}}
    \label{fig:a2dep_B20_compare}
\end{figure*}

\begin{figure*}
    \centering
    \includegraphics[width=\textwidth]{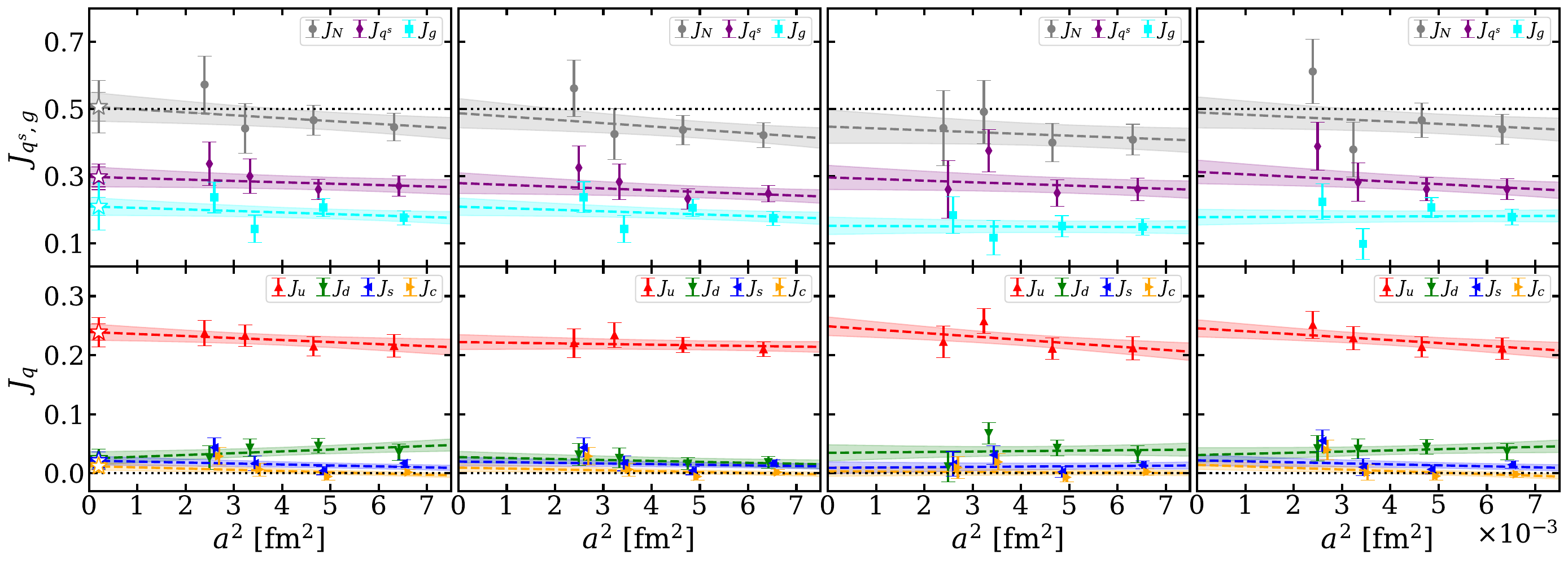}
    \caption{Continuum extrapolation of  the different contributions to the nucleon angular momentum, $J$ and the total $J_N$. The notation is the same as in \cref{fig:a2dep_A20_compare}}
    \label{fig:a2dep_J_compare}
\end{figure*}

\begin{figure*}
    \centering
    \includegraphics[width=\textwidth]{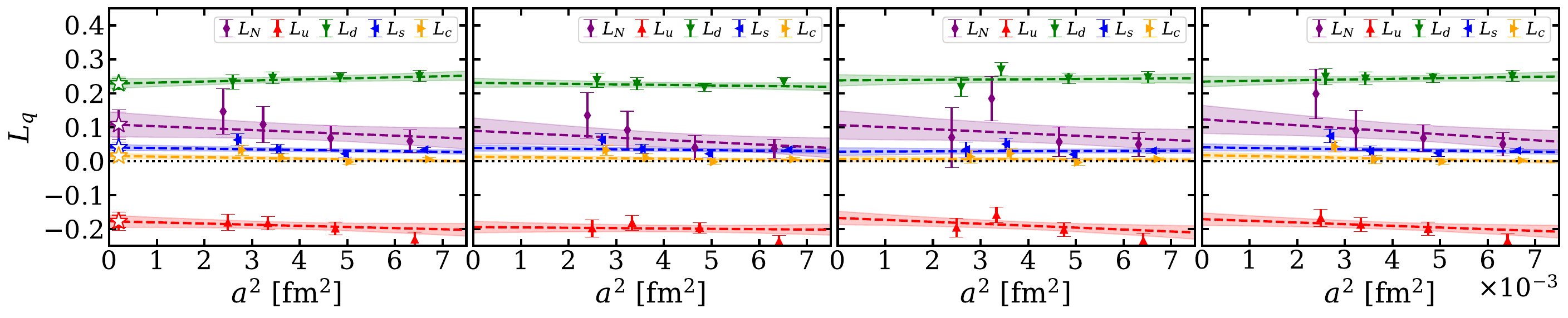}
    \caption{Continuum extrapolation of the quark contributions to the nucleon orbital angular momentum $L_q$ and for the total $L_N$. The notation is the same as in \cref{fig:a2dep_A20_compare}}
    \label{fig:a2dep_L_compare}
\end{figure*}

The renormalized flavor singlet GFFs acquire a dependence
on $n_s$  only through the mixing with the gluon operator. Since the
corresponding mixing coefficients are relatively small, the $n_s$-dependence is
found to be negligible for the five values of $n_s$ investigated in this
work, yielding compatible results on the continuum limit,  as can be seen  for the singlet momentum fraction in Fig.~\ref{fig:jointLinear_quark}. Thus, for consistency we take  $n_s=10$ for determining our results on the renormalized singlet quark
combination.

We perform the same analysis for the gluon contribution to the GFF  $B_{20}^g(0)$ as done for $A_{20}^g(0)$. We show the results 
in \cref{fig:jointLinear_avgx}.  As for the case of 
$\braket{x}_g$, we find consistent results either performing a simultaneous fit using \cref{eq:joint_fit_xg}, or averaging them for each ensembles and then taking the continuum limit.
For this case, the slopes $c_{n_s}$  are not significantly different for different $n_s$, yielding almost equal values in the continuum limit. Therefore, we choose $n_s=10$ and perform a constant  and linear in $a^2$  extrapolation combining them with the AIC, resulting in the point   at $a^2=0$ shown in \cref{fig:jointLinear_avgx}, which is compatible with the results from the other continuum extrapolation procedures.

Using  results for the three nonsinglet
 and  the singlet quark combinations at the continuum limit, we
reconstruct the individual quark-flavor contributions $u$, $d$, $s$, and
$c$. We show in
\cref{fig:a2dep_A20_compare} and \cref{fig:a2dep_B20_compare},   the quark and gluon contributions to the nucleon momentum  in the $\overline{\rm MS}$ at 2~GeV, as well as, their sum in the continuum limit.
In
\cref{fig:a2dep_J_compare}, we  show the corresponding continuum extrapolation of the quark flavor and gluon decomposition of the nucleon angular momentum, $J_{q,g}=\left(A_{20}^{q,g}(0)+B_{20}^{q,g}(0)\right)/2$, as well as their sum.
In \cref{fig:a2dep_L_compare}, we show the continuum extrapolation of the quark flavor decomposition of the nucleon orbital angular momentum $L_q=J_q - \frac{1}{2}\Delta\Sigma_q$ in the $\overline{\rm MS}$ at 2~GeV, as well as, their sum, where $\frac{1}{2}\Delta\Sigma_q=\frac{1}{2}g_A$ is taken from Ref.~\cite{christos}.
Our final  renormalized values of $\braket{x}_{q,g}$, $B_{20}^{q,g}(0)$,
$J_{q,g}$ and $L_{q}$ for each ensemble and in the continuum limit are collected in
\cref{tab:A20_re,tab:B20_re,tab:J_re,tab:L_re}.
For the values in the continuum limit given in these tables, the first uncertainty is 
statistical  obtained from the superjackknife analysis. The second, third, and fourth errors are systematic uncertainties from the excited-state
analysis, the $Q^2$ analysis, and the gluon stout smearing, respectively, as described in
\cref{sec:systematics}.

\subsection{Systematic uncertainties} \label{sec:systematics}
We consider systematics arising from the analysis of excited states,  the fits to the $Q^2$-dependence and  the variation of the stout smearing steps as follows:
\begin{itemize}
\item {\it Systematic due to the excited state analysis.} Instead of using
the two-state fit result for the connected combinations, we use the value from the summation method with
$t_s^{\rm low}\approx 1.0~{\rm fm}$. The rest of the analysis is kept
unchanged. The resulting values and continuum limit is shown in the second columns of \cref{fig:a2dep_A20_compare,fig:a2dep_B20_compare,fig:a2dep_J_compare,fig:a2dep_L_compare}.
For the disconnected contributions, where we do one-state fits, instead of performing a
single constant fit  to all selected values of $t_s$ 
simultaneously, we  fit to a constant at each $t_s$ value and take the weighted average after convergence, which is typically observed for the same values of $t_s^{\rm low}$ as the common fit to all $_s$. With everything else kept fixed, we perform the continuum extrapolation yielding the results shown in the third columns of \cref{fig:a2dep_A20_compare,fig:a2dep_B20_compare,fig:a2dep_J_compare,fig:a2dep_L_compare}. 

\item {\it $Q^2$-dependence.} For cases including the connected $A_{20}^{u\pm d}(Q^2)$ and $B_{20}^{u-d}(Q^2)$, where we fit to the dipole Ansatz, we change the fit ranges from up to $Q^2_{\rm cut}=1~{\rm GeV}^2$ to up to $Q^2_{\rm cut}=0.8~{\rm GeV}^2$. 
For cases including the connected $B_{20}^{u+d}(Q^2)$ and all disconnected  contributions to $B^{q,g}(Q^2)$, where we fit to a constant, we change the fit ranges from up to $Q^2_{\rm cut}=0.5~{\rm GeV}^2$ to up to $Q^2_{\rm cut}=0.4~{\rm GeV}^2$. 
For disconnected contributions to $\braket{x}_{q,g}$ where we use a boosted frame, we extract the forward matrix element directly and thus there is no $Q^2$-dependence.
The corresponding results and continuum extrapolations are shown in the last columns in \cref{fig:a2dep_A20_compare,fig:a2dep_B20_compare,fig:a2dep_J_compare,fig:a2dep_L_compare}.

\item {\it Stout smearing dependence.} The stout smearing dependence is negligible for the quark isosinglet. Therefore, we only consider the systematic in the gluon contribution. We estimate the systematic by using   the  value obtained in the continuum limit
from the joint fit to \cref{eq:joint_fit_xg} instead of the one with $n_s=10$.
\end{itemize}

The systematic uncertainty associated with each alternative procedure outlined above 
is computed by taking the absolute difference between the central value of the preferred analysis and the ones when changing the procedure as described above, after continuum extrapolation.
The systematic uncertainties due to the excited state analysis of connected and disconnected are added in quadrature resulting
in one systematic.

\subsection{Final values} 
Our  final values are collected in
\cref{tab:A20_re,tab:B20_re,tab:J_re}, where the first error is statistical and  the second, third and fourth errors,
respectively,  are the systematic uncertainty coming form the exited state analysis, the $Q^2$-dependence  and stout smearing.
This systematic uncertainty is also shown on the errors  of the continuum values in 
\cref{fig:a2dep_A20_compare,fig:a2dep_B20_compare,fig:a2dep_J_compare},   where the errors are added in quadrature.

\begin{table*}
    \caption{Results on the renormalized momentum fractions $\braket{x}$ for each ensemble and in the continuum limit.
    For the continuum values, the first error is statistical, the second  the systematic  due to the excited state analysis, the third  the range used in the  $Q^2$ extrapolation when applicable,  and the fourth  the stout smearing analysis applicable  only for $\braket{x}_g$ and $\braket{x}_N$. 
    }\label{tab:A20_re}
    \centering
    \renewcommand\arraystretch{1.5}
    \begin{ruledtabular}
\begin{tabular}{ccccccc}
 & $\braket{x}_{u}$ & $\braket{x}_{d}$ & $\braket{x}_{s}$ & $\braket{x}_{c}$ & $\braket{x}_{g}$ & $\braket{x}_{N}$ \\
\hline
B64 & 0.357(24) & 0.199(21) & 0.0313(66) & 0.0034(53) & 0.382(27) & 0.973(59) \\
C80 & 0.359(20) & 0.189(17) & 0.0156(98) & -0.0121(81) & 0.430(39) & 0.981(67) \\
D96 & 0.377(23) & 0.214(21) & 0.048(17) & 0.023(15) & 0.303(56) & 0.96(10) \\
E112 & 0.355(33) & 0.191(30) & 0.072(24) & 0.047(21) & 0.373(69) & 1.04(13) \\
\hline
$a=0$ & 0.357(17)(11)(3) & 0.192(16)(2)(2) & 0.049(12)(5)(3) & 0.025(10)(6)(3) & 0.372(30)(8)(0)(16) & 0.995(60)(24)(1)(16) \\
\end{tabular}
\vspace{2mm}
\begin{tabular}{cccccc}
 & $\braket{x}_{u+d+s+c}$ & $\braket{x}_{u-d}$ & $\braket{x}_{u+d-2s}$ & $\braket{x}_{u+d+s-3c}$ & $\braket{x}_{u+d}$ \\
\hline
B64 & 0.590(47) & 0.158(15) & 0.493(40) & 0.577(41) & 0.556(42) \\
C80 & 0.551(48) & 0.170(11) & 0.517(29) & 0.599(30) & 0.548(35) \\
D96 & 0.661(71) & 0.164(11) & 0.495(24) & 0.571(26) & 0.591(42) \\
E112 & 0.66(10) & 0.1638(98) & 0.401(38) & 0.477(41) & 0.546(62) \\
\hline
$a=0$ & 0.623(46)(21)(1) & 0.1649(71)(96)(1) & 0.450(29)(11)(11) & 0.523(32)(10)(11) & 0.549(32)(13)(5) \\
\end{tabular}
    \end{ruledtabular}
\end{table*}

\begin{table*}
    \caption{Results for the renormalized $B_{20}(0)$ for each ensemble and the continuum limit. The notation is the same as in \cref{tab:A20_re}.}\label{tab:B20_re}
    \centering
    \renewcommand\arraystretch{1.5}
    \begin{ruledtabular}
\begin{tabular}{ccccccc}
 & $B_{20}^{u}$ & $B_{20}^{d}$ & $B_{20}^{s}$ & $B_{20}^{c}$ & $B_{20}^{g}$ & $B_{20}^{N}$ \\
\hline
B64 & 0.074(28) & -0.127(23) & 0.0031(89) & 0.0002(70) & -0.031(33) & -0.081(61) \\
C80 & 0.071(21) & -0.096(18) & -0.0068(100) & 0.0011(84) & -0.016(36) & -0.047(61) \\
D96 & 0.089(23) & -0.127(22) & -0.014(18) & -0.010(16) & -0.019(61) & -0.08(10) \\
E112 & 0.120(28) & -0.138(27) & 0.015(21) & 0.011(19) & 0.099(61) & 0.11(10) \\
\hline
$a=0$ & 0.122(19)(29)(10) & -0.142(18)(19)(10) & -0.0064(99)(212)(38) & -0.0008(84)(142)(74) & 0.047(44)(108)(64)(15) & 0.019(62)(109)(33)(15) \\
\end{tabular}
\vspace{2mm}
\begin{tabular}{cccccc}
 & $B_{20}^{u+d+s+c}$ & $B_{20}^{u-d}$ & $B_{20}^{u+d-2s}$ & $B_{20}^{u+d+s-3c}$ & $B_{20}^{u+d}$ \\
\hline
B64 & -0.050(43) & 0.201(40) & -0.059(27) & -0.051(29) & -0.053(32) \\
C80 & -0.031(40) & 0.167(30) & -0.012(16) & -0.035(17) & -0.025(24) \\
D96 & -0.062(70) & 0.216(24) & -0.009(18) & -0.023(21) & -0.038(38) \\
E112 & 0.008(81) & 0.258(24) & -0.048(39) & -0.035(41) & -0.018(49) \\
\hline
$a=0$ & -0.028(37)(16)(31) & 0.264(29)(30)(0) & -0.008(17)(80)(12) & -0.025(18)(60)(1) & -0.021(22)(40)(20) \\
\end{tabular}
    \end{ruledtabular}
\end{table*}

\begin{table*}
    \caption{Results for the renormalized angular momentum for each ensemble and the continuum limit. The notation is the same as in \cref{tab:A20_re}.}\label{tab:J_re}
    \centering
    \renewcommand\arraystretch{1.5}
    \begin{ruledtabular}
\begin{tabular}{ccccccc}
 & $J_{u}$ & $J_{d}$ & $J_{s}$ & $J_{c}$ & $J_{g}$ & $J_{N}$ \\
\hline
B64 & 0.215(19) & 0.036(14) & 0.0172(55) & 0.0018(44) & 0.176(21) & 0.446(41) \\
C80 & 0.215(16) & 0.047(12) & 0.0044(70) & -0.0055(61) & 0.207(27) & 0.467(45) \\
D96 & 0.233(18) & 0.043(15) & 0.017(13) & 0.006(11) & 0.142(41) & 0.442(74) \\
E112 & 0.237(22) & 0.027(20) & 0.044(16) & 0.029(15) & 0.236(46) & 0.573(85) \\
\hline
$a=0$ & 0.239(14)(20)(6) & 0.025(11)(10)(6) & 0.0215(76)(123)(4) & 0.0121(66)(93)(22) & 0.209(26)(58)(32)(0) & 0.507(43)(63)(17)(0) \\
\end{tabular}
\vspace{2mm}
\begin{tabular}{cccccc}
 & $J_{u+d+s+c}$ & $J_{u-d}$ & $J_{u+d-2s}$ & $J_{u+d+s-3c}$ & $J_{u+d}$ \\
\hline
B64 & 0.270(31) & 0.180(22) & 0.217(23) & 0.263(24) & 0.251(25) \\
C80 & 0.260(31) & 0.168(19) & 0.252(17) & 0.282(17) & 0.261(21) \\
D96 & 0.300(51) & 0.190(16) & 0.243(15) & 0.274(16) & 0.276(30) \\
E112 & 0.337(65) & 0.211(14) & 0.177(27) & 0.221(29) & 0.264(39) \\
\hline
$a=0$ & 0.298(30)(19)(15) & 0.214(16)(20)(0) & 0.221(16)(45)(11) & 0.249(18)(35)(6) & 0.264(20)(24)(12) \\
\end{tabular}
    \end{ruledtabular}
\end{table*}

\begin{table*}
    \caption{Results on the renormalized orbital angular momentum for each ensemble and the continuum limit. The notation is the same as in \cref{tab:A20_re}.}\label{tab:L_re}
    \centering
    \renewcommand\arraystretch{1.5}
    \begin{ruledtabular}
\begin{tabular}{ccccc}
 & $L_{u}$ & $L_{d}$ & $L_{s}$ & $L_{c}$ \\
\hline
B64 & -0.231(22) & 0.251(17) & 0.0337(61) & 0.0055(47) \\
C80 & -0.199(19) & 0.247(14) & 0.0215(89) & -0.0026(67) \\
D96 & -0.183(20) & 0.245(17) & 0.036(13) & 0.010(11) \\
E112 & -0.181(24) & 0.233(22) & 0.062(17) & 0.032(15) \\
\hline
$a=0$ & -0.177(18)(20)(6) & 0.228(15)(10)(6) & 0.0401(92)(123)(4) & 0.0153(73)(93)(22) \\
\end{tabular}
\vspace{2mm}
\begin{tabular}{cccccc}
 & $L_{u+d+s+c}$ & $L_{u-d}$ & $L_{u+d-2s}$ & $L_{u+d+s-3c}$ & $L_{u+d}$ \\
\hline
B64 & 0.059(34) & -0.482(29) & -0.048(24) & 0.037(25) & 0.019(27) \\
C80 & 0.067(36) & -0.446(20) & 0.005(17) & 0.078(20) & 0.048(25) \\
D96 & 0.108(53) & -0.428(19) & -0.009(16) & 0.070(20) & 0.063(31) \\
E112 & 0.146(67) & -0.414(19) & -0.073(30) & 0.018(30) & 0.052(40) \\
\hline
$a=0$ & 0.108(37)(19)(15) & -0.407(22)(20)(0) & -0.029(19)(45)(11) & 0.049(22)(35)(6) & 0.053(25)(24)(12) \\
\end{tabular}
    \end{ruledtabular}
\end{table*}

\section{Discussion of results}\label{sec:discussion}
The continuum  results for the momentum fractions, 
 and angular momentum contributions, shown in
\cref{fig:a2dep_A20_compare,fig:a2dep_J_compare} and tabulated in Tables~\ref{tab:A20_re} and \ref{tab:J_re}, satisfy both the momentum and spin sums.  Specifically, we find  $\braket{x}_N =  0.995(60)(24)(1)(16)$ and $J_N = 0.507(43)(63)(17)(0)$. 
We find sizeable non-valence contributions from sea quarks and
gluons. The gluon contribution is approximately $40\%$ in both the nucleon  momentum fraction and angular momentum, while  the sea-quark contribution is about $10\%$.

The continuum extrapolated value of $ B^{q}_{20}(0)$ and $B^{g}_{20}(0)$ are shown in Fig.~\ref{fig:a2dep_B20_compare} and listed  in \cref{tab:B20_re}. As can be seen, $ B^{q^s}_{20}(0)+B^{g}_{20}(0)$ is consistent with zero although is non-zero for the individual  $u$ and $d$ quarks, which are of approximately  the same absolute value but opposite in sign. The rest of the contributions are consistent with zero individually.

Our results show that,  while $\braket{x}_u$ is approximately twice 
$\braket{x}_d$, as  expected from the valence structure of the proton, $J_u$ is more than twice  $J_d$.  This behavior is consistent with the results for
$B_{20}(0)$ shown in \cref{fig:a2dep_B20_compare}, where $B_{20}^u$ and
$B_{20}^d$ have opposite signs and comparable magnitudes.  This leads to an increase in $J_u$ and to a decrease in $J_d$. The remaining
flavor contributions are small within uncertainties.

 The nucleon axial charges, which are connected to the  intrinsic quark spin contributions,
$\frac{1}{2}\Delta\Sigma_q=\frac{1}{2}g_A^q$, are available for all four  ensembles and at the continuum limit. These are  updates of our previous analysis~\cite{Alexandrou:2024ozj,Iona:2025okj,Alexandrou:2026soz}. The results from Refs.~\cite{Alexandrou:2024ozj,Iona:2025okj,Alexandrou:2026soz}, show that, like for the momentum fraction,  the  u-quark contribution is positive with an absolute value that is   twice the value of the d-quark contribution which is negative,  consistent with the valence quark structure
of the proton. The strange contribution is small and negative, while the charm
contribution is consistent with zero.
Using the updated values of Ref.~\cite{christos}, we can determine
 the orbital
angular momenta, $L_q=J_q-\frac{1}{2}\Delta\Sigma_q$. The values are given in Table~\ref{tab:L_re}.
The results show that  the orbital
angular momenta for the u-quark  is negative, reducing  the u-quark 
contribution to $J_u$, whereas the down-quark contribution is positive and largely
compensates the negative intrinsic spin contribution, yielding a comparatively
small positive contribution to $J_d$.

\section{Summary and Conclusions}
\label{sec:conclusion}
We determine the complete flavor decomposition of the  momentum  and  angular momentum of the proton using, for the first time, a continuum extrapolation at the physical value of the pion mass. This avoids any uncontrolled systematics due to the chiral extrapolation.  We observe that, for comparable statistics, the connected contributions have similar statistical errors across all four ensembles, unlike the disconnected contributions, where the statistical error increases with the lattice volume. This is due to loss of correlations between the loops and the nucleon two-point function in the construction of the disconnected three-point functions.  A possible technique to  address this is the Cluster Decomposition Error Reduction~\cite{Liu:2017man,Liang:2019xdx,Wang:2021vqy}, that involves truncating the Fourier transform at a certain radius. This approach  becomes prohibitively expensive for the larger lattices used in this work. Alternative approaches will be explored in the future.

Since we use the twisted mass fermion formulation, the values at a given finite lattice spacing  only have   corrections of  ${\cal O}(a^2)$. Using ensembles with $a< 0.09$~fm, we find  that cut-off effects are mild, which allows us to take the continuum limit using a model average of linear in $a^2$ and constant fits. However, although the $a^2$-dependence is mild, extrapolating to $a=0$ has an effect and a continuum extrapolation  is, thus, important for obtaining the physical values with a comprehensive  account of systematic effects. Such an analysis requires access to leadership computers for multiple years, not only for the production of the gauge ensembles but also for their subsequent analysis. 

The most important conclusion of this work is that we provide a full decomposition of the momentum fraction and angular momentum of the proton confirming from first principles the momentum and spin sums, thus resolving the so-called proton spin puzzle. Performing a compete flavor decomposition for  $B_{20}(0)$ and the orbital angular momentum with a full budget of the systematics  provides a reliable comparison with results from phenomenological analyses. 

\section*{Acknowledgments}
We would like to thank all members of the Extended Twisted Mass Collaboration for a very enjoyable cooperation.  This project is partly funded by the European Union’s Horizon 2020 Research and Innovation Programme ENGAGE under the Marie Sklodowska-Curie COFUND scheme with grant agreement No. 101034267. C.A., S.B., C.I., G.K., Y.L., and G.S. acknowledge partial support from the projects 3D-nucleon (EXCELLENCE/0421/0043), IMAGE-N (EXCELLENCE/0524/0459), MuonHVP (EXCELLENCE/0524/0017), PulseQCD (EXCELLENCE/0524/0269), StrongILA (EXCELLENCE/0524/0001), RobustSigma (EXCELLENCE/0925/0325), and partonWF (VISION ERC/0525/0010) funded by the European Regional Development Fund and the Republic of Cyprus through the Cyprus Research and Innovation Foundation. C.A. and C.K. are supported by European Union’s HORIZON MSCA Doctoral Networks programme, under Grant Agreement No. 101072344, project AQTIVATE (Advanced computing, QuanTum algorIthms and data-driVen Approaches for science, Technology and Engineering). 
This project received funding from the European Research Council (ERC) via the project ”LEEX” grant agreement 101170304. Funded by the European Union. Views and opinions expressed are however those of the author(s) only and do not necessarily reflect those of the European Union or the European Research Council Executive Agency (ERCEA). Neither the European Union nor the ERCEA can be held responsible for them.
The authors gratefully acknowledge the Gauss Centre for Supercomputing e.V. (www.gauss-centre.eu) for funding this project by providing computing time through the John von Neumann Institute for Computing (NIC) on the GCS Supercomputers JUWELS, JUWELS Booster~\cite{JUWELS-BOOSTER} and JUPITER Booster at J\"ulich Supercomputing Centre (JSC). We also acknowledge computing
time granted on Piz Daint at Centro Svizzero di Calcolo Scientifico (CSCS) via the projects s849, s982, s1045, s1133 and s1197.
The authors also acknowledge the Texas Advanced Computing Center (TACC) at University of Texas at Austin for providing HPC resources.
\bibliography{refs}

\appendix
\section{Superjackknife with overlapping gauge configurations}
\label{app:superjackknife}

The superjackknife was introduced in Ref.~\cite{LHPC:2010jcs} to combine the analysis of  statistically independent observables computed on different gauge ensembles. 
We generalize the construction to allow for the analysis of quantities computed on gauge configurations that have overlapping sets  in a given gauge ensemble and show that the method reproduces the corresponding mean and covariance estimators up to controlled corrections.

Consider quantities $A^\alpha$, $\alpha=1,\ldots,n$, measured on gauge configuration sets $S_\alpha$ with the corresponding number of configurations denoted by $N_\alpha$ and estimators $\bar A^\alpha=N_\alpha^{-1}\sum_{i\in S_\alpha}A^\alpha_i$. Different gauge configurations are assumed statistically independent, while quantities measured on the same configuration may be correlated.

Let $\mathcal S$ denote a set of superjackknife labels satisfying $\cup_\alpha S_\alpha\subseteq\mathcal S$ with the number of total superjackknife samples denoted by $N$. For each $s\in\mathcal S$, define the superjackknife sample
\begin{align}
A^{\alpha,(s)}=
\begin{cases}
\bar A^\alpha_{(-s)}, & s\in S_\alpha,\\
\bar A^\alpha, & s\notin S_\alpha,
\end{cases}
\label{eq:sj_sample}
\end{align}
where
\begin{align}
\bar A^\alpha_{(-s)}=\frac{1}{N_\alpha-1}\sum_{i\neq s,\, i\in S_\alpha}A^\alpha_i
\end{align}
is the leave-one-out jackknife estimator. Applying the usual jackknife mean formula,
\begin{align}
\bar A^\alpha_{\rm JK}=\frac1{N}\sum_{s\in\mathcal S}A^{\alpha,(s)},
\end{align}
and using $\sum_{s\in S_\alpha}\bar A^\alpha_{(-s)}=N_\alpha\bar A^\alpha$, one immediately finds $\bar A^\alpha_{\rm JK}=\bar A^\alpha$. Thus, the superjackknife reproduces to the original estimator.

To determine the covariance structure,
\begin{align}
\mathrm{Cov}(\bar A^\alpha,\bar A^\beta)
&=\frac1{N_\alpha N_\beta}\sum_{i\in S_\alpha}\sum_{j\in S_\beta}\mathrm{Cov}(A^\alpha_i,A^\beta_j) \nonumber \\
&=\frac{N_{\alpha\beta}}{N_\alpha N_\beta}\mathrm{Cov}(A^\alpha,A^\beta),
\label{eq:target_cov}\end{align}
where $N_{\alpha\beta}$ is the number of configurations of the set $S_\alpha\cap S_\beta$, the independence of different configurations has been used to eliminate terms with $i\neq j$, and the remaining terms are identical as each configuration is generated from the same underlying distribution. The corresponding standard  covariance estimator is
\begin{align}
\widehat\Sigma^{\alpha\beta}
=\frac{N_{\alpha\beta}}{N_\alpha N_\beta}\frac{1}{N_{\alpha\beta}-1}
\sum_{i\in S_\alpha\cap S_\beta}(A^\alpha_i-\bar A^\alpha)(A^\beta_i-\bar A^\beta).
\label{eq:ordinary_cov}
\end{align}

Defining $\delta A^{\alpha,(s)}=A^{\alpha,(s)}-\bar A^\alpha_{\rm JK}$, the jackknife covariance of the superjackknife samples is
\begin{align}
\Sigma^{\alpha\beta}_{\rm JK}
=\frac{N-1}{N}\sum_{s\in\mathcal S}\delta A^{\alpha,(s)}\delta A^{\beta,(s)}.
\end{align}
Using $\delta A^{\alpha,(s)}=-(A^\alpha_s-\bar A^\alpha)/(N_\alpha-1)$ for $s\in S_\alpha$ and $\delta A^{\alpha,(s)}=0$ otherwise, we find
\begin{align}
\Sigma^{\alpha\beta}_{\rm JK}
=\frac{N-1}{N}&\frac1{(N_\alpha-1)(N_\beta-1)}\times \nonumber\\
&\sum_{s\in S_\alpha\cap S_\beta}(A^\alpha_s-\bar A^\alpha)(A^\beta_s-\bar A^\beta).
\label{eq:sj_cov}
\end{align}
Comparing Eqs.~\eqref{eq:ordinary_cov} and \eqref{eq:sj_cov}, we get
\begin{align}
\Sigma^{\alpha\beta}_{\rm JK}
=\widehat\Sigma^{\alpha\beta}\left[1+O(N_{\min}^{-1})\right]\frac{N_{\alpha\beta}-1}{N_{\alpha\beta}},
\label{eq:sj_ratio}
\end{align}
where $N_{\min}=\min_\alpha N_\alpha$. Thus the overlap-aware superjackknife reproduces the standard covariance estimator up to corrections of order $N_{\min}^{-1}$. The additional factor $(N_{\alpha\beta}-1)/N_{\alpha\beta}$ differs appreciably from unity only when $N_{\alpha\beta}$ is small. In that case the covariance itself is $O(N_{\min}^{-1})$ suppressed relative to the diagonal variances. In particular, the complete-overlap limit reproduces the standard jackknife covariance exactly.

For smooth derived quantities $P^a=f^a(A^1,\ldots,A^n)$, we define the standard estimator $P^a=f^a(\bar A^1,\ldots,\bar A^n)$, the superjackknife samples $P^{a,(s)}=f^a(A^{1,(s)},\ldots,A^{n,(s)})$, and the jackknife mean $P^a_{\rm JK}=N^{-1}\sum_s P^{a,(s)}$. Writing $J^a{}_\alpha=\partial f^a/\partial A^\alpha$ and $H^a{}_{\alpha\beta}=\partial^2f^a/\partial A^\alpha\partial A^\beta$, a Taylor expansion gives
\begin{align}
P^{a,(s)}=P^a+J^a{}_\alpha\delta A^{\alpha,(s)}
+\tfrac12H^a{}_{\alpha\beta}\delta A^{\alpha,(s)}\delta A^{\beta,(s)}+\cdots.
\end{align}
Since $\delta A^{\alpha,(s)}=O(N_\alpha^{-1})$ and $\sum_s\delta A^{\alpha,(s)}=0$, one finds $P^a_{\rm JK}=P^a+O(N_{\min}^{-2})$. So the jackknife mean  approximates well the standard estimator.

Defining $\delta P^{a,(s)}=P^{a,(s)}-P^a_{\rm JK}$, we find
\begin{align}
\delta P^{a,(s)}&=J^a{}_\alpha\delta A^{\alpha,(s)}\left[1+O(N_{\min}^{-1})\right].
\end{align}
Applying the jackknife covariance formula, we get
\begin{align}
\Sigma^{ab}_{P,\rm JK}
&=\frac{N-1}{N}\sum_{s\in\mathcal S}\delta P^{a,(s)}\delta P^{b,(s)} \nonumber\\
&=J^a{}_\alpha\Sigma^{\alpha\beta}_{\rm JK}J^b{}_\beta\left[1+O(N_{\min}^{-1})\right].
\label{eq:sj_derived}
\end{align}
The corresponding standard covariance estimator is
\begin{align}
\widehat\Sigma^{ab}_P=J^a{}_\alpha\widehat\Sigma^{\alpha\beta}J^b{}_\beta.
\end{align}
Combining Eqs.~\eqref{eq:sj_ratio} and \eqref{eq:sj_derived}, shows that the overlap-aware superjackknife reproduces the standard covariance estimator for derived quantities up to corrections of order $N_{\min}^{-1}$, with any larger relative corrections confined to covariance matrix elements associated with small overlap and hence small numerical significance.

Thus, the overlap-aware superjackknife reproduces the mean exactly for the original quantities, reproduces the corresponding covariance matrix up to corrections of order $N_{\min}^{-1}$, and propagates these properties to arbitrary smooth derived quantities at the same order $N_{\min}^{-1}$.

\vspace{1em}
\section{Expressions for GFFs}\label{app:gffs}
The following expressions are provided in Euclidean space. We suppress
the $Q^2$ argument of the GFFs, $E_N$ is the
nucleon energy for three-momentum $\vec{q}$, for the case $\vec{p}\,'=\vec{0}$, the kinematic factor
$\mathcal{K} = \sqrt{2m_N^2/[E_N(E_N+m_N)]}$ and Latin indices ($k$,
$n$, and $j$) take values 1, 2, and 3 with $k\neq j$ while $\rho$
takes values 1, 2, 3, and 4.

\begin{widetext}
\begin{align}
  \Pi^{00}(\Gamma^0, \vec q) =& A_{20}\,\mathcal{K}\,\left(-\frac{3\,E_N}{8} - \frac{E_N^2}{4\,m_N} -
  \,\frac{m_N}{8} \right) +
  B_{20}\,\mathcal{K}\,\left( -\,\frac{E_N }{8} +
  \,\frac{E_N^3}{8\,m_N^2} + \frac{E_N^2}{16\,m_N} - \frac{m_N}{16}
  \right)\nonumber\\
  +&   C_{20}\,\mathcal{K}\,\left(\,\frac{E_N}{2} - \frac{E_N^3}{2\,m_N^2} +
  \frac{E_N^2}{4\,m_N} - \frac{m_N}{4} \right) , \label{eq:gff1} \\ 
  \Pi^{00}(\Gamma^n, \vec q) =& 0 , \label{eq:gff2}\\
  \Pi^{kk}(\Gamma^0, \vec q) =&  A_{20}\,\mathcal{K}\,\left(
  \frac{E_N}{8} + \frac{m_N}{8} + \frac{q_k^2}{4\,m_N} \right)  +
  B_{20}\,\mathcal{K}\,\left( -\frac{E_N^2}{16\,m_N} + \frac{m_N}{16} - \frac{q_k^2\,E_N}{8\,m_N^2} +
  \frac{q_k^2}{8\,m_N} \right)  \nonumber \\
  +&  C_{20}\,\mathcal{K}\,\left( -\frac{E_N^2}{4\,m_N} + \frac{m_N}{4} + \frac{q_k^2\,E_N}{2\,m_N^2} +
  \frac{q_k^2}{2\,m_N} \right)  , \label{eq:gff3}\\
  \Pi^{kk}(\Gamma^n, \vec q) =&  A_{20}\,\mathcal{K}\,\left(-i\,\frac{\epsilon_{k\,n\,0\,\rho}\,
    q_k\,q_\rho }{4\,m_N}\right) +
  B_{20}\,\mathcal{K}\,\left( -i\,\frac{\epsilon_{k\,n\,0\,\rho}\,
    q_k\,q_\rho }{4\,m_N}\right), \label{eq:gff4} \\
  \Pi^{k0}(\Gamma^0, \vec q) =&  A_{20}\,\mathcal{K}\,\left(-i\,\frac{q_k}{4} -i\,\frac{q_k\,E_N}{4\,m_N} \right)+
  B_{20}\,\mathcal{K}\,\left(-i\,\frac{q_k}{8} +i\, \frac{q_k\,E_N^2}{8\,m_N^2}  \right)+
  C_{20}\,\mathcal{K}\,\left(i\,\frac{q_k}{2}-i\, \frac{q_k\,E_N^2}{2\,m_N^2}  \right), \label{eq:gff5} \\
  \Pi^{k 0}(\Gamma^n, \vec q) =&  A_{20}\,\mathcal{K}\,\left(\,-\epsilon_{k\,n\,0\,\rho}\,\left(\frac{
    q_\rho}{8} +\frac{q_\rho\,E_N}{8\,m_N}  \right) \right) +
  B_{20}\,\mathcal{K}\,\left(\,-\epsilon_{k\,n\,0\,\rho}\,\left(\frac{q_\rho}{8} +
  \frac{q_\rho\,E_N}
       {8\,m_N}   \right)\right)  \label{eq:gff6}
        \end{align}
\begin{align}
       \Pi^{kj}(\Gamma^0, \vec q) =&    A_{20}\,\mathcal{K}\,\frac{q_k\,q_j}
          {4\,m_N} + B_{20}\,\mathcal{K}\,\left(-\frac{
            q_k\,q_j \,E_N}{8\,m_N^2} +
          \frac{q_k\,q_j}{8\,m_N} \right)  +
          C_{20}\,\mathcal{K}\,\left( \frac{q_k\,
            q_j\,E_N}{2\,m_N^2} +
          \frac{q_k\,q_j}{2\,m_N} \right) , \label{eq:gff7}\\
          \Pi^{kj}(\Gamma^n, \vec q) =&  A_{20}\,\mathcal{K}\,\left(-i\,\frac{\epsilon_{k\,n\,0\,\rho}\,
            q_j\,q_\rho }{8\,m_N}-i\,\frac{\epsilon_{j\,n\,0\,\rho}\,
            q_k\,q_\rho }{8\,m_N}\right) +
          B_{20}\,\mathcal{K}\,\left(-i\,\frac{\epsilon_{k\,n\,0\,\rho}\,
            q_j\,q_\rho }{8\,m_N}-i\,\frac{\epsilon_{j\,n\,0\,\rho}\,
            q_k\,q_\rho }{8\,m_N}\right) .\label{eq:gff8}
\end{align}
\end{widetext}

\section{Evolution and conversion functions}
\label{app:evolution}

In this Appendix, we provide the perturbative expressions for the evolution and conversion functions relevant to the renormalization of EMT operators. The evolution functions are expressed in terms of the $\beta$-function and the anomalous dimension $\gamma_{ij}$ ($i,j \in \{q,g,c\}$)
of the EMT operators in both the RI$'$-MOM and $\overline{\rm MS}$ renormalization schemes. 

Using standard notation, we define (in terms of the renormalization scale $\mu$) the Callan-Symanzik equations satisfied by the renormalized gauge coupling $g^R$ and the renormalized EMT operators in an arbitrary renormalization scheme, denoted by ``R'',\footnote{It is understood that in Eqs. \eqref{gamma_ij} and \eqref{scheme-conversion}, the singlet operator $T^{\mu\nu;s}_{q}$ is employed when $i=q$ or $j=q$. To distinguish from the nonsinglet anomalous dimension, we use the notation $\gamma_{ij}^{s,R}$. \phantom{throughout the remainder of this section.}}
\begin{eqnarray}
    \mu \frac{d}{d\mu} g^R (\mu) &=& \beta^R (g^R(\mu)), \label{beta}\\
    \mu \frac{d}{d\mu} (T^{\mu\nu;ns}_{q})^R &=& -\gamma_{qq}^{R} (g^R(\mu)) \ (T^{\mu\nu;ns}_{q})^R, \\
    \mu \frac{d}{d\mu} (T^{\mu\nu}_{i})^R &=& -\sum_j {\gamma}_{ij}^{R} (g^R(\mu)) \ (T^{\mu\nu}_{j})^R. \label{gamma_ij}
\end{eqnarray}
The perturbative expansions of the $\beta$ function and the anomalous dimensions $\gamma_{ij}$ are defined below in terms of $a (\mu)\equiv g^2/(16 \pi^2)$,
\begin{eqnarray}
    \beta^R (g) &=& -g^3 \, [\beta^R_0 + \beta^R_1 a + \beta^R_2 a^2 + \beta^R_3 a^3 + \mathcal{O} (a^4)], \quad \\
    \gamma^R_{ij} (g) &=& a \, [(\gamma_0)^R_{ij} + (\gamma_1)^R_{ij} \, a + (\gamma_2)^R_{ij} \, a^2 + (\gamma_3)^R_{ij} \, a^3 \nonumber \\
    && \quad  + \mathcal{O} (a^4)]. \label{pertexpansion}
\end{eqnarray}
The coefficients $\beta^R_k$ and $(\gamma^R_k)_{ij}$ in the RI$'$-MOM and $\overline{\rm MS}$ schemes are obtained by combining the results of~\cite{Gracey:2003mr,Baikov:2015tea,Chetyrkin:2017bjc,Panagopoulos:2020qcn,Gehrmann:2023ksf}. The following scheme-conversion formulae, derived from Eqs.(\ref{beta}--\ref{gamma_ij}), are used:
\begin{eqnarray}
    \beta^{{\rm RI}'} (g^{{\rm RI}'}(\mu)) &=& \beta^{\overline{\rm MS}} (g^{\overline{\rm MS}}(\mu)) \frac{\partial g^{{\rm RI}'}(\mu)}{\partial g^{\overline{\rm MS}}(\mu)}, \\
    \gamma^{{\rm RI}'}_{qq} (g^{{\rm RI}'}(\mu)) &=& \gamma^{\overline{\rm MS}}_{qq} (g^{\overline{\rm MS}}(\mu)) \nonumber \\
    && \hspace{-1.5cm} + g^{\overline{\rm MS}}(\mu) \,\beta^{\overline{\rm MS}} (g^{\overline{\rm MS}}(\mu)) \,\frac{\partial \ln{C_{qq}^{\overline{\rm MS},{\rm RI}'} (\mu,\mu)}}{\partial g^{\overline{\rm MS}}(\mu)}, \\
    \gamma^{{\rm RI}'}_{ij} (g^{{\rm RI}'}(\mu)) &=&  \nonumber \\
    && \hspace{-2cm}\overline{C}_{ik}^{{\rm RI}',\overline{\rm MS}} (\mu,\mu) \, \gamma^{\overline{\rm MS}}_{kl} (g^{\overline{\rm MS}}(\mu)) \,[\overline{C}^{{\rm RI}',\overline{\rm MS}} (\mu,\mu)]^{-1}_{lj} \\
    && \hspace{-2cm} - g^{\overline{\rm MS}}(\mu) \,\beta^{\overline{\rm MS}} (g^{\overline{\rm MS}}(\mu)) \times \nonumber \\
    && \hspace{-1.3cm} \frac{\partial \overline{C}_{ik}^{{\rm RI}',\overline{\rm MS}} (\mu,\mu)}{\partial g^{\overline{\rm MS}}(\mu)} [\overline{C}^{{\rm RI}',\overline{\rm MS}} (\mu,\mu)]^{-1}_{kj}, \label{scheme-conversion}
\end{eqnarray}
where $C_{qq}^{\overline{\rm MS},{\rm RI}'}$ and $\overline{C}_{ij}^{{\rm RI}',\overline{\rm MS}}$ are defined in Eqs. (\ref{Cqq}--\ref{Cbar_ij}). To simplify the expressions, we give the coefficients in the Landau gauge and in SU(3). 

\begin{align}
\beta_0^{\overline{\rm MS}} &= \beta_0^{{\rm RI}'} = 11 - \frac{2}{3}N_f, \\
\beta_1^{\overline{\rm MS}} &= \beta_1^{{\rm RI}'} = 102 - \frac{38}{3}N_f, \\
\beta_2^{\overline{\rm MS}} &= \beta_2^{{\rm RI}'} = \frac{2857}{2} - \frac{5033}{18}N_f + \frac{325}{54}N_f^2, \\
\beta_3^{\overline{\rm MS}} &= \beta_3^{{\rm RI}'} = \frac{149753}{6} + 3564\,\zeta(3) \nonumber \\
&\quad - \left(\frac{1078361}{162} + \frac{6508}{27}\zeta(3)\right)N_f \nonumber \\
&\quad + \left(\frac{50065}{162} + \frac{6472}{81}\zeta(3)\right)N_f^2
+ \frac{1093}{729}N_f^3.
\end{align}
\begin{widetext}
\begin{align}
(\gamma_0)_{qq}^{\overline{\rm MS}}
&= (\gamma_0)_{qq}^{{\rm RI}'} = \frac{64}{9}, \\[0.5ex]
(\gamma_1)_{qq}^{\overline{\rm MS}}
&= \frac{23488}{243} - \frac{512}{81}N_f, \\[0.5ex]
(\gamma_2)_{qq}^{\overline{\rm MS}}
&= \frac{11028416}{6561}
+ \frac{2560}{81}\zeta(3)  - \left(\frac{334400}{2187}
+ \frac{2560}{27}\zeta(3)\right)N_f
- \frac{1792}{729}N_f^2, \\[0.5ex]
(\gamma_3)_{qq}^{\overline{\rm MS}}
&= \frac{6200738288}{177147}
+ \frac{52121728}{6561}\zeta(3) - \frac{14080}{27}\zeta(4)
- \frac{2498560}{243}\zeta(5) \nonumber\\
&\quad
+ \Big(
-\frac{334439344}{59049}
- \frac{12645952}{2187}\zeta(3)  + \frac{129280}{81}\zeta(4)
+ \frac{29440}{9}\zeta(5)
\Big)N_f \nonumber\\
&\quad
+ \left(
\frac{2169808}{19683}
+ \frac{5120}{27}\zeta(3)
- \frac{2560}{27}\zeta(4)
\right)N_f^2  + \left(
-\frac{8192}{6561}
+ \frac{1024}{243}\zeta(3)
\right)N_f^3, \\[0.5ex]
(\gamma_1)_{qq}^{{\rm RI}'}
&= \frac{48040}{243} - \frac{112}{9}N_f, \\[0.5ex]
(\gamma_2)_{qq}^{{\rm RI}'}
&= \frac{59056304}{6561}
- \frac{103568}{81}\zeta(3) - \left(
\frac{2491456}{2187}
+ \frac{416}{27}\zeta(3)
\right)N_f
+ \frac{19552}{729}N_f^2, \\[0.5ex]
(\gamma_3)_{qq}^{{\rm RI}'}
&= \frac{358802599559}{708588}
- \frac{859904531}{6561}\zeta(3) + \frac{4710620}{243}\zeta(5) \nonumber \\
& \quad + \Big(
-\frac{11334165445}{118098}
+ \frac{4974662}{729}\zeta(3) 
+ \frac{119320}{81}\zeta(5)
\Big)N_f \nonumber \\
& \quad + \left(
\frac{96897860}{19683}
+ \frac{4256}{81}\zeta(3)
\right)N_f^2 - \frac{432160}{6561}N_f^3, \\
(\gamma_0)_{qq}^{s,\overline{\rm MS}}
&= (\gamma_0)_{qq}^{s,{\rm RI}'} = \frac{64}{9}, \\[0.5ex]
(\gamma_1)_{qq}^{s,\overline{\rm MS}}
&= \frac{23488}{243} - \frac{832}{81}N_f, \\[0.5ex]
(\gamma_2)_{qq}^{s,\overline{\rm MS}}
&= \frac{11028416}{6561}
+ \frac{2560}{81}\zeta(3)
+ \left(
-\frac{269776}{2187}
- \frac{5120}{27}\zeta(3)
\right)N_f
- \frac{2272}{243}N_f^2,
\\[1ex]
(\gamma_1)_{qq}^{s,{\rm RI}'}
&= \frac{48040}{243} - \frac{1424}{81}N_f, \\[0.5ex]
(\gamma_2)_{qq}^{s,{\rm RI}'}
&= \frac{59056304}{6561}
- \frac{103568}{81}\zeta(3)
+ \left(
-\frac{5600992}{3645}
+ \frac{752}{15}\zeta(3)
\right)N_f
+ \frac{13712}{243}N_f^2, \\[0.5ex]
\gamma_{gq}^{\overline{\rm MS}} &= -\gamma_{qq}^{s,\overline{\rm MS}}, \\[0.5ex]
(\gamma_0)_{gq}^{{\rm RI}'}
&= -\frac{64}{9}, \\[0.5ex]
(\gamma_1)_{gq}^{{\rm RI}'}
&= -\frac{37420}{243} + \frac{896}{81}N_f, \\[0.5ex]
(\gamma_2)_{gq}^{{\rm RI}'}
&= -\frac{213070066}{32805}
+ \frac{191816}{135}\zeta(3)
+ \left(
\frac{252364}{243}
- \frac{1136}{27}\zeta(3)
\right)N_f
- \frac{8960}{243}N_f^2, \\[0.5ex]
\gamma_{cq}^{\overline{\rm MS}} &= 0, \\[0.5ex]
(\gamma_0)_{cq}^{{\rm RI}'}
&= 0, \\[0.5ex]
(\gamma_1)_{cq}^{{\rm RI}'}
&= \frac{68}{27}
+ \frac{16}{9}N_f, \\[0.5ex]
(\gamma_2)_{cq}^{{\rm RI}'}
&= -\frac{1490831}{3645}
- \frac{57608}{405}\zeta(3)
+ \left(
\frac{49028}{405}
- \frac{1088}{135}\zeta(3)
\right)N_f
- \frac{512}{81}N_f^2,
\end{align}
\begin{align}
(\gamma_0)_{gg}^{\overline{\rm MS}}
&= (\gamma_0)_{gg}^{{\rm RI}'} = \frac{4}{3}N_f, \\[0.5ex]
(\gamma_1)_{gg}^{\overline{\rm MS}}
&= \frac{1222}{81}N_f, \\[0.5ex]
(\gamma_2)_{gg}^{\overline{\rm MS}}
&= \left(
\frac{670871}{2187}
- \frac{5200}{27}\zeta(3)
\right)N_f
- \frac{17660}{729}N_f^2,
\\[1ex]
(\gamma_1)_{gg}^{{\rm RI}'}
&= -\frac{173}{4}
+ \frac{3433}{81}N_f
- \frac{40}{27}N_f^2, \\[0.5ex]
(\gamma_2)_{gg}^{{\rm RI}'}
&= \frac{116317}{60}
- \frac{25461}{10}\zeta(3)
+ \left(
\frac{14176921}{10935}
+ \frac{89062}{135}\zeta(3)
\right)N_f 
+ \left(
-\frac{102640}{729}
- \frac{464}{9}\zeta(3)
\right)N_f^2
+ \frac{400}{243}N_f^3, \nonumber \\
\gamma_{qg}^{\overline{\rm MS}} &= -\gamma_{gg}^{\overline{\rm MS}}, \\[0.5ex]
(\gamma_0)_{qg}^{{\rm RI}'}
&= -\frac{4}{3} N_f, \\[0.5ex]
(\gamma_1)_{qg}^{{\rm RI}'}
&= -\frac{2389}{81}N_f
+ \frac{40}{27}N_f^2, \\[0.5ex]
(\gamma_2)_{qg}^{{\rm RI}'}
&= \left(
-\frac{9251356}{10935}
- \frac{23374}{45}\zeta(3)
\right)N_f 
+ \left(
\frac{405329}{3645}
+ \frac{2248}{45}\zeta(3)
\right)N_f^2
- \frac{400}{243}N_f^3, \\[0.5ex]
\gamma_{cg}^{\overline{\rm MS}} &= 0, \\[0.5ex]
(\gamma_0)_{cg}^{{\rm RI}'}
&= 0, \\[0.5ex]
(\gamma_1)_{cg}^{{\rm RI}'}
&= -\frac{399}{4}
+ \frac{16}{9}N_f, \\[0.5ex]
(\gamma_2)_{cg}^{{\rm RI}'}
&= -\frac{759371}{80}
+ \frac{25461}{10}\zeta(3)
+ \left(
\frac{909943}{972}
- \frac{3788}{27}\zeta(3)
\right)N_f 
+ \left(
-\frac{7471}{405}
+ \frac{8}{5}\zeta(3)
\right)N_f^2, \\[0.5ex]
(\gamma_0)_{qc}^{\overline{\rm MS}} &= (\gamma_0)_{qc}^{{\rm RI}'}=0, \\[0.5ex]
(\gamma_1)_{qc}^{\overline{\rm MS}}
&= -\frac{8}{3}N_f, \\[0.5ex]
(\gamma_2)_{qc}^{\overline{\rm MS}}
&= \left(
-\frac{89}{243}
- \frac{11}{3}\zeta(3)
\right)N_f
- \frac{67}{162}N_f^2,
\\[1ex]
(\gamma_1)_{qc}^{{\rm RI}'}
&= \frac{35}{27}N_f
- \frac{4}{9}N_f^2, \\[0.5ex]
(\gamma_2)_{qc}^{{\rm RI}'}
&= \left(
\frac{163093}{729}
- \frac{794}{3}\zeta(3)
\right)N_f 
+ \left(
-\frac{26219}{810}
+ \frac{136}{5}\zeta(3)
\right)N_f^2
+ \frac{80}{81}N_f^3, \\[0.5ex]
(\gamma_0)_{gc}^{\overline{\rm MS}} &= (\gamma_0)_{gc}^{{\rm RI}'}=
= -3, \\[0.5ex]
(\gamma_1)_{gc}^{\overline{\rm MS}}
&= -\frac{45}{2}
+ \frac{14}{3}N_f, \\[0.5ex]
(\gamma_2)_{gc}^{\overline{\rm MS}}
&= -\frac{15355}{288}
- \frac{9}{4}\zeta(3)
+ \left(
\frac{9937}{1944}
+ \frac{25}{3}\zeta(3)
\right)N_f
+ \frac{11}{18}N_f^2,
\\[1ex]
(\gamma_1)_{gc}^{{\rm RI}'}
&= -\frac{167}{2}
+ \frac{280}{27}N_f, \\[0.5ex]
(\gamma_2)_{gc}^{{\rm RI}'}
&= -\frac{3191231}{1440}
- \frac{14931}{20}\zeta(3)
+ \left(
\frac{2779663}{5832}
+ \frac{808}{3}\zeta(3)
\right)N_f 
+ \left(
-\frac{2855}{162}
- 24\zeta(3)
\right)N_f^2. \\[0.5ex]
\gamma_{cc}^{\overline{\rm MS}} &= -(\gamma_{qc}^{\overline{\rm MS}}+\gamma_{gc}^{\overline{\rm MS}}), \\[0.5ex]
(\gamma_0)_{cc}^{{\rm RI}'}
&= 3, \\[0.5ex]
(\gamma_1)_{cc}^{{\rm RI}'}
&= \frac{87}{4} - N_f, \\[0.5ex]
(\gamma_2)_{cc}^{{\rm RI}'}
&= -\frac{1667929}{1440}
+ \frac{14931}{20}\zeta(3)
+ \left(
\frac{73817}{648}
- \frac{14}{3}\zeta(3)
\right)N_f 
+ \left(
-\frac{1658}{405}
- \frac{16}{5}\zeta(3)
\right)N_f^2.
\end{align}

\end{widetext}

In the case of nonsinglet operators, which are multiplicatively renormalizable, the evolution function $U^R (\mu_2,\mu_1)$ from scale $\mu_1$ to scale $\mu_2$ in the scheme $R$ is defined by
\begin{eqnarray}
    {(T^{\mu\nu;ns}_{q})}^R (\mu_2) &=& U^{R}_{qq}(\mu_2,\mu_1) \, {(T^{\mu\nu;ns}_{q})}^R (\mu_1), \label{evol_nomix}\\ 
    U^{R}_{qq}(\mu_2,\mu_1) &=& \frac{W_{qq}^R (a^R(\mu_1))}{W^R_{qq}(a^R(\mu_2))}, \\
    W^R_{qq}(x) &=& \exp \left[ \int^{x} dx' \frac{\gamma^R_{qq} (x')}{\beta^R (x')}\right].
\end{eqnarray}
For the singlet operators, where mixing is present, the evolution function takes a $3\times3$ matrix form~\cite{Papinutto:2016xpq},
\begin{eqnarray}
    {(T^{\mu\nu}_{i})}^R (\mu_2) &=& U^{R}_{ij}(\mu_2,\mu_1) \, {(T^{\mu\nu}_{j})}^R (\mu_1), \\
    U^{R}_{ij}(\mu_2,\mu_1) &=& [W^R (a^R(\mu_2))]^{-1}_{ik} [W^R(a^R(\mu_1))]_{kj}, \\
    W^R (x) &=& x^{-\frac{\gamma_0^R}{2 \beta_0^R}} \Big[\openone + x J_1^R + x^2 J_2^R + \mathcal{O}(x^3)\Big], \qquad \,\label{evol_mix}
\end{eqnarray}
where $J_1^R$ and $J_2^R$ can be obtained by solving the following matrix equations,
\begin{align}
\label{eq:J1}
2 \beta^R_0 J^R_1-\left[\gamma^R_0,J^R_1\right] &= \frac{\beta^R_1}{\beta^R_0}\gamma^R_0-\gamma^R_1\,,\\
\label{eq:J2}
4\beta^R_0 J^R_2-\left[\gamma^R_0, J^R_2\right]
&= J^R_1\left(\frac{\beta^R_1}{\beta^R_0}\gamma^R_0-\gamma^R_1\right) \nonumber \\
&\hspace{-1.5cm}+\left(\frac{\beta^R_2}{\beta^R_0}-\frac{{\beta^R_1}^2}{{\beta^R_0}^2}\right)\gamma^R_0+\frac{\beta^R_1}{\beta^R_0}\gamma^R_1-\gamma^R_2\,.
\end{align}

For completeness, we also provide the perturbative expressions for the conversion coefficients $C_{qq}^{\overline{\rm MS},{\rm RI}'} (\mu_2,\mu_1)$ and $\overline{C}_{ij}^{{\rm RI}',\overline{\rm MS}} (\mu_2,\mu_1)$ from the RI$'$-MOM to the $\overline{\rm MS}$ scheme, taken from Refs.~\cite{Gracey:2003mr,Panagopoulos:2020qcn}. In our analysis, we set $\mu_1=\mu_2$.
\begin{eqnarray}
C_{qq}^{\overline{\rm MS},{\rm RI}'} &=& 1 + (c_1)_{qq} \, a^{\overline{\rm MS}} + (c_2)_{qq} \, (a^{\overline{\rm MS}})^2 \nonumber \\
&& \quad + (c_3)_{qq} \, (a^{\overline{\rm MS}})^3 + \mathcal{O} ((a^{\overline{\rm MS}})^4), \\
\overline{C}_{ij}^{{\rm RI}',\overline{\rm MS}} &=& 1 + (\overline{c}_1)_{ij} \, a^{\overline{\rm MS}} + (\overline{c}_2)_{ij} \, (a^{\overline{\rm MS}})^2 \nonumber \\
&& \quad + \mathcal{O} ((a^{\overline{\rm MS}})^3),
\end{eqnarray}
where
\begin{align}
(c_1)_{qq}
&= -\frac{124}{27}, \\[0.5ex]
(c_2)_{qq}
&= -\frac{98072}{729}
+ \frac{268}{9}\zeta(3)
+ \frac{2668}{243}N_f, \\[0.5ex]
(c_3)_{qq}
&= -\frac{849683327}{157464}
+ \frac{7809041}{4374}\zeta(3)
- \frac{640}{81}\zeta(4) \nonumber \\
& - \frac{36410}{81}\zeta(5) 
\nonumber \\
& + \left(
\frac{14433520}{19683}
- \frac{4184}{81}\zeta(3)
+ \frac{640}{27}\zeta(4)
\right)N_f \notag\\
&
+ \left(
-\frac{105992}{6561}
- \frac{256}{243}\zeta(3)
\right)N_f^2, \\[0.5ex]
(\bar{c}_1)^s_{qq} 
&= \frac{124}{27}, \\[0.5ex]
(\bar{c}_1)_{qg} 
&= -\frac{4}{9} N_f, \\[0.5ex]
(\bar{c}_1)_{qc} 
&= \frac{1}{3} N_f, \\[0.5ex]
(\bar{c}_1)_{gq} 
&= -\frac{88}{27}, \\[0.5ex]
(\bar{c}_1)_{gg} 
&= -\frac{5}{4}+\frac{10}{9} N_f, \\[0.5ex]
(\bar{c}_1)_{gc} 
&= -\frac{5}{2}, \\[0.5ex]
(\bar{c}_1)_{cq} 
&= -\frac{4}{3}, \\[0.5ex]
(\bar{c}_1)_{cg} 
&= -\frac{21}{4}, \\[0.5ex]
(\bar{c}_1)_{cc} 
&= -\frac{3}{4}, \\[0.5ex]
(\bar{c}_2)^s_{qq} &= \frac{37816}{243} - \frac{268}{9}\zeta(3)
- \frac{4420}{243}N_f, \\[0.5ex]
(\bar{c}_2)_{qg} &= \left(
-\frac{2413}{270}
- \frac{272}{15}\zeta(3)
\right) N_f, \\[0.5ex]
(\bar{c}_2)_{qc} &= 
+ \left(
\frac{11963}{1620}
- \frac{36}{5}\zeta(3)
\right)N_f -\frac{10}{27}N_f^2,
\\[0.5ex]
(\bar{c}_2)_{gq} &= -\frac{23197}{243} + \frac{212}{9}\zeta(3)
+ \frac{3208}{243}N_f, \\[0.5ex]
(\bar{c}_2)_{gg} &= \frac{2251}{30}
- \frac{621}{10}\zeta(3)
+ \left(
\frac{49}{3}
+ \frac{58}{3}\zeta(3)
\right)N_f, \\[0.5ex]
(\bar{c}_2)_{gc} &= -\frac{2869}{120}
- \frac{423}{20}\zeta(3)
+ \left(
\frac{1085}{324}
+ 6\zeta(3)
\right)N_f,
\\[0.5ex]
(\bar{c}_2)_{cq} &= -\frac{2806}{81} + \frac{56}{9}\zeta(3)
+ \frac{64}{27}N_f, \\[0.5ex]
(\bar{c}_2)_{cg} &= -\frac{16411}{80} + \frac{621}{10}\zeta(3)
+ \left(
\frac{5711}{540}
- \frac{6}{5}\zeta(3)
\right)N_f, \\[0.5ex]
(\bar{c}_2)_{cc} &= -\frac{701}{20}
+ \frac{423}{20}\zeta(3)
+ \left(
\frac{223}{180}
+ \frac{6}{5}\zeta(3)
\right)N_f.
\end{align}

\end{document}